\documentclass[useAMS,usenatbib]{mn2e}
\usepackage{times,mathptm}
\usepackage{lscape}
\usepackage{psfig}
\usepackage{subfigure}
\usepackage{multirow}

\usepackage{psfig}
\usepackage{times}
\usepackage{amssymb}
\usepackage{graphicx}
\usepackage{graphics}
\usepackage{rotating}
\usepackage{lscape}

\def\apj{ApJ}
\def\apjl{ApJL}
\def\mnras{MNRAS}

\def\araa{ARAA}
\def\aap{A\&A}
\def\aj{AJ}
\def\apjs{APJS}

\def\nat{Nature}

\def\gs{\mathrel{\raise0.35ex\hbox{$\scriptstyle >$}\kern-0.6em
\lower0.40ex\hbox{{$\scriptstyle \sim$}}}}
\def\ls{\mathrel{\raise0.35ex\hbox{$\scriptstyle <$}\kern-0.6em
\lower0.40ex\hbox{{$\scriptstyle \sim$}}}}

\def\kms{\,\hbox{km}\,\hbox{s}^{-1}}

\def\Wm2{\,\hbox{W}\,\hbox{m}^{-2}}
\def\gsim{\mathrel{\raise0.35ex\hbox{$\scriptstyle >$}\kern-0.6em\lower0.40ex\hbox{{$\scriptstyle \sim$}}}}
\def\lsim{\mathrel{\raise0.35ex\hbox{$\scriptstyle <$}\kern-0.6em\lower0.40ex\hbox{{$\scriptstyle \sim$}}}}
\def\ltsima{$\; \buildrel < \over \sim \;$}
\def\simlt{\lower.5ex\hbox{\ltsima}}
\def\gtsima{$\; \buildrel > \over \sim \;$}
\def\simgt{\lower.5ex\hbox{\gtsima}}

\begin{document}

\title[IFU spectroscopy of high-$z$ ULIRGs hosting AGN]{
Energetic galaxy-wide outflows in high-redshift ultra-luminous
  infrared galaxies hosting AGN activity}

\author[C.\ M.\ Harrison et al.]
{ \parbox[h]{\textwidth}{ 
C.\ M.\ Harrison,$^{\! 1\, *}$ D.\ M.\ Alexander,$^{\! 1}$ A.\ M.\
Swinbank,$^{\! 2}$ Ian Smail,$^{\! 2}$ S.\ Alaghband-Zadeh,$^{\! 3}$
F.\ E.\ Bauer,$^{\! 4,5}$ S.\ C.\
Chapman,$^{\! 3}$ A.\ Del Moro,$^{\! 1}$ R.\ C.\ Hickox,$^{\! 6,1}$
R.\ J.\ Ivison,$^{\! 7,8}$ Kar\'in Men\'endez-Delmestre,$^{\! 9,10}$
J.\ R.\ Mullaney$^{\! 11}$ and N.\ P.\ H.\ Nesvadba$^{\! 12}$
}
\vspace*{6pt} \\
$^1$Department of Physics, Durham University, South Road, Durham, DH1 3LE, UK \\
$^2$Institute for Computational Cosmology, Durham University, South
Road, Durham, DH1 3LE, UK \\
$^3$Institute of Astronomy, Madingley Road, Cambridge, CB3 0HA, UK\\
$^4$Space Science Institute, 4750 Walnut Street, Suite 205, Boulder,
CO 80301, USA \\
$^5$Pontifica Universidad Cat\'olica de Chile, Departamento de
Astronom\'ica y Astrof\'isica, Casilla 306, Santiago 22, Chile \\
$^{6}$Department of Physics and Astronomy, Dartmouth College, 6127
Wilder Laboratory, Hanover, NH 03755, USA\\
$^{7}$UK Astronomy Technology Centre, Royal Observatory, Blackford
Hill, Edinburgh, EH9 3HJ, UK\\
$^{8}$Institute for Astronomy, University of Edinburgh, Blackford
Hill, Edinburgh, EH9 3HJ, UK\\ 
$^{9}$Observat\'orio do Valongo, Universidade Federal do Rio de
Janeiro, Ladeira Pedro Ant\^onio, 43, Sa\'uder CEP 20080-090 Rio de
Janeiro, RJ, Brazil\\
$^{10}$California Institute of Technology, MC 249-17, Pasadena, CA
91125, USA\\
$^{11}$Laboratoire AIM-Paris-Saclay, CEA/DSM/Irfu - CNRS, Universit\'e
Paris Diderot, CE-Saclay, pt courrier 121, 91191 Gif-sur-Yvette,
France\\
$^{12}$Institut d'Astrophysique Spatiale, Universit\'e Paris-Sud,
Bat. 120-121, 91405 Orsay, France \\
$^*$Email: c.m.harrison@durham.ac.uk \\
}
\maketitle
\begin{abstract}
We present integral field spectroscopy observations, covering the
[O~{\sc iii}]$\lambda\lambda4959,5007$ emission-line doublet of eight high-redshift
($z$=1.4--3.4) ultra-luminous infrared galaxies (ULIRGs) that host Active Galactic
Nuclei (AGN) activity, including known sub-millimetre luminous galaxies (SMGs).  The targets have moderate radio luminosities that are
typical of high-redshift ULIRGs ($L_{1.4\rm{GHz}}$=10$^{24}$--10$^{25}$\,W\,Hz$^{-1}$) and therefore
are not radio-loud AGN. We de-couple kinematic components due to
the galaxy dynamics and mergers from those due to outflows. We
find evidence in the four most luminous systems ($L_{\rm{[O III]}}\gsim10^{43}$\,erg\,s$^{-1}$)
for the signatures of large-scale energetic outflows: extremely broad [O~{\sc iii}] emission (FWHM$\approx$700--1400\,km\,s$^{-1}$)
  across $\approx$4--15\,kpc, with high
  velocity offsets from the systemic redshifts (up to
  $\approx$850\,km\,s$^{-1}$). The four less luminous systems have
  lower quality data displaying weaker evidence for spatially extended outflows.
  We estimate that these outflows are potentially
 depositing energy into their host galaxies at considerable rates
 ($\dot{E}\approx$10$^{43}$--10$^{45}$\,erg\,s$^{-1}$); however, due
   to the lack of constraints on the density of the outflowing
   material and the structure of the outflow, these estimates should
   be taken as illustrative only. Based on the measured maximum
   velocities ($v_{\rm{max}}\approx$\,400-1400\,km\,s$^{-1}$) the outflows observed are likely to
   unbind some fraction of the gas from their host galaxies, but are unlikely to completely
   remove gas from the galaxy haloes. By using a combination of
   energetic arguments and a comparison to ULIRGs without clear
   evidence for AGN activity, we show that the AGN activity could be the dominant power
   source for driving all of the observed outflows, although star formation may also play a significant role
   in some of the sources.
\end{abstract}


\section{Introduction}
Many of the most successful models of galaxy formation require energetic
outflows over galaxy scales (i.e. $\approx$1--10\,kpc) to reproduce properties of local massive
galaxies (e.g., black-hole-spheroid mass relationship;
the bright-end of the galaxy luminosity function; e.g., \citealt{Silk1998}; \citealt{Springel2005}; \citealt{DiMatteo2005};
\citealt{Hopkins2006,Hopkins2008a};
\citealt{DeBuhr2012}). Observations show that galaxy-wide
outflows can be powered by star formation and/or active galactic nuclei
(AGN) activity (e.g., \citealt{Heckman1990}; \citealt{Crenshaw2003}; \citealt{Rupke2005a,Rupke2005b}); however, most models predict that AGN-driven
outflows are required in the most massive galaxies. These
simulated outflows inhibit further black-hole accretion and star formation by injecting considerable kinetic energy into
the interstellar medium (ISM) of the host galaxies. While
theoretically attractive, these models can only be judged to be successful
if supported by observational data.

AGN-driven outflows are initially launched from the accretion
disk or dusty torus surrounding the black hole (BH), either in the form of a radio
jet/lobe or a radiatively-driven wind. Radio jets/lobes are
thought to be most effective at {\em heating} gas in radio-loud
systems through the so-called {\em radio mode} (\citealt{Croton2006}; \citealt{Bower2006}) and evidence of this is observed in low-redshift
massive galaxies, where mechanical heating appears to be preventing gas from
cooling (e.g., \citealt{Best2005,Best2006}; \citealt{Smolcic2009}; \citealt{Danielson2012}). In
contrast, many models implement the more rapid {\it quasar mode},
where the radiation field produced by AGN launches a high-velocity wind which
sweeps up gas in the ISM and results in a kpc-scale ``outflow''  (e.g., \citealt{Springel2005}; \citealt{King2005}; \citealt{DiMatteo2005};
\citealt{Hopkins2006}; \citealt{DeBuhr2012}). These radiatively-driven
outflows are predicted to be most prevalent at high redshift, ($z\approx$\,2), in luminous AGN with high
accretion rates (e.g., \citealt{DiMatteo2005}; \citealt{King2005}).

Observationally, AGN-driven winds are known to be prevalent in luminous
AGN at low and high redshift: X-ray and UV
absorption-line spectroscopy reveal high-velocity winds of up to $v\approx$\,0.1$c$ (e.g., \citealt{Reeves2003};
\citealt{Blustin2003}; \citealt{Trump2006};
\citealt{Gibson2009}; \citealt{Gofford2011}; \citealt{Page2011}) and may be a ubiquitous property of all
AGN (\citealt{Ganguly2008}). However, these high-velocity winds are
most likely to be produced close to the accretion disk ($<$ 1\,pc scale;
e.g., \citealt{Crenshaw2003}; \citealt{Tombesi2012}). For these winds to have
a global effect on the host galaxy, they must couple to the
ISM and drive the gas over galaxy-wide ($\approx$~1--10~kpc) scales. To test the
impact of AGN-driven outflows on the formation and evolution of
galaxies, it is therefore necessary to search for kpc-scale energetic outflows. 

Spatially resolved spectroscopy provides a particularly direct method to search for and characterise
kpc-scale outflows. Such observations of low-redshift
galaxies have identified kpc-scale outflows in both ionised (e.g.,
\citealt{Holt2008}; \citealt{Fu2009}; \citealt{Rupke2011}; \citealt{Westmoquette2012}) and
molecular gas (e.g., \citealt{Feruglio2010}; \citealt{Alatalo2011}). At high redshift, where accretion-related outflows are
predicted to be most prevalent, integral-field unit (IFU) observations of high-redshift radio galaxies
(HzRGs, at $z\approx$~2--3; \citealt{Nesvadba2006,Nesvadba2008})
have showed that kpc-scale energetic outflows are
present in at least a fraction of the high-redshift galaxy
population. AGN-driven outflows were revealed in these systems through
the presence of broad (FWHM$>$1000\,km\,s$^{-1}$), high-velocity ($>$300\,km\,s$^{-1}$) and
spatially extended [O~{\sc iii}] emission.\footnote{Emission from
[O~{\sc iii}] cannot be produced in high density environments, such as the broad-line
region (BLR) of AGN, without being collisionally de-excited (e.g.,
\citealt{Osterbrock1989}; \citealt{Robson1996}). Luminous ($L>10^{42}$~erg~s$^{-1}$),
broad (FWHM$\simgt$~1000~km$^{-1}$), and extended ($\approx$
a few kiloparsecs) [O~{\sc iii}] emission is therefore not associated with
the BLR.} With implied velocities of greater than a few hundred
km\,s$^{-1}$ these outflows could potentially drive gas out of the
host galaxy, inhibiting future BH growth and star formation. Recent
IFU observations of the [O~{\sc iii}], H$\alpha$ and [C~{\sc ii}]
emission lines in high-redshift quasars also reveal similar outflow
signatures (\citealt{CanoDiaz2012}; \citealt{Maiolino2012}). Due to the small number of
sources and the biased selection of the high-redshift galaxies studied with IFUs so far, the prevalence and significance
of such galaxy-wide outflows remains poorly constrained.

Potentially the best place to search for AGN-driven outflows are
high-redshift far-IR (FIR) luminous galaxies with
$L_{\rm{FIR}}>10^{12}$L$_ {\sun}$ (Ultra-Luminous Infra-Red Galaxies -
ULIRGs), such as sub-millimetre galaxies
(SMGs). These systems are the most intensely star-forming
galaxies known with star formation rates (SFRs) of
$\gsim$100--1000\,M$_{\sun}$\,yr$^{-1}$ and are thought to
represent a rapid star-forming phase that every massive galaxy goes
through (e.g., \citealt{Swinbank2006}; \citealt{Tacconi2008}). They have a peak space density at $z\approx$\,2--2.5 (\citealt{Chapman2004,Chapman2005};
\citealt{Wardlow2011}) which roughly coincides
with the peak epoch of BH
growth (e.g., \citealt{Cattaneo2003}; \citealt{Hopkins2007a};
\citealt{Silverman2008}). Using optical,
mid-infrared and X-ray diagnostics, high-redshift ULIRGs are found to have emission dominated by starburst activity
with $\approx$30\%  of SMGs also revealing significant AGN activity
(e.g., \nocite{Alexander2003,Alexander2005,Alexander2008}Alexander
et~al. 2003, 2005, 2008;
\citealt{Chapman2005}; \citealt{Ivison2010};
\citealt{Rafferty2011}). Indeed, clustering analysis
of SMGs and quasars supports the idea that starburst and AGN activity
occur in the same systems (\citealt{Hickox2012}) and that SMGs may
evolve into massive elliptical galaxies by the present day. Therefore,
high-redshift ULIRGs that host AGN activity could be in transition from a star-formation dominated to an AGN dominated
phase. This transition phase, which is potentially facilitated by
kpc-scale outflows, is an evolutionary
period required by many models (e.g., \citealt{Sanders1988}; \citealt{Silk1998};
\citealt{Springel2005}; \citealt{DiMatteo2005};
\citealt{Hopkins2006,Hopkins2008a}; see \citealt{Alexander2012} for a
review). 

In this paper we present Gemini-North Near-Infrared Integral Field
Spectrometer (NIFS) and VLT Spectrograph for Integral Field
Observations in the Near Infrared (SINFONI) IFU
observations of $z\approx$\,2 ULIRGs with known
optical AGN activity. These sources have moderate radio luminosities
($L_{1.4\rm{GHz}}=$10$^{24}$--10$^{25}$\,W\,Hz$^{-1}$) and can
therefore be considered to be radio-quiet AGN. In \S2 we give details of the
IFU observations and data reduction, in \S4 we present the analysis and results, in \S5 we
discuss the results and their implications for galaxy evolution models
and in \S6 we give our conclusions. We provide background information
and detailed results on individual sources in Appendix A. We have adopted
$H_{0}=71\kms$, $\Omega_{M}=0.27$ and $\Omega_{\Lambda}=0.73$ throughout; in this
cosmology, 1$''$ corresponds to 8.5\,kpc at $z=2.0$.


\section{Observations and data reduction}

\begin{table*}
\begin{center}
{\centerline{\sc Gemini-North NIFS and VLT SINFONI Observations}}
\begin{tabular}{lllcccccccccc}
\hline
\smallskip
ID & Source         & R. A.         &
Dec      & $z_{\rm [OIII]}$   &$S_{1.4}$ & $L_{1.4\rm{GHz}}$ & AGN    & $t_{\rm{exp}}$ &
\multicolumn{2}{c}{Average seeing}&Instrument&Quality\\
    &&(J2000)               &(J2000)     &             &($\mu$Jy)&(10$^{24}$\,W\,Hz$^{-1}$)&  &(ks)&($^{\prime\prime}$)&(kpc)&&\\
(1)&(2)&(3)&(4)&(5)&(6)&(7)&(8)&(9)&(10)&(11)&(12)&(13)\\
\hline
1&SMM\,J0217$-$0503& 02:17:38.68 & $-$05:03:39.5        &2.021   &207$\pm$16   &5.0  &O& 4.8&0.47&4.0&S&1\\
2&SMM\,J0302+0006   & 03:02:27.73 & +00:06:53.5            &1.405   &217$\pm$9     &2.2  &O,M?& 11.4 &0.42&3.6&N&2\\
3&RG\,J0302+0010      & 03:02:58.94 & +00:10:16.3            &2.239   &55$\pm$10     &1.7  &O,U& 9.6   &0.38&3.2&N&1\\
4&RG\,J0332$-$2732   & 03:32:56.75 & $-$27:32:06.3        &2.315   &160$\pm$20   &5.3  &O& 2.4&0.53&4.4&S&1\\ 
5&SMM\,J0943+4700$^{\star}$ & 09:43:04.08 & +47:00:16.2 &3.351   &60                   &4.7  &O,U,M& 14.4 &0.42&3.2&N&1\\
6&SMM\,J1235+6215   & 12:35:49.41 & +62:15:36.9            & 2.199   &93$\pm$6      &2.8  &O,X,M?& 7.8   &0.45&3.8&N&2\\
7&SMM\,J1237+6203   & 12:37:15.97 & +62:03:23.1            & 2.075   &179$\pm$9    &4.6  &X,U,B& 7.8 &0.30&2.5&N&1\\
8&SMM\,J1636+4057   & 16:36:50.43 & +40:57:34.5            & 2.385   &242$\pm$11  &8.7  &O,U,M,B& 13.2 &0.33&2.7&N&1\\
9&SMM\,J2217+0010   & 22:17:37.39 & +00:10:25.1            & (2.610) &179$\pm$19  &7.9  &O,M& 3.6  &0.40 &3.2&N &3\\
10&SMM\,J2217+0017 & 22:17:42.25 & +00:17:02.0            & (2.278) &114$\pm$27  &3.7  &O& 9.6&0.56&4.7&S&3\\   
\hline
\hline
\end{tabular}
\caption{\label{Tab:otherProps}
{\protect\sc Notes:}\protect\\
Column (1): Source ID. Column (2): Source name. Columns (3)--(4): The co-ordinates corresponding to the radio positions from: \protect\cite{Ledlow2002}; \protect\cite{Swinbank2004};
\protect\cite{Chapman2005}; \protect\cite{Miller2008};
\protect\cite{Morrison2010}; \protect\cite{Ivison2011}; Arumugam
et~al. (in prep); R.\,J.\,Ivison
(priv. comm). Column (5) The redshifts based on the narrow
[O~{\protect\sc iii}] emission lines using the galaxy-integrated
spectra presented in Fig.~\protect\ref{Fig:int_specs}. For
SMM\,J2217+0010 and SMM\,J2217+0017 we do not
  detect [O~{\protect\sc iii}] and we therefore quote the H$\alpha$ redshift from
 \protect\cite{Takata2006} and \protect\cite{AlaghbandZadeh2012},
 respectively. Column (6): Radio flux densities taken from
 the same references as for the radio positions. Column (7):
 Rest-frame radio luminosities using the standard
 $k$-correction and a spectral index of $\alpha$=0.8. Column (8): Evidence for AGN activity in the sources on
the basis of: O: optical emission-line ratios
(\protect\citealt{Swinbank2004}; \protect\citealt{Takata2006}; \protect\citealt{AlaghbandZadeh2012}; this work); M: excess in the mid-infrared continuum
from infrared spectra
(\protect\citealt{Valiante2007};
\protect\citealt{MenendezDelmestre2009}; the evidence for IR-AGN
activity in the sources with a ``?'' is tentative); U: rest-frame UV spectral
signatures (\protect\citealt{Ledlow2002}; \protect\citealt{Smail2003};
\protect\citealt{Chapman2004}; \protect\citealt{Chapman2005}); X:
X-ray observations (\protect\citealt{Alexander2005}); B: Broad-line
AGN (\citealt{Swinbank2005}; \citealt{Coppin2008}; see Appendix A for
more details on the classifications of individual sources). Column (9): The total on-source exposure times for the data
presented here. Columns (10)--(11): The average seeing for the observations, based on standard star observations.
Column (12): The instrument used for the observations, either SINFONI
(S) or NIFS (N). Column (13) A data quality flag as follows: 1: data
which have sufficiently high signal-to-noise ratios
to study the spatially resolved kinematics; 2. data
with lower signal-to-noise ratios and limited spatially resolved information;
3. undetected in [O~{\sc iii}]$\lambda$5007 in this study.
$^{\star}$ For SMM\,J0943+4700 we observed the radio counter-part H6
(\protect\citealt{Ledlow2002}) and the 1.4\,GHz flux density has been corrected for an amplification
factor of 1.2 (\protect\citealt{Cowie2002}).
}
\end{center}
\end{table*}

\subsection{Target selection}
For this study we selected ten radio-detected $z=$~1.4--3.4 star-forming galaxies that host AGN activity to study their
spatially resolved dynamics.  The sample consists of eight
SMGs (labelled `SMM' with 850\,$\mu$m flux densities $S_{850}>4$\,mJy)
and two sources that are undetected at 850\,$\mu$m (labelled
`RG'). We will collectively refer to the
sample as high-redshift ULIRGs throughout. We detected
[O~{\sc iii}] emission in eight of the sources (see \S4 for details). All our detected sources are [O~{\sc iii}] luminous
($L_{{\rm[O~III]}}$=10$^{42}$--10$^{44}$\,erg\,s$^{-1}$), and have
high [N~{\sc ii}]/H$\alpha$ and [O~{\sc iii}]/H$\beta$ emission-line
ratios, or an AGN-dominated continuum, 
indicating the presence of AGN activity (e.g.,
\citealt{Kewley2006}), with most of the sources having additional
observational evidence for AGN activity (see
Table~\ref{Tab:otherProps} for details); see Appendix A for background
information on the individual sources. We provide details
of the target selection here.

Seven of our targets are selected from \cite{Takata2006} who presented near-IR spectroscopy around the redshifted [O~{\sc iii}]$\lambda\lambda$4959,5007
emission-line doublet of 22 high-redshift ULIRGs. In order to select sources
that might contain AGN-driven outflows, we chose seven targets with broad (FWHM $\gsim$ 700\,km\,s$^{-1}$) and bright ($F_{\rm{[O III]}}
\gsim$~few~$\times$~10$^{-16}$\,erg\,s$^{-1}$) [O~{\sc
  iii}] emission. We also include in our sample three,
$z\approx$\,2--2.5 ULIRGs selected from \cite{AlaghbandZadeh2012} which were
identified as hosting AGN activity on the basis of their rest-frame optical
emission-line ratios. These targets were initially
selected on the basis of their H$\alpha$ luminosities and are therefore
not pre-selected to have luminous and broad [O~{\sc iii}]
emission. One of the sources was presented in a pilot
study (SMM\,J1237+6203; \citealt{Alexander2010}) and we present the
whole sample here.

Details of our observations and the
targets' 1.4\,GHz flux densities, obtained from the literature, are provided in
Table~1. The 1.4\,GHz radio luminosities ($L_{1.4\rm{GHz}}$) are calculated
using Equation (2) from \cite{Alexander2003}, with the
1.4\,GHz flux densities shown here and an assumed spectral index of
$\alpha$=0.8.\footnote{This value is approximately the mean value
  for radio-identified SMGs with an observed range of roughly
  $\alpha$=0.2--1.5, which includes AGN (\citealt{Ibar2010}). Here we define a spectral index $\alpha$ such that flux
  density, $S_{\nu}$, and frequency, $\nu$, are related by $S_{\nu}
  \propto \nu^{-\alpha}$. } We estimate AGN luminosities for
  our sources using the SED fitting procedures outlined in \S\ref{Sec:SEDfitting}
  and these values are provided in Table~\ref{Tab:SEDpoints}.

\begin{figure*}
\centerline{\psfig{figure=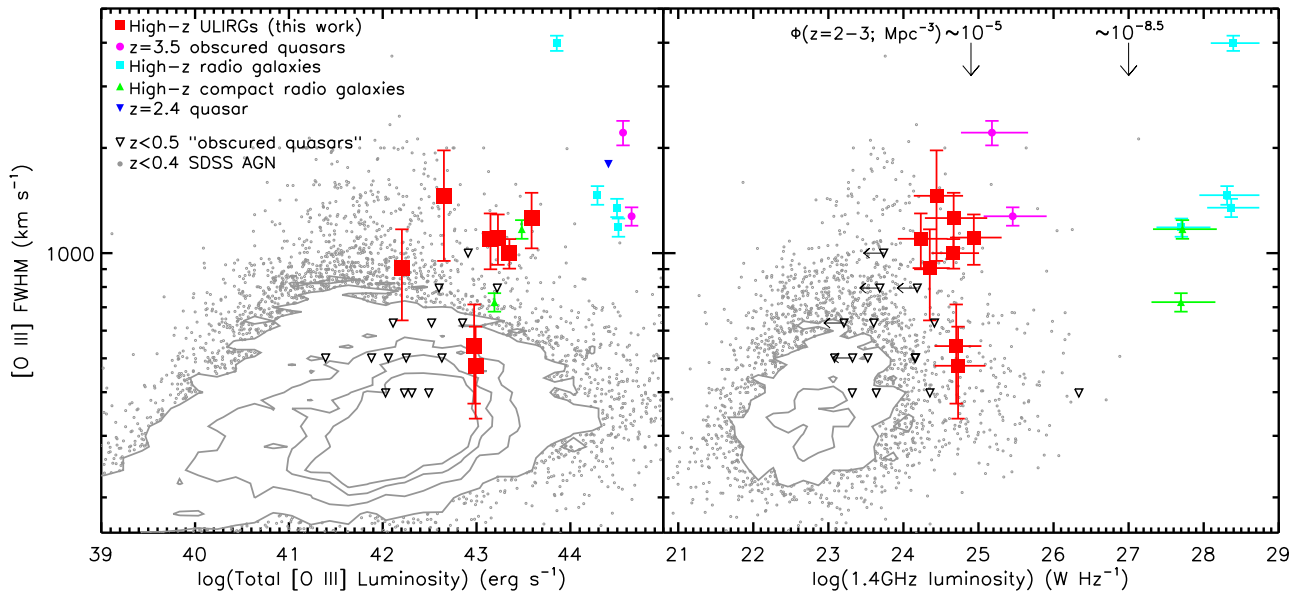,width=7in,angle=0}}
\caption{ FWHM versus
  total [O~{\sc iii}]$\lambda 5007$ luminosity ({\it left}) and FWHM
  verses 1.4\,GHz
  luminosity ({\it right}) for our eight
  [O~{\sc iii}] detected targets. Shown for comparison
  are low redshift ($z$$<$0.4) AGN from the Sloan Digital Sky Survey
  (Mullaney et~al. in prep) and $z<0.5$ obscured quasars with long-slit data
  (\citealt{Greene2011}). Also shown are other high-redshift sources with
  IFU data: $z$$\approx$2--3 radio galaxies (\citealt{Nesvadba2006,Nesvadba2008}), compact
  radio galaxies (\citealt{Nesvadba2007}a), $z$$\approx$3.5
  obscured quasars (\citealt{Nesvadba2011}) and a $z$=2.4
  quasar (\citealt{CanoDiaz2012}). Our targets have very
broad [O~{\sc iii}] emission, comparable with high-redshift
radio galaxies but having 3--4 orders of magnitude lower radio luminosities
such that they are $\approx$3 orders of magnitude more common at $z\approx$2--3. When the [O~{\sc iii}] profile is
  decomposed into multiple components, the FWHMs of the broadest components have been
  plotted. All rest-frame radio luminosities are calculated using the
  standard $k$-correction, on the basis
  of their measured 1.4\,GHz flux densities. Radio
spectral indices are taken to be $\alpha$=0.8; however, the horizontal error bars indicate the range of luminosities
using spectral indices in the range $\alpha$=0.2--1.5. The arrows in
the right panel correspond to space densities of $\Phi\approx$10$^{-5}$\,Mpc$^{-3}$
  (\citealt{Simpson2012}) and $\Phi\approx$10$^{-8.5}$\,Mpc$^{-3}$
  (\citealt{Willott1998}) from the radio luminosity function at $z\approx2.5$.
}
\label{fig:Selection}
\end{figure*}

Although this sample is heterogeneous, it is more representative of the
overall AGN population than previous IFU studies searching for AGN-driven
outflows in high-redshift radio galaxies (with $L_{\rm
    {1.4GHz}}\gsim10^{27}$\,W\,Hz$^{-1}$; e.g., \citealt{Nesvadba2008}). For
example the radio luminosities of our sources
($L_{{\rm 1.4GHz}}$=10$^{24}$--10$^{25}$\,W\,Hz$^{-1}$) imply a space density of $\Phi \approx$10$^{-5}$\,Mpc$^{-3}$ compared to
$\Phi \lesssim$10$^{-8}$ \,Mpc$^{-3}$ for high-redshift radio galaxies
(Fig.~\ref{fig:Selection}; \citealt{Simpson2012}; \citealt{Willott1998}). This is important if we are to
understand the prevalence of AGN-driven outflows in high-redshift galaxies. In Fig.~\ref{fig:Selection} we compare the integrated [O~{\sc
  iii}] properties and radio luminosities of our sample with low- and
high-redshift AGN. 

\subsection{Gemini-North NIFS observations}

Seven of the sources in our sample (those selected from \citealt{Takata2006}) were observed using Gemini-North
NIFS (Table~1). The NIFS IFU uses an image slicer to reimage a 3.0\,$\times$\,3.0 arcsec$^{2}$ field into 29 slices of width
0.103$''$ with a pixel scale of 0.04'' along the slices. The dispersed spectra from the slices are reformatted on
the detector to provide two-dimensional spectro-imaging. Depending on
the target redshift, we used the $J-$, $H-$ or $K-$band grisms to
ensure that we cover the rest-frame [O~{\sc
  iii}]$\lambda\lambda$4959,5007 emission-line doublet. These grisms
have an approximate spectral resolutions of $\lambda/\Delta\lambda\approx$ 6040, 5290 and
5290, respectively (FWHM $\approx$ 2, 3, 4
\AA, or $\Delta v \approx$\,50, 57, 57\,km\,s$^{-1}$). For our
  anaylsis we measured the instrumental dispersion more accurately, by measuring the widths of
several bright sky-lines close to the observed wavelengths of the
emission lines for the observations of each source. We found that the uncertainties in the
instrumental dispersion (by measuring the scatter from the widths of several
sky-lines) to be insignificant ($\approx$\,5\%) compared to the uncertainties
from our fitting procedures. We consequently do not include the
uncertainties on the instrumental resolution in our measurements. These observations were taken between 2008 March 25 and
May 19 and 2009 August 5 and December 30. All observations were taken in $<$0.5$''$ seeing. 

The observations were performed using repetitions of the standard ABBA configuration in
which we chopped either 6$''$ to blank sky or, if the source was
compact (emission-line region $<$1.5$^{\prime\prime}$), we chopped
within the IFU. Individual exposures were 600s. Final on-source exposure times and seeing measurements, calculated using the
corresponding standard star observations, are shown in Table~1.

We reduced the data with the Gemini {\sc iraf nifs} pipeline which includes
sky subtraction, wavelength calibration, and flat fielding. Attempts
to remove residual OH sky emission lines were made
using the sky-subtraction techniques described in \cite{Davies2007}. We further corrected for variations in the night sky emission by masking the source (when clearly detected) and taking a median
of each pixel-column and consequently subtracting the
(empty sky) background separately from each wavelength. 

\subsection{VLT SINFONI observations}
Three of the sources in our sample were observed using VLT SINFONI (Table~1). The full details
of these SINFONI observations are given in
\cite{AlaghbandZadeh2012} with only basic information given
here. We used the widest available field-of-view, 8.0\,$\times$\,8.0 arcsec$^{2}$, which is
divided into 32 slices of width 0.25$^{\prime\prime}$ with a
  pixel scale 0.125$^{\prime\prime}$ along the slices . The $H$+$K$
grating was used, which gives an approximate resolution of $\lambda/\Delta\lambda\approx$1500 (FWHM
$\approx$\,13\,\AA \, or $\Delta v \approx$\,200\,km\,s$^{-1}$). 
  We made accurate measurements of the instrumental dispersion for
  the observations for each source in the same manner as outlined for
  the NIFS observations in the previous section. Again we found the
  uncertainties on the instrumental dispersion to be insignificant
  compared to the uncertainties from our fitting procedures. The targets
were chopped around quadrants of the IFU for sky subtraction. All
observations were taken in $<$0.6$^{\prime\prime}$ seeing. Individual exposures were 600s. Final on-source
exposure times and seeing measurements, calculated using the
corresponding standard star observations, are shown in Table~1.

The data were reduced using the SINFONI ESOREX pipeline,
which includes sky subtraction, wavelength calibration and flat fielding. 

\subsection{Flux calibration and stacking}
Each night's data were flux calibrated separately using observations of standard
stars, at a similar airmass to the target galaxies, which were reduced
in the same manner as the targets. Since the [O~{\sc
  iii}]$\lambda\lambda$4959,5007 emission-line doublets typically
lie in regions of $>95$\% sky transparancy, no corrections were made for
telluric absorption. To find spatial centroids for the flux-calibrated data cubes, white light (wavelength
collapsed) images around the redshifted [O~{\sc iii}]$\lambda$5007
emission line (NIFS targets) or
H$\alpha$ emission line (SINFONI targets) were produced. The data cubes for which we could not
measure a centroid were discarded to improve the reliability of the
spatially resolved data. Data cubes were only discarded
  for SMM\,J0302+0006 (2 discarded out of the 21 observed), RG\,J0302+0010 (4
discarded out of 20) and SMM\,J1235+6215 (8 discarded out of 21). We then spatially aligned and co-added the individual data
cubes to create the final mosaic, using a median with a 3$\sigma$ clipping threshold to
remove remaining cosmetic defects and cosmic rays. The total on-source
exposure times used in the final stacks are given in Table~1. 

In all of the following sections, the effect of instrumental dispersion was
corrected for by subtracting it in quadrature from the observed line
dispersions. 

\section{Star-formation rates and AGN luminosities}
\label{Sec:SEDfitting}

\begin{figure*}
\centerline{\psfig{figure=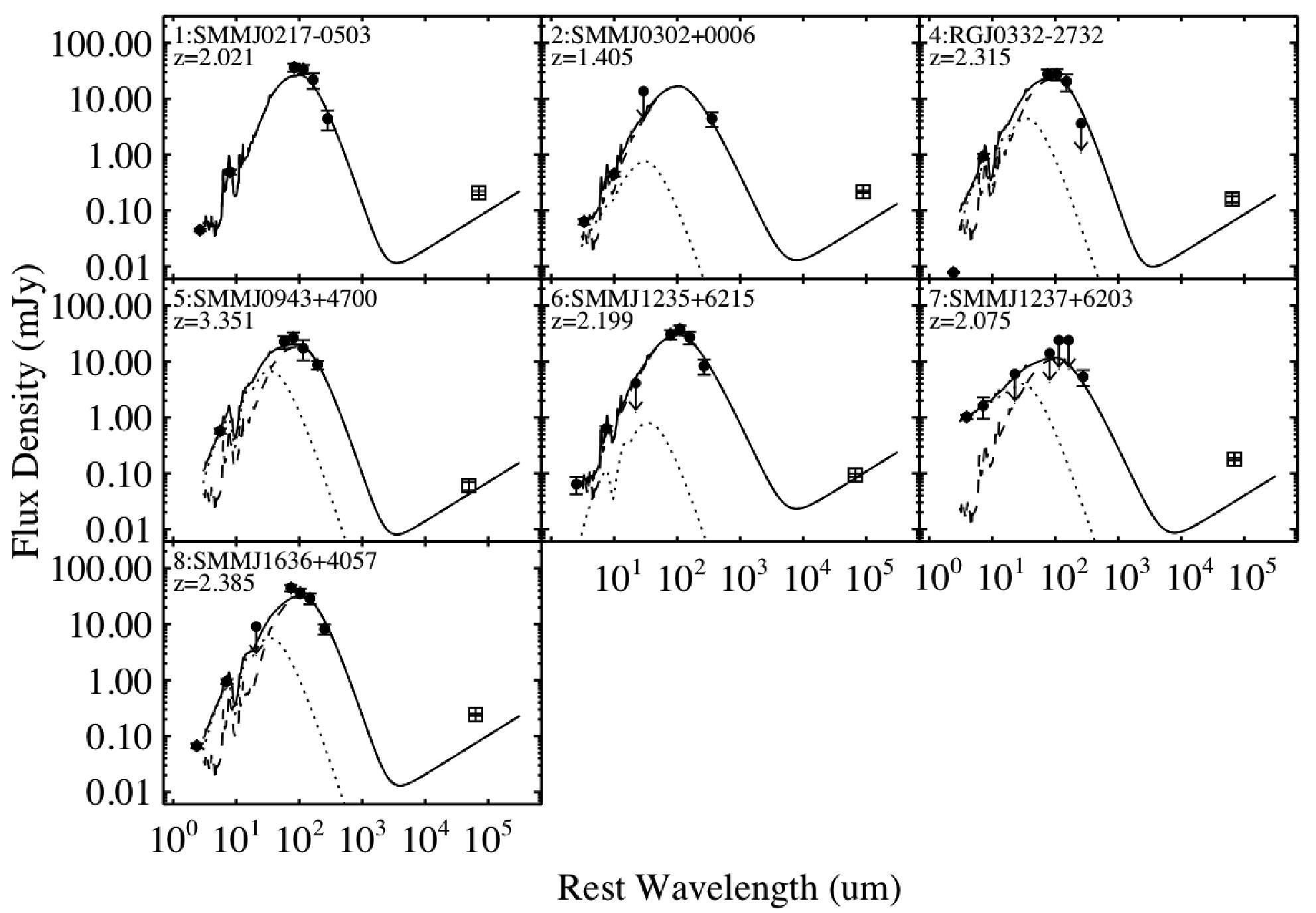,width=6in,angle=0}}
\caption{The infrared flux densities (filled circles) in the rest
  frame for the seven sources for which we performed SED fitting (see
  \S\ref{Sec:SEDfitting}, Table~\ref{Tab:SEDpoints} and Del Moro
  et~al. 2012 for more details). The source
  RG\,0302+0010 (source ID number 3) does not appear
  due to insufficient infrared data available. For guidance, also shown are the
  overall best-fit SEDs. The total SEDs are shown as solid curves, the AGN
  templates are shown as dotted curves and the starburst templates are
  shown as dashed curves. The 1.4\,GHz flux densities (open squares) were not
  included in the SED fitting process. The source RG\,J0302+0010 does
  not appear due to insufficient available infrared data. These SED
  fits were used to constrain the contributions to the
  infrared luminosities from AGN activity and star formation activity
  for each of the sources (Table~\ref{Tab:SEDpoints}).
}
\label{fig:SEDfits}
\end{figure*}

To assist in the analyses of this work we need to constrain the
star-formation rates and AGN luminosities of the eight [O~{\sc iii}]-detected sources in a consistent manner. To achieve this we obtained infrared flux densities
from the literature (Table~\ref{Tab:SEDpoints}) and fit these data using AGN and
star-forming galaxy templates with the $\chi^2$ minimisation spectral
energy distribution (SED) fitting procedure outlined in Del Moro et~al. (2012). Briefly, we fit the infrared data using the empirical
AGN template defined in \cite{Mullaney2011}, allowing for
$A_{V}$=0--5\,mag of extinction, with each of the ``SB2''--``SB5''
star-forming galaxy templates; the ``SB1''
star-formation template did not fit the data
well in any of the sources. This provides four best-fitting solutions (i.e., one for each of the four
star-forming galaxy templates) with minimum $\chi^2$ solutions. We
used these solutions to find the star-formation rates and AGN luminosities
for each of the sources using the methods described below. The
infrared data points and the overall best-fit solutions are shown in Fig.~\ref{fig:SEDfits}.

We found convincing evidence for both star formation and AGN activity at infrared
wavelengths for all of the sources except for RG\,J0302+0010 (for
which there is insufficient infrared data available) and
SMM\,J0217$-$0503, which had no significant AGN component providing
only an upper limit on the AGN infrared luminosity. The
AGN components in SMM\,J0943+4700 and SMM\,J1237+6203 were found to be
particularly bright and dominated the mid-infrared emission; the SED
fitting results for SMM\,J0943+4700 also agreed well with that found
from mid-IR spectral fitting (\citealt{Valiante2007}) but no mid-IR
spectroscopy exists for SMM\,J1237+6203. The AGN components for the
other sources are weaker at mid-IR wavelengths. However, three of the
sources have published mid-infrared spectroscopy (SMM\,J0302+0010,
SMM\,J1235+6215, and SMM\,J1636+4057; \citealt{MenendezDelmestre2009}), and we found good agreement between
the strength of the AGN component calculated from our SED-fitting
procedure and the strength of the AGN derived from the mid-infrared
spectroscopy.

We used the SED-fitting results to calculate the infrared luminosities
of the star-formation components ($L_{\rm{IR,SF}}$; integrated over
8-1000$\mu$m) and consequently calculated star formation rates
following \cite{Kennicutt1998}; see Table~\ref{Tab:SEDpoints}. This was
achieved by taking the mean luminosity of the best-fitting solutions
that used the four star-formation templates described above. The quoted uncertainties in the star-formation luminosity and
star-formation rates are the average of the difference between these
mean values and the models with the highest and lowest values. We
found that all of the sources have $L_{\rm{IR,SF}}>10^{12}$\,L$_{\sun}$,
confirming that they are ULIRGs with SFRs in the range
$\approx$300--1400\,M$_{\sun}$\,yr$^{-1}$. For the source
RG\,J0302+0010 we estimated the infrared luminosity using the
radio-infrared relationship for star-forming galaxies
(e.g., \citealt{Helou1985}) with $q=2.1$, which is a typical value for
high-redshift ULIRGs (\citealt{Kovacs2006}). For this source we take
the value of $L_{\rm{IR,SF}}$ as an upper limit because the AGN will
be contributing some unknown fraction of the radio luminosity 
(e.g., see Fig.~\ref{fig:SEDfits} which shows that some of the sources
have excess radio emission relative to that expected from
star-formation alone; also see Del Moro et~al. 2012).

To determine the bolometric AGN luminosities for each source we first
calculated the AGN continuum luminosity [rest-frame $\nu L_{\nu}$(6$\mu$m)] by taking the average AGN
contribution to the 6$\mu$m flux of the four models described above. These average infrared AGN luminosities are then converted
into AGN bolometric luminosities using the 6$\mu$m-to-2--10~keV
luminosity relationship found for AGNs (conversion factor of
$\approx$0.3; \citealt{Lutz2004}) and the 2--10~keV-to-bolometric
luminosity ratio in \cite{Elvis1994} (conversion factor of
$\approx$25; see Table~\ref{Tab:SEDpoints}). This overall correction
factor of 7.5 that we apply to convert between between $\nu L_{\nu}$(6$\mu$m) and AGN bolometric luminosity is
in excellent agreement with the correction factor of $\approx$8$\pm$2
found by \cite{Richards2006} from their mean SED of SDSS quasars. We found that the two
sources with the largest fractional uncertainties on their SED-derived
AGN luminosities are also those which only have tentative evidence for
IR-AGN activity in their mid-infrared spectra (see
Table~\ref{Tab:SEDpoints}). However, we compared all the SED-derived AGN
bolometric luminosities to those predicted using the measured [O~{\sc
iii}] luminosities and the locally derived relationship of
\cite{Heckman2004} ($L_{\rm{AGN}}$=3500$L_{\rm{[O III]}}$). We found
that the $L_{\rm{[O III]}}$ derived AGN luminosities are on average a
factor of $\approx$~2.5 higher than the SED derived values, with a
remarkably small scatter (all sources have a factor of 2.5$\pm$1.5),
giving evidence that (1) our derived AGN bolometric luminosities are
reasonable, and (2) the [O~{\sc iii}] luminosities are dominated by AGN
activity. To predict the AGN luminosity for the source RG\,J0302+0010,
for which we were unable to perform the SED fitting, we therefore take the
$L_{\rm{[O III]}}$ derived AGN luminosity and divide by a factor of
2.5 (see Table~\ref{Tab:SEDpoints}).

\section{Analysis and Results}
We have studied the galaxy-integrated and spatially resolved properties of the [O~{\sc
  iii}]$\lambda\lambda$4959,5007 emission-line doublet in our sample. Of
the ten sources in our sample, six have spatially extended [O~{\sc
  iii}] emission, two have lower quality [O~{\sc iii}] data and limited
spatially-resolved information and two were undetected in [O~{\sc
  iii}] (see Table~\ref{Tab:otherProps}). In this section we provide the details of our analyses and give an overview of the
results of the sample as a whole. We give specific results for
individual sources in Appendix A. In this section we first describe the
galaxy-integrated properties of the eight [O~{\sc iii}] detected targets before
providing a detailed analysis of the
kinematics for the six spatially resolved sources. We defer discussion of the
results and their implications for galaxy evolution to \S5.

\subsection{Galaxy-integrated spectra}
\label{Sec:intspecs}

\begin{figure*}
\centerline{\psfig{figure=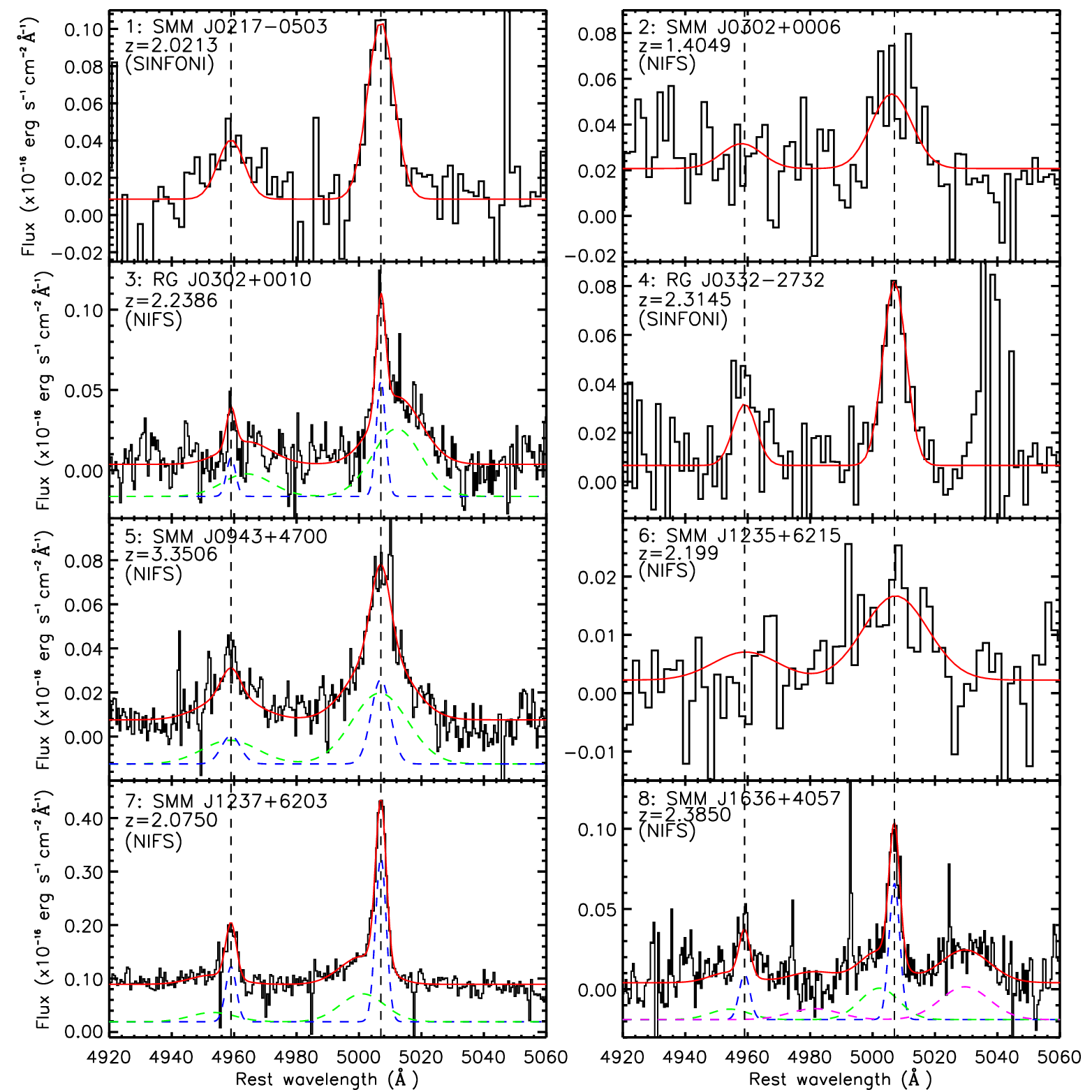,width=7in}}
\caption{Galaxy-integrated spectra, shifted
  to the rest-frame, around the
  [O~{\sc iii}]$\lambda\lambda 4959,5007$ emission-line doublet. A large
  variety in the [O~{\sc iii}] emission-line profiles are found with
  both blueshifted and redshifted extremely broad (FWHM$\approx$900--1500\,km\,s$^{-1}$)
  emission-line components, in addition to narrow emission-line components. Of particular note is
  SMM\,J1636+4057 which exhibits two broad components separated by
  $\approx$1700\,km\,s$^{-1}$, one blueshifted and one redshifted with
  respect to the narrow component (see Appendix A8 for
  details). The dashed curves show the individual Gaussian components
  (with an arbitrary flux offset) of the fits when two or three
  components are required. The solid curves show the
  best fitting overall emission-line profiles. The vertical dashed lines show the rest-frame wavelengths of
  redshifted [O~{\sc iii}]$\lambda 4959$ and [O~{\sc iii}]$\lambda
  5007$ based on the redshift of the narrowest Gaussian component. The spectra for
  SMM\,J0302+0006 and SMM\,1235+6215 are of lower signal-to-noise than the other
  spectra and have been binned by a factor of
  four for clarity. The parameters of the
  emission-line fits and their uncertainties are given in Table~\ref{Tab:OIIIprops}. 
}
\label{Fig:int_specs}
\end{figure*}

In Fig.~\ref{Fig:int_specs} we show the galaxy-integrated spectra around the [O~{\sc
  iii}]$\lambda\lambda$4959,5007 emission-line doublet for the eight
sources with detected [O~{\sc iii}] emission. The two sources
that have an [O~{\sc iii}] flux below
our detection threshold (SMM\,J2217+0010 and
SMM\,J2217+0017; see Appendix A9 and A10) are not shown. The spectra were created by
integrating over the full spatial extent of the observed [O~{\sc iii}]
line emission; i.e., the aperture sizes to create the spectra were chosen by increasing the
sizes until the [O~{\sc iii}] fluxes reached a maximum. 

To fit the emission-line profiles we employed a $\chi^2$ minimisation
procedure, down-weighting at the positions of the strongest sky lines. We fit the [O~{\sc
  iii}]$\lambda\lambda4959,5007$ emission-line doublet with Gaussian
components using a fixed wavelength separation. The intensity ratio was fixed at
[O~{\sc iii}]$\lambda$4959/[O~{\sc iii}]$\lambda$5007=0.33 (e.g., \citealt{Dimitrijevi2007}) and the width of
the [O~{\sc iii}]$\lambda$4959 line was fixed to be the same as the [O~{\sc
  iii}]$\lambda$5007 line. In several cases the emission-line profiles
are complex, with the presence of asymmetric wings. To characterise this complexity, we first fit a
single Gaussian profile followed by a double Gaussian profile, only accepting the double Gaussian
profile if it resulted in a significant improvement ($\Delta\chi^{2} > 25
\sim 5\sigma$). In one source, SMM\,J1636+4057, we find evidence for a third Gaussian
component at +1350\,km\,s$^{-1}$
from the narrow component; we confidently attribute this to [O~{\sc
  iii}] as there are other emission lines found at the same velocity
as this component (\citealt{Smail2003}; also see Appendix A8). Uncertainties on the parameters are calculated by perturbing one parameter at a time, allowing the others to
find their optimum values, until the $\chi^{2}$ value increases by
1. The mean of these upper and lower uncertainties are quoted
throughout. The parameters for
all of the fits and their uncertainties are given in Table~\ref{Tab:OIIIprops}. We note that the true absolute uncertainties on the flux
  measurements will be a factor of $\approx$2--3 higher, due to unknown
  uncertainties on the absolute flux calibrations. We note that there are some discrepancies between the fluxes shown here and those
given in \cite{Takata2006} and \cite{Alexander2010}. We attribute this to the more
reliable flux calibration procedures used in our work.

 In Fig.~\ref{Fig:int_specs} it can be seen that the [O~{\sc iii}] emission-line profiles are diverse across the sample, with both narrow (FWHM~$\approx$~few hundred km\,s$^{-1}$) and broad components
(FWHM$\approx$900--1500\,km\,s$^{-1}$) which can be
blueshifted and/or redshifted with respect to the narrow emission ($-350$
$\lesssim$~$\Delta v$~$\lesssim$\,1350\,km\,s$^{-1}$). We will discuss
  in \S\ref{Sec:NarrowLines} that the narrow [O~{\sc iii}] components are likely to be
tracing the host galaxy dynamics and merger remnants while the broad
components are most likely due to energetic outflows (see
  Appendix A for a discussion on individual objects). We will also argue that the observed
diversity across the sample is due to a combination of orientation and
obscuration effects. 

The four sources for which we were able to decompose the line profiles into
multiple Gaussian components are those which have the highest flux and
consequently highest signal-to-noise (Table~\ref{Tab:OIIIprops}). It is possible
that the other four detected sources also
have both narrow and broad emission line components but we were unable to
decompose them in these spectra due to their lower signal-to-noise ratios. For example, in one source
(SMM\,J0217$-$0503), we identify a broad [O~{\sc iii}] component in a
sub-region of the IFU datacube that is not identified in the
galaxy-integrated spectrum (\S\ref{Sec:regions}).

In the following sections we discuss the spatially resolved properties
of all of the [O~{\sc iii}] detected sources except SMM\,J0302+0006
and SMM\,J1235+6215. SMM\,J0302+0006 is spatially unresolved
in our data while SMM\,J1235+6215 only displays tentative evidence for
extended emission (see Appendix A2 and A6 for further details on these two sources). Due to the low
signal-to-noise ratios of the data for these two sources it is
difficult to determine if they are intrinsically compact or if any extended
emission lies below our detection threshold. We do
not discuss these two sources any further in this section. 

\subsection{Flux, velocity and FWHM maps}
\label{Sec:velmaps}

\begin{figure*}
\centerline{\psfig{figure=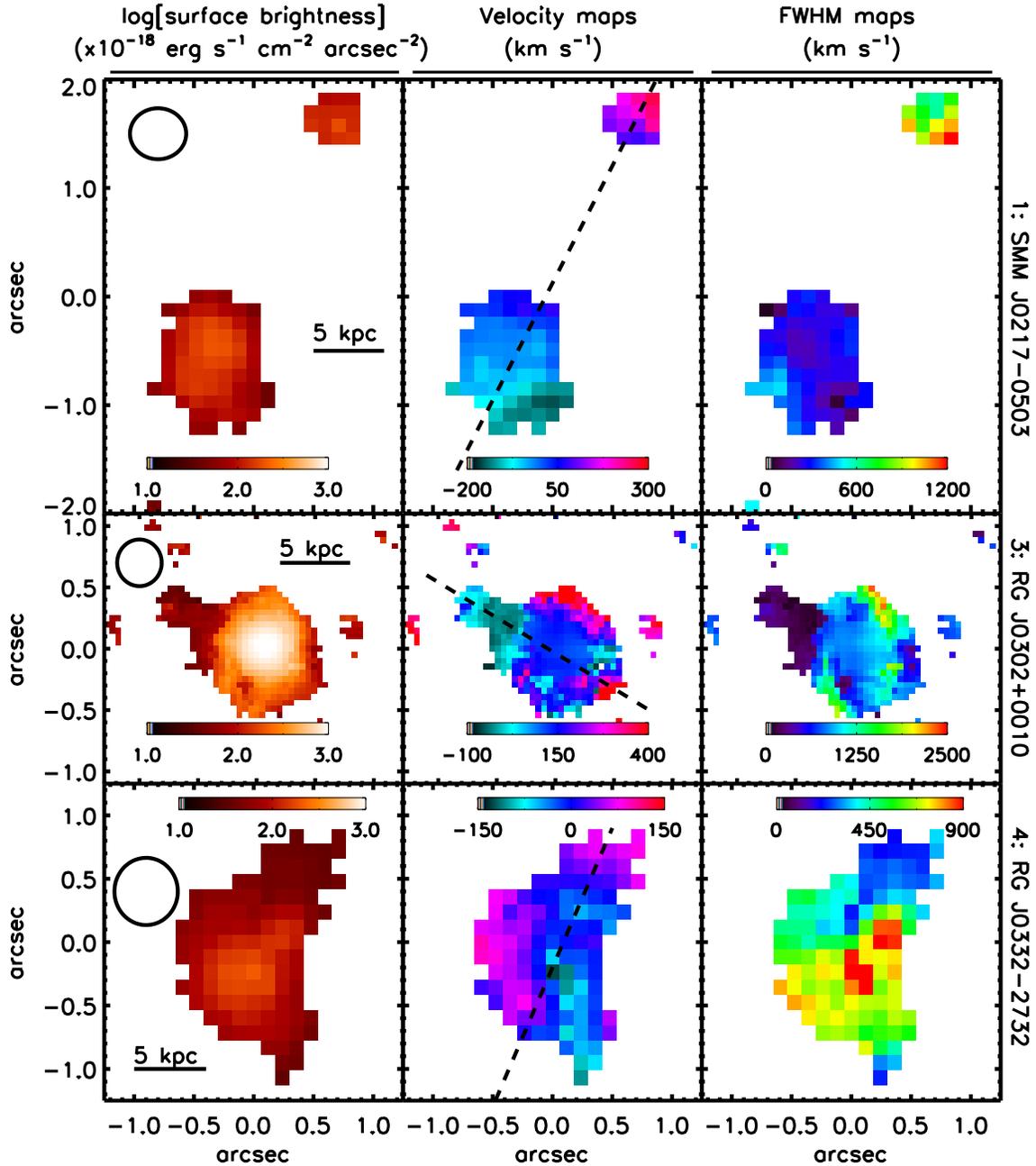,width=6.2in,angle=0}}
\caption{[O~{\sc iii}] flux, velocity and
  FWHM fields, from fitting single Gaussian components to individual
  pixels for our six sources with spatially resolved emission. The [O~{\sc
    iii}] emission is
  evidently very diverse and complex with large velocity offsets
  (up to $\approx$\,1700\,km\,s$^{-1}$) and regions of both narrow (FWHM
  $\approx$ few 100\,km\,s$^{-1}$) and broad
  (FWHM~$\approx$\,700--2500\,km\,s$^{-1}$) emission-line components. We attribute the
  narrow emission to galaxy dynamics and merger remnants and the
  broad emission to energetic outflows. Zero velocity is defined at the redshift of the
  narrowest component shown in the galaxy-integrated spectra
  (Fig.~\ref{Fig:int_specs}). No corrections have been made for
  unknown orientation
effects or dust extinction. The open circles denote the average seeing disc
  during the observations of each target and the solid bars indicate
  5\,kpc in extent. The dashed lines show the
  axes through which we produced the velocity and FWHM profiles shown
  in Fig.~\ref{fig:velcurves}. North is up and east is left in all panels.}
\label{fig:velmaps}
\end{figure*}

\addtocounter{figure}{-1}
\begin{figure*}
\centerline{\psfig{figure=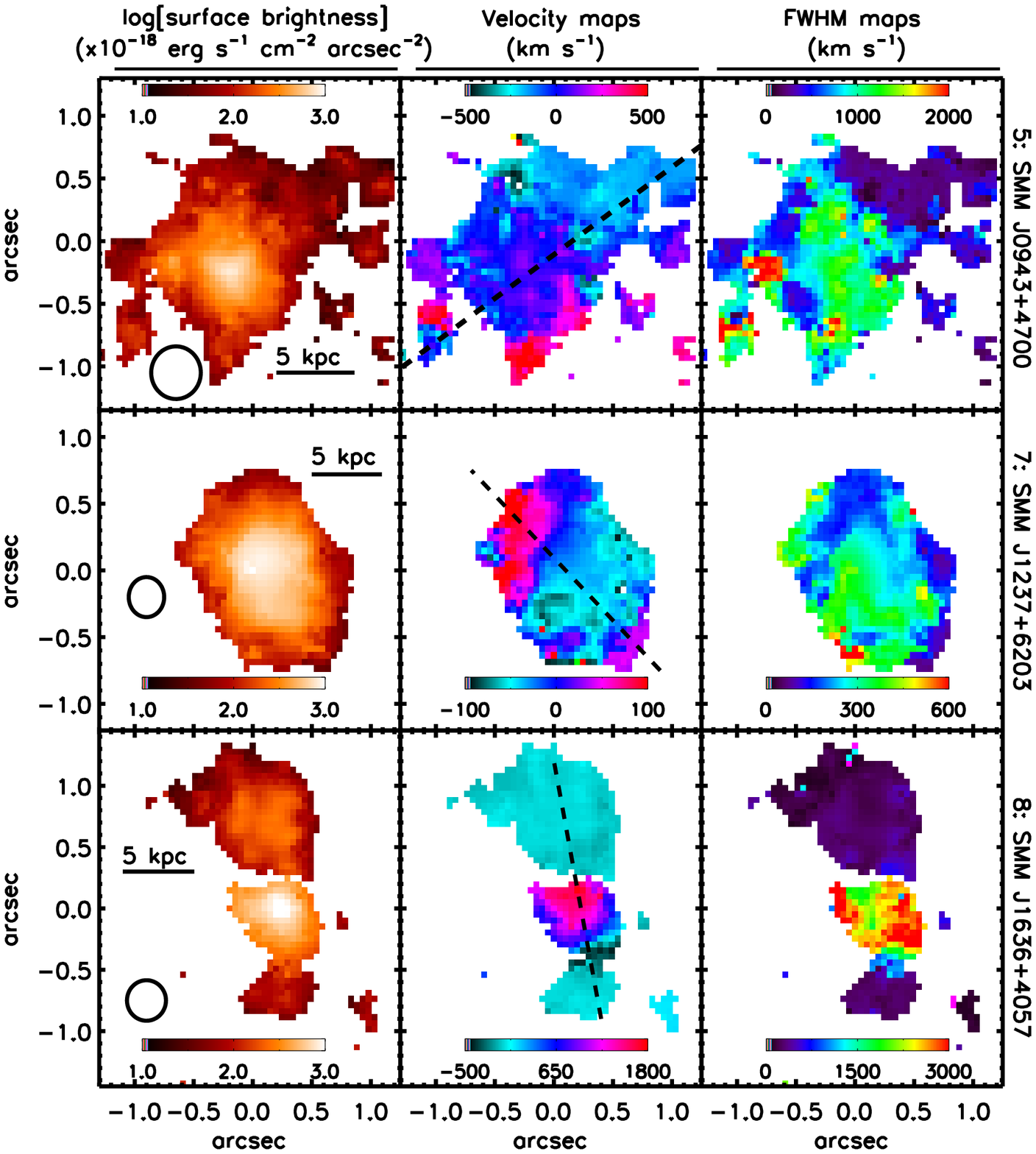,width=6.2in,angle=0}}
\caption{continued.}
\end{figure*}

In Fig.~\ref{fig:velmaps} we show flux maps, velocity fields and FWHM
maps for the six sources for which we were able to spatially resolve the
[O~{\sc iii}] emission. These maps were produced using a adaptive
  binning technique such that the high surface brightness regions are
  averaged over smaller regions than the outer low surface brightness
  regions. We do this by taking a
  mean spectrum over 3\,$\times$\,3 spatial pixels, increasing
to 5\,$\times$\,5 pixels (maximum for SINFONI data) and ultimately to
7\,$\times$\,7 pixels (maximum for NIFS data) if the signal was too low to produce a sufficiently high $\chi^2$ improvement in a
continuum+emission-line fit over a simple continuum fit ($\Delta\chi^2
> 16$; equivalent to $\approx 4\sigma$). The continuum level was taken
to be the median of line-free continuum in the vicinity of the emission-lines. When this
criterion is met, we fit the [O~{\sc iii}]$\lambda\lambda$4959,5007 emission-line doublet with a
single-Gaussian profile, allowing the normalisation, width and central
wavelength to vary. We defined zero velocity at the central wavelength
of the narrowest Gaussian component in the galaxy-integrated
spectra.

The flux maps in Fig.~\ref{fig:velmaps} show that the observed emission-line regions of these sources are
diverse and irregular, with at least two sources showing spatially distinct emission-line regions, indicating multiple interacting or merging 
systems (SMM\,J0217$-$0503 and SMM\,J1636+4057). The [O~{\sc iii}] emission is
observed to be extended up to $\approx$20\,kpc. The velocity fields and FWHM maps in
Fig.~\ref{fig:velmaps} show that there are regions dominated by narrow
[O~{\sc iii}] emission ($\approx$ few hundred km\,s$^{-1}$) with modest
velocity gradients ($\Delta v \lesssim$200\,km\,s$^{-1}$). There are also regions which are kinematically chaotic, extended up to 15\,kpc in spatial extent, with extreme widths (700\,$\lesssim$\,FWHM\,$\lesssim$\,2500\,km\,s$^{-1}$) and
large velocity offsets ($\Delta v$ up to $\approx$1700\,km\,s$^{-1}$).

The velocity fields and FWHM maps in Fig.~\ref{fig:velmaps} were
created using single Gaussian components and do not separately trace
the broad and narrow [O~{\sc iii}] emission line components observed
in the galaxy-integrated spectra (Fig.~\ref{Fig:int_specs}). Instead
they roughly trace the brightest [O~{\sc iii}] kinematic component. For example, we will
show that although the source SMM\,J1237+6203 looks to have fairly
quiescent kinematics in Fig.~\ref{fig:velmaps}, the extremely
broad and high-velocity [O~{\sc iii}] component seen in the
galaxy-integrated spectrum (FWHM$\approx$1000\,km\,s$^{-1}$; $\Delta v \approx
-350$\,km\,s$^{-1}$; Fig.~\ref{Fig:int_specs}) is found over the central 5--8\,kpc (see also \citealt{Alexander2010}). In the following sections, we therefore use further
analysis methods to give a more complete
description of the kinematics of these sources.

\subsection{Regions dominated by narrow and broad emission lines}
\label{Sec:regions}

\begin{figure*}
\centerline{\psfig{figure=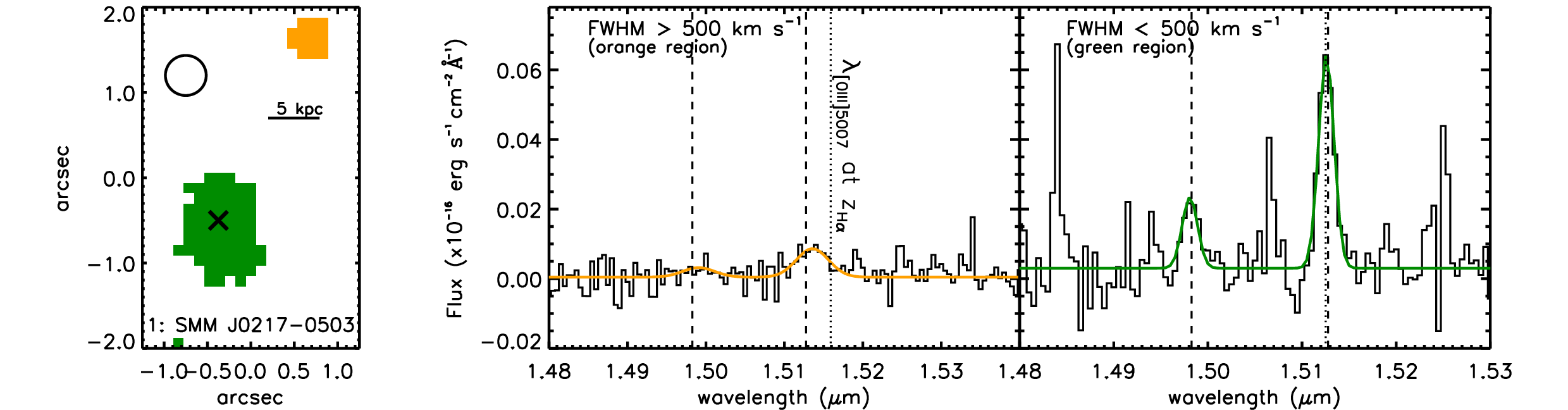,width=6.05in,angle=90}}
\centerline{\psfig{figure=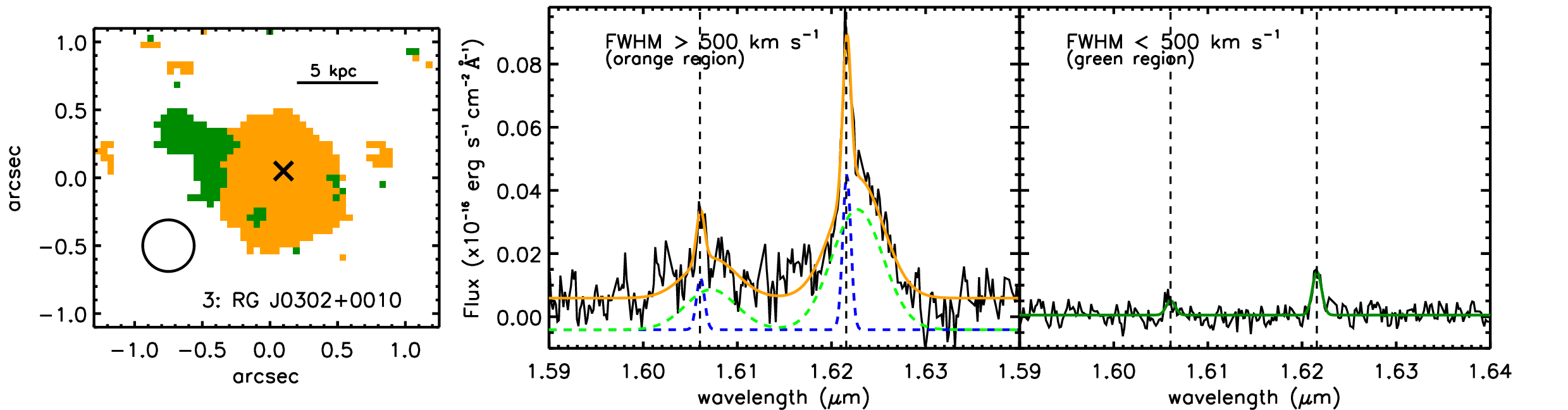,width=6.05in,angle=90}}
\centerline{\psfig{figure=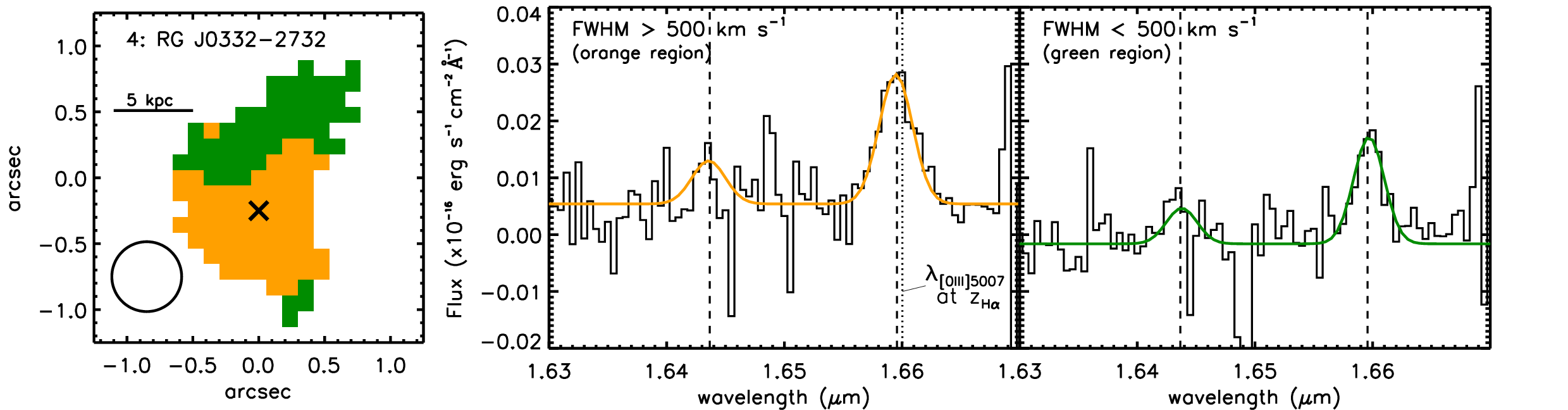,width=6.05in,angle=90}}
\centerline{\psfig{figure=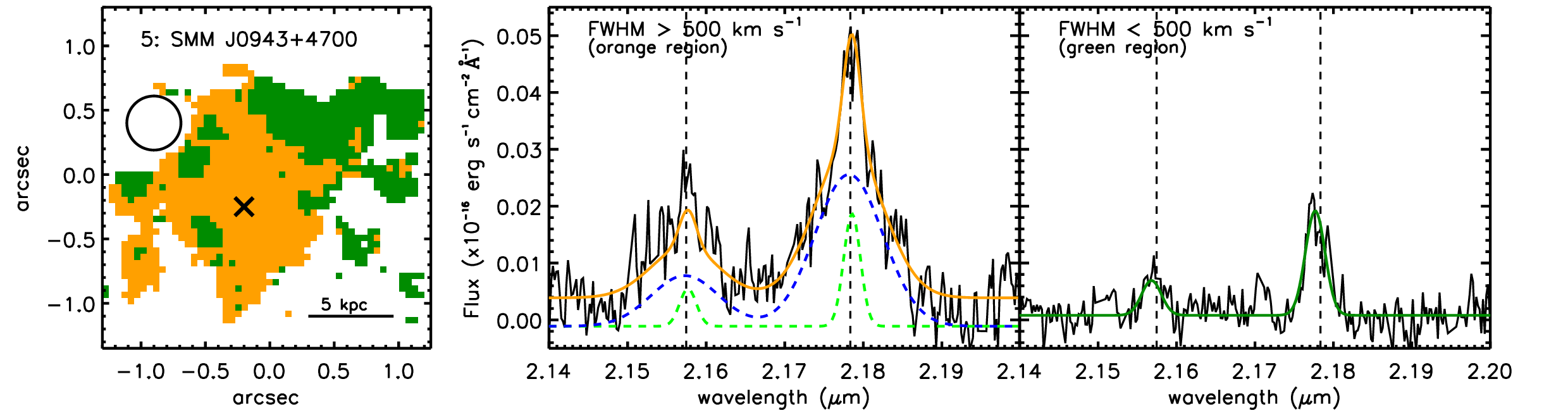,width=6.05in,angle=90}}
\centerline{\psfig{figure=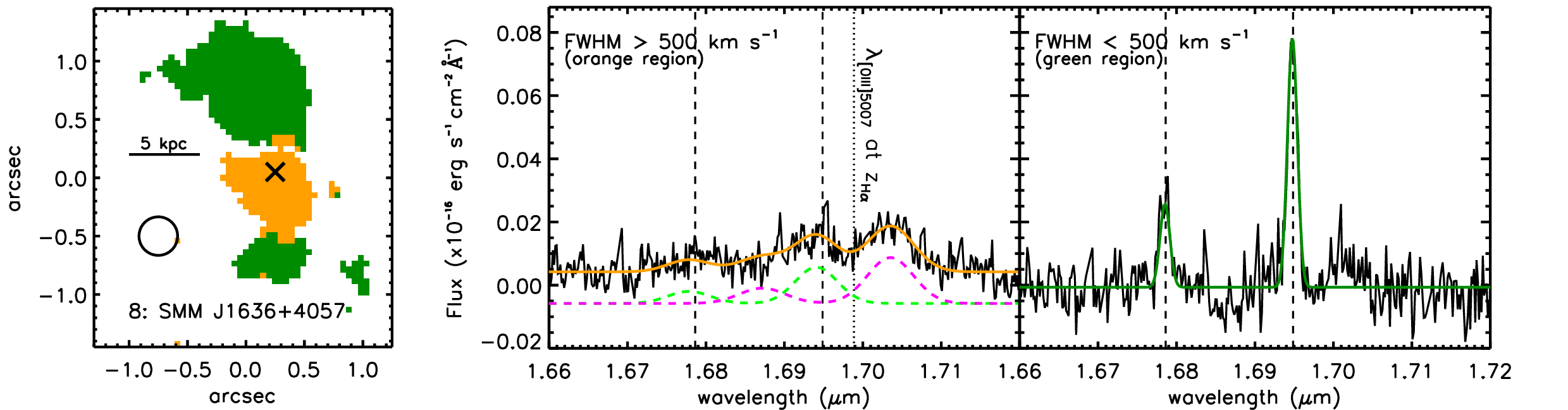,width=6.05in,angle=90}}
\caption{{\it Left:} Maps of our spatially resolved sources colour-coded to
  separate regions dominated by narrow [O~{\sc iii}] emission lines
  (green: FWHM $< 500$\,km\,s$^{-1}$) or broad [O~{\sc iii}] emission-lines (orange: FWHM $>
500$\,km\,s$^{-1}$) determined from the FWHM maps shown in
Fig.~\ref{fig:velmaps}. For the source SMM\,J1237+6203 the
  narrow component dominates the [O~{\sc iii}] flux throughout and therefore we
  are unable to define the separate spatial regions using this method (see \S\ref{Sec:regions} for details). The narrow emission lines are observed over the largest
extents but the broad emission lines are still observed in extended regions ($\approx$4--15\,kpc). The open circles denote the average seeing discs during the
observations, the solid bars indicate 5\,kpc in extent and the crosses
indicate the peak in [O~{\sc iii}] flux. North is up and
  east is left. {\it Middle and Right:} The
spatially-integrated spectra around the [O~{\sc iii}]$\lambda\lambda
  4959,5007$ emission-line doublet extracted from the regions indicated in the left
  panels.  The central regions exhibit complex emission-line
  profiles. The dashed curves show the individual Gaussian components
  (with an arbitrary flux shift) of the fits when two components
  were required. The solid curves show the best fitting overall emission-line
  profiles; the parameters and their uncertainties are shown in
  Table~\ref{Tab:OIIIregions}. The vertical dashed lines shows the
  centroids of the [O~{\sc iii}]$\lambda\lambda$4959,5007 narrow line
  components in the galaxy-integrated spectra (Fig.~\ref{Fig:int_specs}). The vertical
  dotted lines for SMM\,J0217$-$0503 and RG\,J0332$-$2732 show the
  expected position of [O~{\sc iii}] given the
  redshift of the H$\alpha$ emission-line from the same
  spatial region (Alaghband-Zadeh priv. comm.). For the source
  SMM\,J1636+4057 the vertical dotted line indicates the expected
  wavelength of the [O~{\sc iii}]5007 line using the redshift of
  the H$\alpha$ broad-line region (\citealt{Swinbank2005}).}
\label{fig:regions}
\end{figure*}

In Fig.~\ref{fig:regions} we  show spatially-integrated spectra from regions dominated by broad or narrow emission lines. To produce these
spectra we took our [O~{\sc iii}] FWHM maps (Fig.~\ref{fig:velmaps}) and integrated the spectra over the pixels
that have FWHM$<$500\,km\,s$^{-1}$ (narrow [O~{\sc iii}]) and those that have
FWHM$>$500\,km\,s$^{-1}$ (broad [O~{\sc iii}]). We do not do this for SMM\,J1237+6203 as
there are too few pixels that contain emission lines with
FWHM$>$500\,km\,s$^{-1}$. For this source the narrow component of
  the [O~{\sc iii}] emission line seen in Fig.~\ref{Fig:int_specs} dominates the
  whole of the velocity and FWHM maps. The value of
500\,km\,s$^{-1}$ was motivated by the maximum line-widths generally
expected from galaxy dynamics and merger remnants in high-redshift
ULIRGs (see \S\ref{Sec:NarrowLines}; \citealt{AlaghbandZadeh2012}). We fit the spectra using the same method as the
galaxy-integrated spectra (see \S~\ref{Sec:intspecs}) and provide all of
the parameters of the fits and their uncertainties in
Table~\ref{Tab:OIIIregions}. We also fit any observed H$\beta$ emission in
these regions using a single Gaussian profile and give the
parameters and their uncertainties in
Table~\ref{Tab:OIIIregions}. When no H$\beta$ emission lines were
detected we use the standard deviation of the emission-line free continuum and the FWHM of
the [O~{\sc iii}] emission-line components to derive 3$\sigma$ upper
limits on the flux.

By comparing the spatially-integrated spectra from the regions shown in
Fig.~\ref{fig:regions} with the galaxy-integrated spectra shown in
Fig.~\ref{Fig:int_specs} we are able to locate the different kinematic
components in the galaxies. For example, in the majority of the sources there is narrow
emission (FWHM$\approx$ a few 100\,km\,s$^{-1}$) extended over
the full spatial extent of the emission-line regions (up to
$\approx$20\,kpc). The regions dominated by the broadest emission lines
(Fig.~\ref{fig:regions}) are also the regions where the [O~{\sc iii}] emission
peaks in surface brightness and the [O~{\sc iii}]/H$\beta$ ratios are
the largest ($\log$([O~{\sc iii}]/H$\beta$)$>$0.6; see
Table~\ref{Tab:OIIIregions}), indicating excitation dominated by
AGN activity in these regions (e.g., \citealt{Kewley2006}). Further
evidence for the broad emission lines
being associated with AGN activity is found in the source SMM\,J0217$-$0503, which is 
comprised of two separate systems, and the broad emission line is
only observed in the system that is identified as hosting AGN activity by
\cite{AlaghbandZadeh2012} (also see Appendix A1). In addition, SMM\,J1636+4057
displays two broad [O~{\sc iii}] components which are co-spatial with the H$\alpha$ broad-line region of the AGN (Men\'endez-Delmestre
et~al. 2012\nocite{MenendezDelmestre2012}; see Appendix A8).

\subsection{Velocity profiles}
\label{Sec:velprofiles}

\begin{figure*}
\centerline{\psfig{figure=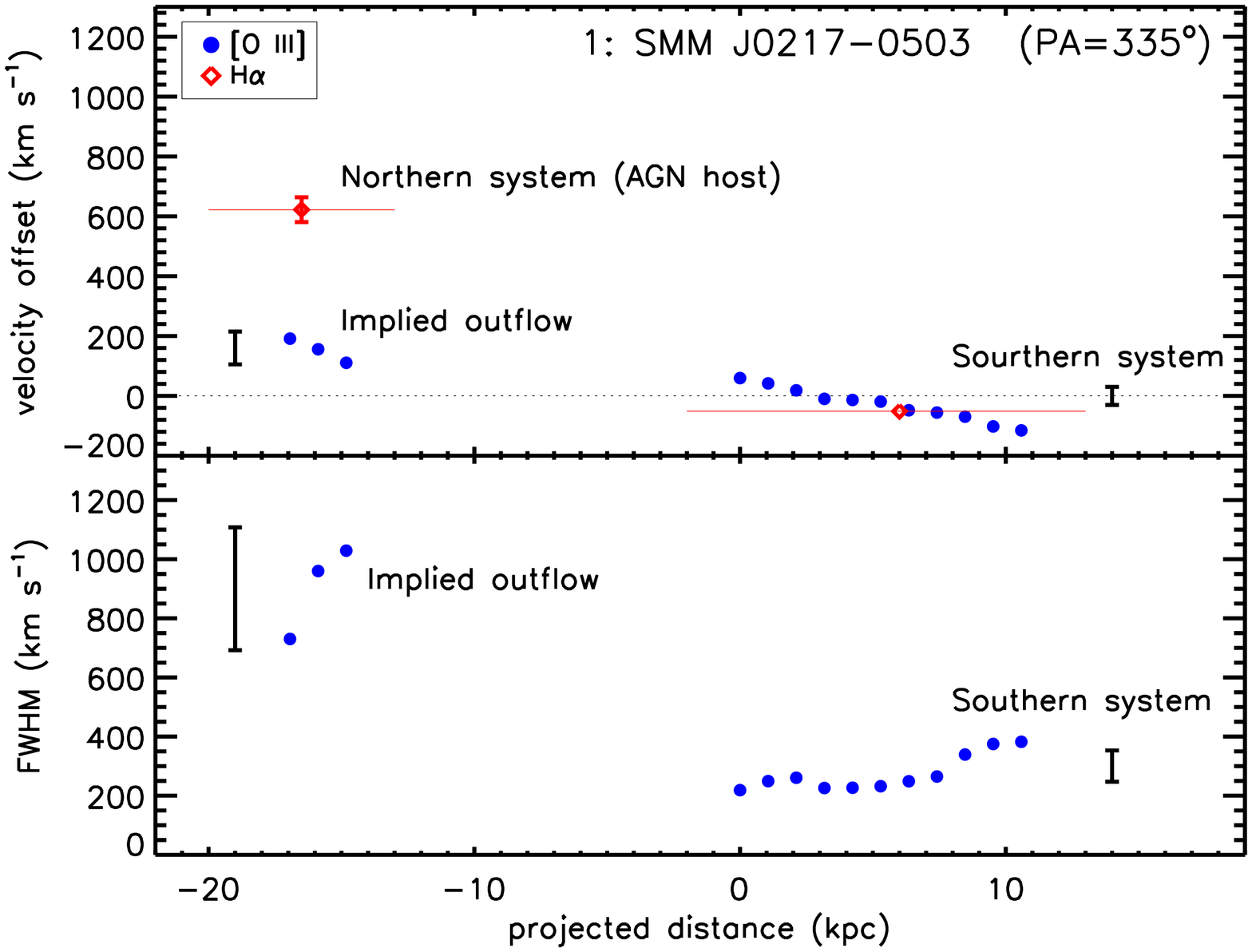,width=3.5in,angle=90}\psfig{figure=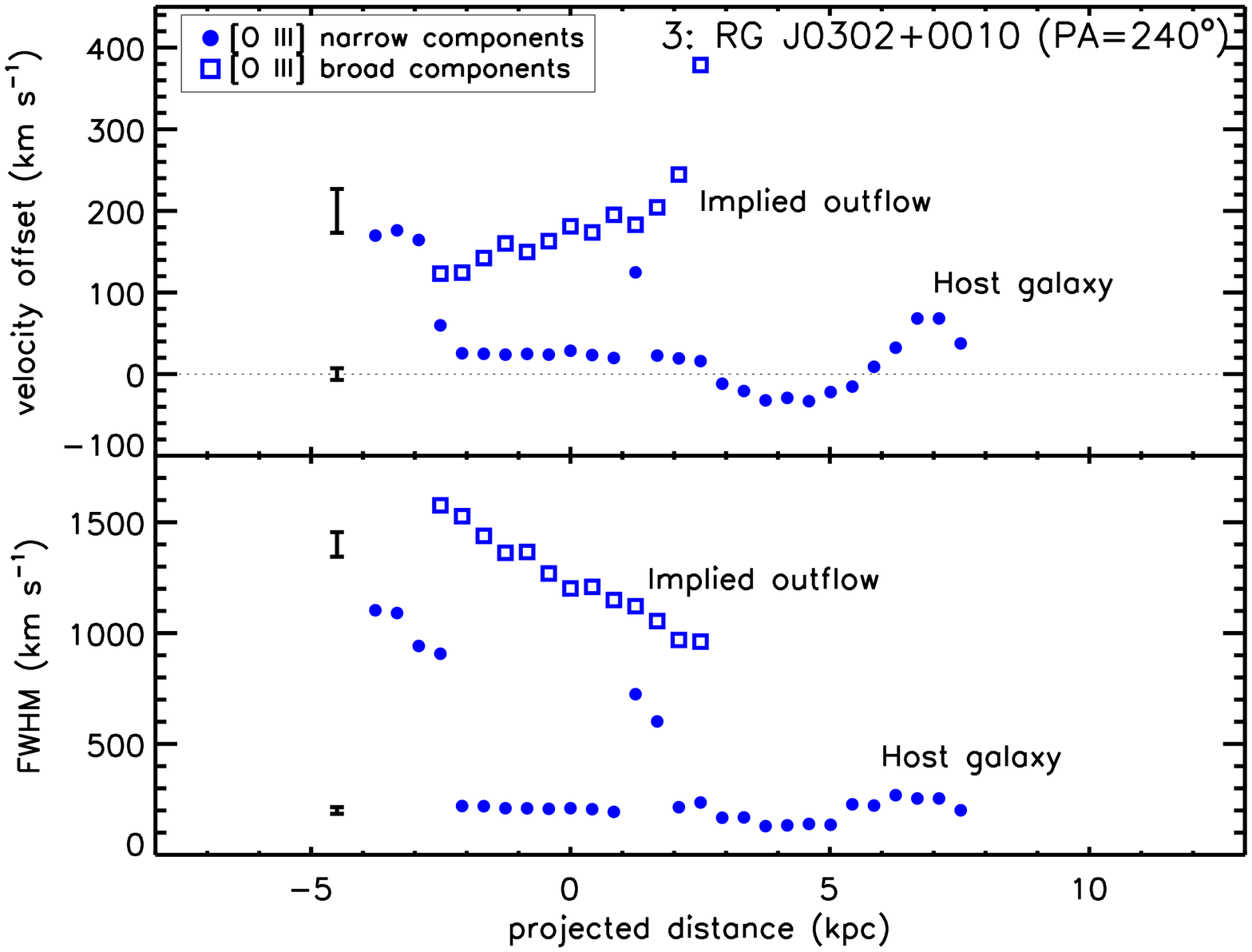,width=3.5in,angle=90}}
\centerline{\psfig{figure=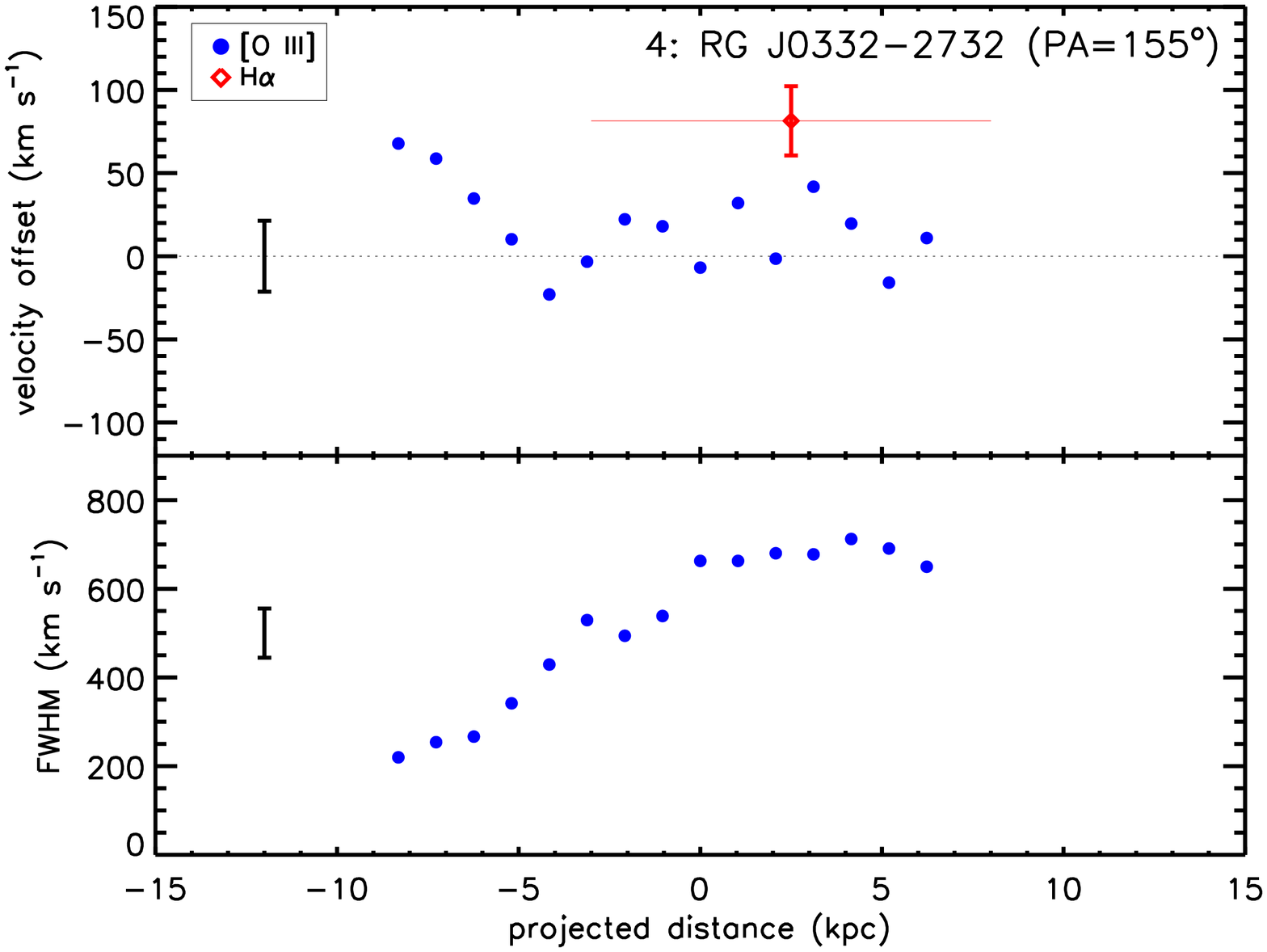,width=3.5in,angle=90}\psfig{figure=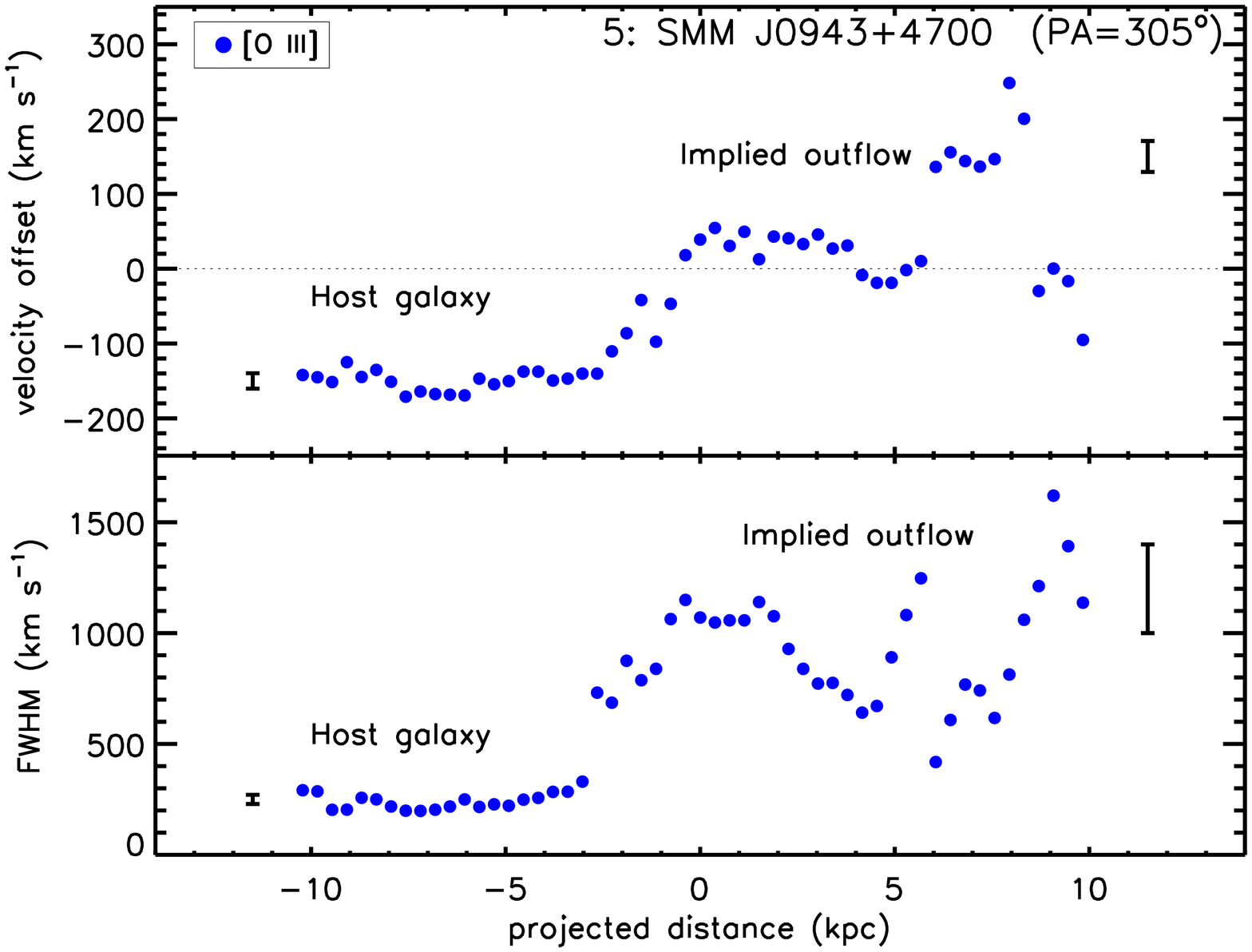,width=3.5in,angle=90}}
\centerline{\psfig{figure=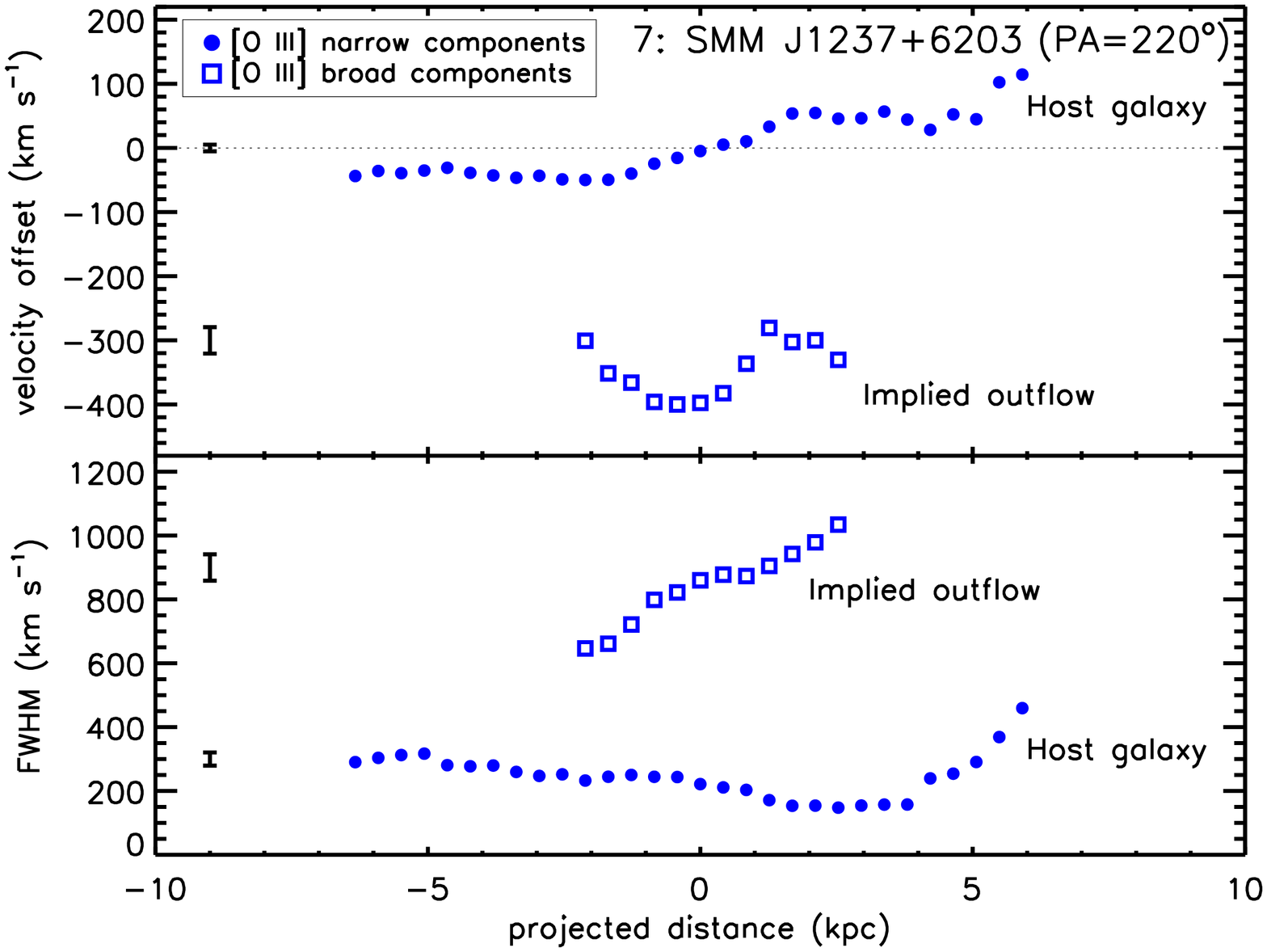,width=3.5in,angle=90}\psfig{figure=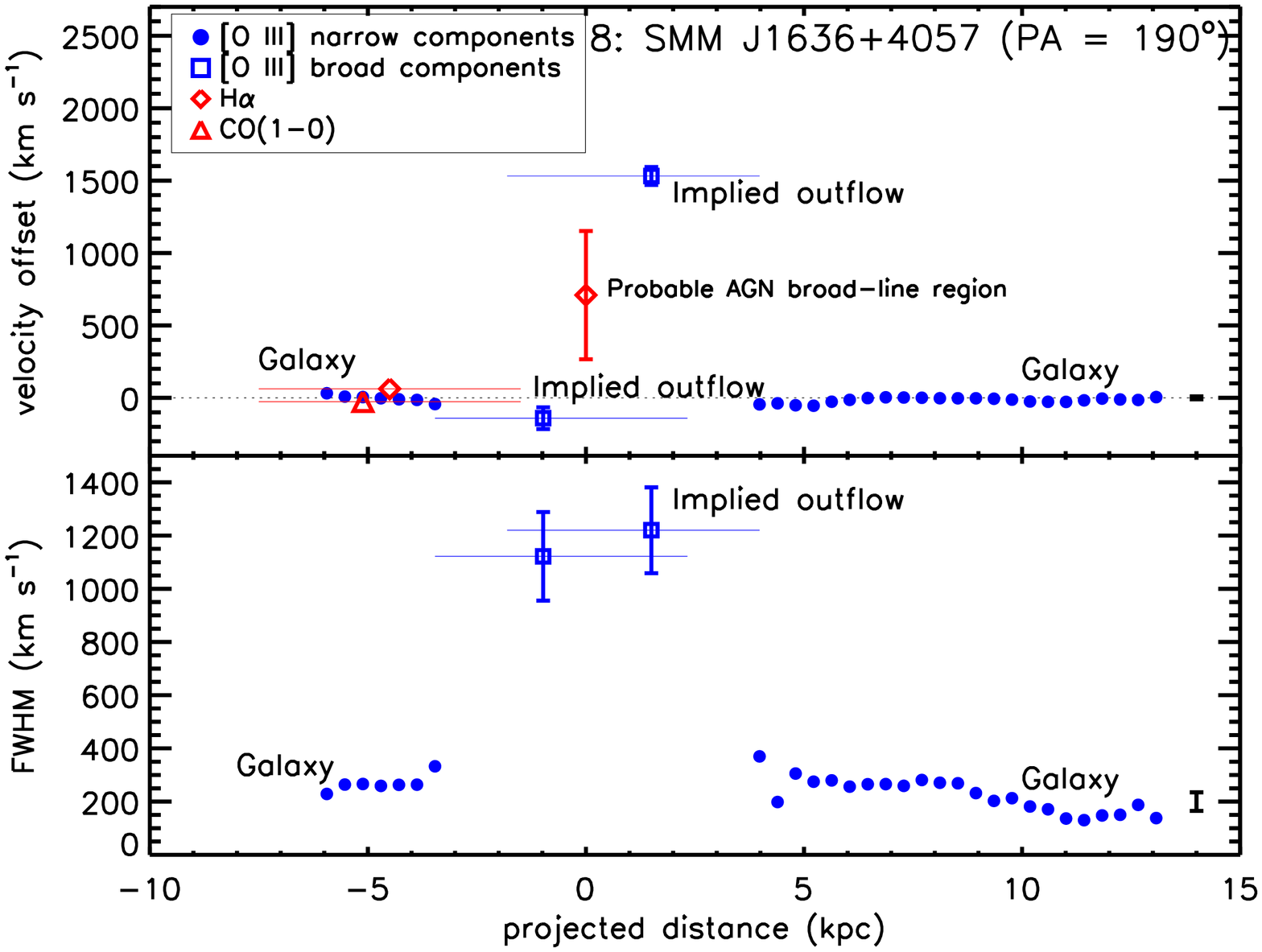,width=3.5in,angle=90}}
\vspace{-0.3cm}
\caption{A velocity and FWHM profile of the [O~{\sc
    iii}] emission lines for our six spatially resolved sources, along
  vectors at the angles indicated in the upper-right of each panel
  (see \S\ref{Sec:velprofiles} for how these were
  produced). When it was possible to de-couple the [O~{\sc iii}] emission-line
  profiles into two components we have shown these as filled circles
  (narrow) and open squares (broad). In some cases H$\alpha$ data and CO(1--0) data are
  plotted, taken from the literature (\citealt{Swinbank2005};
  \citealt{Ivison2011}; \citealt{AlaghbandZadeh2012};
  Men\'endez-Delmestre et~al. 2012). The data points for the broad
  [O~{\sc iii}] components for the source SMM\,J1636+4057 are taken
  from the fits to the spectra shown in Fig.~\ref{fig:regions}. In all cases there are underlying
  narrow [O~{\sc iii}] emission lines (FWHM $\approx$ few hundred
  km\,s$^{-1}$) with small velocity gradients ($\Delta
  v \lesssim$200\,km\,s$^{-1}$) which we associate with galaxy
  dynamics and merger remnants. Broad [O~{\sc iii}] emission lines (FWHM$\approx$700--1400\,km\,s$^{-1}$) are
  offset from these galaxy kinematics which we interpret as
  energetic outflows. The error bars
  indicate are representative 1$\sigma$ uncertainties for the [O~{\sc iii}]
  measurements. The horizontal solid lines indicate the approximate
  uncertainties in the spatial positions of the measurements when
  applicable. The dotted lines indicate zero velocity using the
  redshift of the narrowest [O~{\sc iii}] components in the
  galaxy-integrated spectra (Fig.~\ref{Fig:int_specs}).
}
\label{fig:velcurves}
\end{figure*}

To further quantify the [O~{\sc iii}] velocity structure and spatial extents of the different kinematic components in our
sample we created velocity and FWHM profiles for each of the six spatially
resolved sources and show them in Fig.~\ref{fig:velcurves}. To produce these profiles we first repeated the pixel-by-pixel fitting
routine outlined in \S\ref{Sec:velmaps} with the addition of allowing
a second Gaussian component to be fit at each
pixel, if this resulted in a significant improvement ($\Delta\chi^{2}
> 16$; equivalent to $\gsim4\sigma$) over a single Gaussian component. Only two of the
sources (RG J0302+0010 and SMM J1237+6203) had sufficiently bright
narrow and broad [O~{\sc iii}] emission-line components to
de-compose the profiles using this
method. For these two sources we took these
double-Gaussian component FWHM and velocity maps and took a running median of
all the pixels orthogonal to a vector through
the most extended emission (see dashed lines in Fig.~\ref{fig:velmaps}). For the other four
spatially resolved sources (SMM\,J0217$-$0503; RG\,J0332$-$2732;
SMM\,J0943+4700 and SMM\,J1636+4057) we used the
single-Gaussian component FWHM and velocity maps shown in
Fig.~\ref{fig:velmaps}. For the sources SMM\,J0217$-$0503 and
RG\,J0332$-$2732 we also plot the relative velocities of the
H$\alpha$ emission, extracted from the regions shown in
Fig.~\ref{fig:regions}, on these velocity
profiles. For the source SMM\,J1636+4057 we plot the
[O~{\sc iii}] FWHM and velocity offset values from the spectra extracted from the broad [O~{\sc
  iii}] region
(Fig.~\ref{fig:regions}) and H$\alpha$ and CO(1--0) data from the
literature (\citealt{Swinbank2005}; \citealt{Ivison2011};
Men\'endez-Delmestre et~al. 2012; see Appendix A8 for more details). 

Fig.~\ref{fig:velcurves} enables us to separately trace the
kinematics of the narrow (FWHM $\approx$ a few
hundred km\,s$^{-1}$) and broad (FWHM
$\approx$700-1400\,km\,s$^{-1}$) [O~{\sc iii}] emission-line components. The
narrow emission lines have small velocity gradients ($\Delta v
\lesssim$200\,km\,s$^{-1}$) and appear to be at roughly the same velocity
as the H$\alpha$ emission and molecular gas, when these data are available for
comparison. In contrast, the broad components are offset from the narrow emission
by $\Delta |v| \approx$150--1400\,km\,s$^{-1}$ and are found over
$\approx$4--15\,kpc in observed linear extent (except SMM\,J0217$-$0503
for which the broad emission is spatially unresolved; $\le$4\,kpc). The source RG\,J0332$-$2732 has emission
that is not clearly separated into broad and narrow components;
however, we note that the emission is broadest (FWHM$\approx$700\,km\,s$^{-1}$) in the brightest central region. The source SMM\,J0217$-$0503 consists of two separate systems, which are indicated in
Fig.~\ref{fig:velcurves}. In the northern system we do not observe any
narrow emission lines; however, the [O~{\sc iii}]
emission has a large velocity offset from the peak in the H$\alpha$
emission ($\Delta v=-460\pm60$\,km\,s$^{-1}$; see Appendix A1 for
further details). Of particular note is the
source SMM\,1636+4057 for which the broad components are found at
$v\approx\pm850$\,km\,s$^{-1}$ from the systemic velocity of the the H$\alpha$ broad-line
region of the AGN (\citealt{Swinbank2005}; Men\'endez-Delmestre
et~al. 2012;
see Appendix A8).

Due to the large uncertainties in the velocity offsets and FWHM of the
broad [O~{\sc iii}] components, we do not interpret their kinematic structure any further
than quoting their spatial extent and velocity offsets from the
systemic redshifts. However, we did perform tests to ensure that the
broad emission lines are truly spatially extended and that the observed extents are
not just the result of intrinsically compact regions being extended by
the seeing during our observations. Following \cite{Alexander2010} we
did this by extracting spectra from different regions and we found
velocity structure in the broad emission lines, confirming that they
are spatially extended (see Appendix A for
details on individual sources). 

For the remainder of this work we quote the spatial extent of the
broad [O~{\sc iii}] emission as the extent of
the broad [O~{\sc iii}] regions shown in Fig.~\ref{fig:regions} (see Table~\ref{Tab:OIIIregions}). These
values are consistent, within the quoted uncertainties, with spatial
extents of the broad [O~{\sc iii}] emission found in the velocity profiles shown in
Fig~\ref{fig:velcurves}. The source SMM\,J1237+6203 does not
  appear in Fig.~\ref{fig:regions} (see \S\ref{Sec:regions}), we
  therefore take the spatial extent from Fig.~\ref{fig:velcurves} 
  ($\approx$5\,kpc) which is conservative based on the 4--8\,kpc extent found by \cite{Alexander2010} for this source. The quoted spatial extents are likely to be lower limits as they are
not corrected for orientation and low surface brightness broad [O~{\sc iii}] emission will not be detected
by our fitting procedures (described in \S\ref{Sec:velmaps}). We define the velocity
offsets of the broad emission lines from the systemic, as the velocity
differences between the broad and narrow [O~{\sc iii}] emission lines in the spectra extracted from
the broad [O~{\sc iii}] regions (see Fig.~\ref{fig:regions} and Table~\ref{Tab:OIIIregions}). For
SMM\,J0217$-$0503 and RG\,J0332$-$2732 we use the velocity offset to the
narrow H$\alpha$ emission, as no
narrow [O~{\sc iii}] emission is observed. For SMM\,1636+4057 we use half of the
velocity offset between the two broad components, which is also
consistent with the offset between each component and the velocity of
the H$\alpha$ broad-line region (see Appendix A8).


\section{Discussion}
Our IFU observations of eight $z = 1.4-3.4$
ULIRGs that host AGN activity have revealed
extreme gas kinematics with very broad [O~{\sc
  iii}] emission (FWHM$\approx$700--1400\,km\,s$^{-1}$), extended up to 15\,kpc and
with high-velocity offsets, up to
$\approx$850\,km\,s$^{-1}$. We have also identified narrow [O~{\sc
  iii}] emission lines with moderate widths and velocity gradients (FWHM$\lesssim$500\,km\,s$^{-1}$; $\Delta
v \lesssim$200\,km\,s$^{-1}$) over the full extent of the observed
emission-line regions (up to $\approx$\,20\,kpc). In the following
discussion, we focus on the six sources for which we have
high-quality spatially-resolved data and show that the narrow emission
lines are consistent with tracing galaxy-dynamics and mergers, while the broadest emission lines are consistent
with being predominantly due to energetic outflows. We then go on to discuss the
impact that these outflows could be having on the future growth of the
central BHs and host galaxies and the implications of these results
for the formation of massive galaxies.

\subsection{Tracing galaxy dynamics, mergers and outflows}
\label{Sec:NarrowLines}

\begin{figure}
\centerline{\psfig{figure=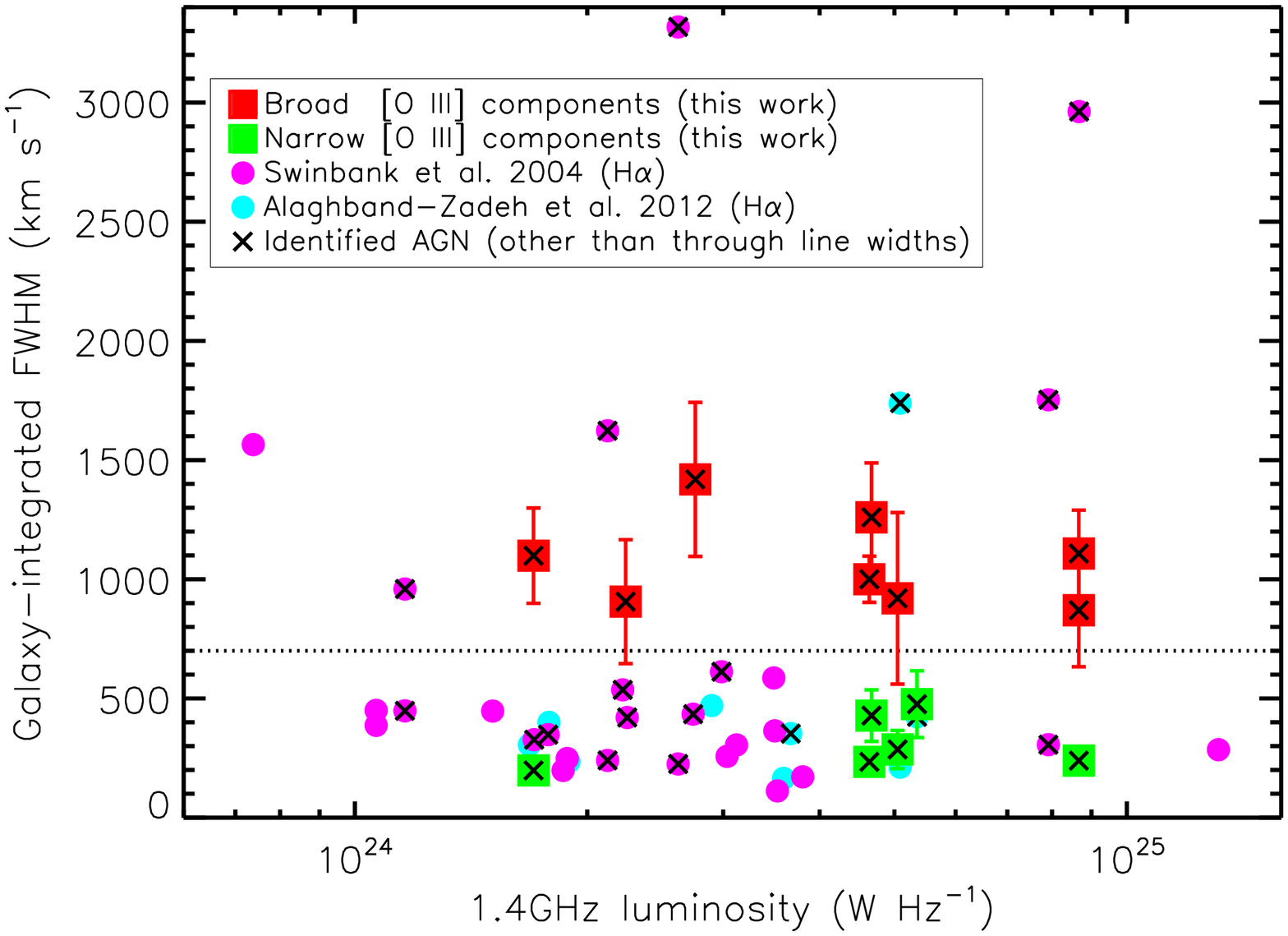,width=3.8in,angle=90}}
\caption{The [O~{\sc iii}] FWHM from the galaxy-integrated
  spectra of our eight [O~{\sc iii}] detected targets and
  high-redshift ULIRGs from the literature versus 1.4\,GHz radio
  luminosity. Both broad components (FWHM$\gsim$700\,km\,s$^{-1}$;
  dashed line) and narrow components of the profile fits have been plotted (see
  Table~\ref{Tab:OIIIprops}). For SMM\,J0217$-$0503 we have plotted the
  northern and southern systems separately
  (see Table~\ref{Tab:OIIIregions}). We also plot the galaxy
  integrated H$\alpha$ FWHMs from other high-redshift ULIRGs with similar radio luminosities (\citealt{Swinbank2004}; \citealt{AlaghbandZadeh2012}). The
  sources that are identified as AGN (using methods not based on
  emission-lines widths), have been highlighted (see \S~\ref{Sec:NarrowLines} for
  details). The non-AGN ULIRGs have FWHMs that lie below
  $\approx$700\,km\,s$^{-1}$. The narrow [O~{\sc iii}] components of
  our sample are consistent with
  the emission-lines widths from non-AGN ULIRGs and presumably trace galaxy dynamics and mergers
  while the broadest components appear to be constrained to the AGN
  host galaxies and are likely to be tracing AGN-driven outflows.}
\label{fig:narrow_comps}
\end{figure}

In six of our sources we find an extended [O~{\sc iii}] emission-line region (up to
$\approx$20\,kpc; Fig.~\ref{fig:velmaps}), which are often found in
gas-rich systems containing luminous AGN (e.g., \citealt{Colina1999}; \citealt{Nesvadba2008}; \citealt{Greene2011};
\nocite{VillarMartin2011a}Villar-Mart\'in et~al. 2011a; \citealt{Matsuoka2012}). The
presence of these emission-line regions enables us to trace the kinematics of the
ionised gas in these galaxies.

We identified narrow [O~{\sc iii}] emission (FWHM up to a few
$\times$100\,km\,s$^{-1}$) with irregular
velocity profiles and small velocity gradients ($\Delta v\lesssim
200$\,km\,s$^{-1}$) over the full extent of the emission-line regions
(Fig.~\ref{fig:velmaps} and Fig.~\ref{fig:velcurves}). The line widths
and velocity offsets observed in this narrow [O~{\sc iii}] emission are
similar to the kinematics of gas found in simulated merger remnants
(\citealt{Cox2006}). The kinematics of
these narrow emission lines are also broadly consistent with other
high-redshift star-forming galaxies that have similar
radio-luminosities, many of which are also undergoing mergers (e.g.,
\citealt{Swinbank2004}; \nocite{Nesvadba2007b}Nesvadba et~al. 2007b; \citealt{Lehnert2009};
\citealt{AlaghbandZadeh2012}) as well as nearby ULIRGs (e.g., \citealt{Colina2005}). To illustrate this, in
Fig.~\ref{fig:narrow_comps}, we compare the galaxy-integrated
emission-line FWHMs of our AGN sample to the H$\alpha$ emission line
FWHM in other high-redshift ULIRGs from \cite{Swinbank2004} and \cite{AlaghbandZadeh2012}. We highlight sources that are identified as AGN
on the basis of their emission-line ratios, X-ray emission, UV
spectral features or their mid-infrared spectra. Fig.~\ref{fig:narrow_comps} shows that the narrow
[O~{\sc iii}] emission lines (FWHM$\lesssim$500\,km\,s$^{-1}$) in our sources are
comparable to the emission-line widths of non-AGN high-redshift ULIRGs and therefore
are likely to be tracing the same kinematic components seen in these
sources (i.e., galaxy dynamics and mergers). The velocity
structure of the emission lines seen in the IFU data of the non-AGN
ULIRGs (\citealt{AlaghbandZadeh2012}) are also similar to the narrow [O~{\sc iii}] components observed here. In particular, for the two
sources where there are IFU observations covering both the [O~{\sc
  iii}] and H$\alpha$ emission lines (SMM\,J0217-0503 and RG\,J0332-2732; this work and \citealt{AlaghbandZadeh2012}) the
kinematics of the narrow [O~{\sc iii}] and H$\alpha$ emission-line
components are roughly consistent (e.g., see Fig.~\ref{fig:regions}). 

In contrast to the narrow emission lines, the identification of broad emission lines (FWHM$\gsim$700\,km\,s$^{-1}$)
in the galaxy integrated spectra of high-redshift ULIRGs
appears to be constrained to galaxies that contain known AGN
activity (Fig.~\ref{fig:narrow_comps}).\footnote{The one source from
\cite{Swinbank2004} with FWHM$>$700\,km\,s$^{-1}$ that is not
identified as an AGN (Fig.~\ref{fig:narrow_comps}) has a complex spectrum and has not been covered
by multi-wavelength observations in the literature. It is
therefore not possible to rule out the presence of a significant AGN in this galaxy.} This is in agreement with
studies of low-redshift ULIRGs where the broadest emission lines are found in
systems hosting powerful AGN activity (e.g., \citealt{Veilleux1999},
\citealt{Zheng2002}; \citealt{Westmoquette2012}). Although broad emission-line components may be found in
non-AGN high-redshift star-forming galaxies, they are generally faint and are difficult to identify other
than by stacking the spectra of multiple sources or with excellent
quality data (e.g., \citealt{Swinbank2004};
\nocite{Nesvadba2007b}Nesvadba et~al. 2007b;
\citealt{LeTiran2011}). However, we note that we are basing these
arguments on small samples of galaxies which are biased in their
selection and to fully address if
bright and broad emission lines, such as those in our sources, are unique to galaxies hosting AGN activity,
requires larger systematic samples of spectra and IFU data. In addition to comparisons with the literature, there is further
evidence that the broadest emission lines are associated with AGN activity
in our sample. The broadest emission lines are found to be spatially co-incident with the most likely
location of the AGN in these galaxies based on very high [O~{\sc
  iii}]/H$\beta$ emission-line ratios (see \S\ref{Sec:regions}). The most striking example of
this is the source SMM\,J1636+4057 where the broadest [O~{\sc iii}] emission lines
are found to be co-spatial with the AGN, and not in the region of unobscured
star formation (details in Appendix A8). 

Broad and asymmetric [O~{\sc  iii}]$\lambda\lambda$4959,5007 emission-line profiles, such as
those found in this work, are most commonly attributed to
high-velocity outflowing gas (e.g., \citealt{Heckman1981}; \citealt{Greene2005}; \citealt{Barth2008};
\citealt{Nesvadba2008}; \citealt{Greene2011}).  In particular IFU
observations of local ULIRGs that host AGN activity reveal broad and spatially
extended emission lines (similar to those that we observe) associated with
energetic outflows (\citealt{Rupke2011}; \citealt{Westmoquette2012}). On the basis of the above arguments the broadest kinematic components
(FWHM$\gsim$700\,km\,s$^{-1}$) appear to be associated with AGN
activity and therefore do not appear to be the result of mergers and
galaxy dynamics or even (as we will further argue below) star-formation driven outflows, and therefore
AGN-driven outflows is the most natural explanation. We find convincing evidence for extended ($\gsim 4$\,kpc) outflows in four of
our sample (RG\,J0302+0010; SMM\,J0943+4700; SMM\,J1237+6203 and
SMM\,J1636+4057). The source RG\,J0332-2732 only has very broad [O~{\sc iii}] emission lines in the
most central regions, and the broad [O~{\sc iii}] emission line in
the source SMM\,J0217-0503 has a low signal-to-noise ratio and is spatially
unresolved. These two sources have lower quality data than the
other four spatially extended sources and the evidence for outflows in
these objects is weaker (see Appendix A1 and A4 for further discussion
on these two sources).

\subsection{Outflow properties}
\label{Sec:impact}
\begin{figure*}
\centerline{\psfig{figure=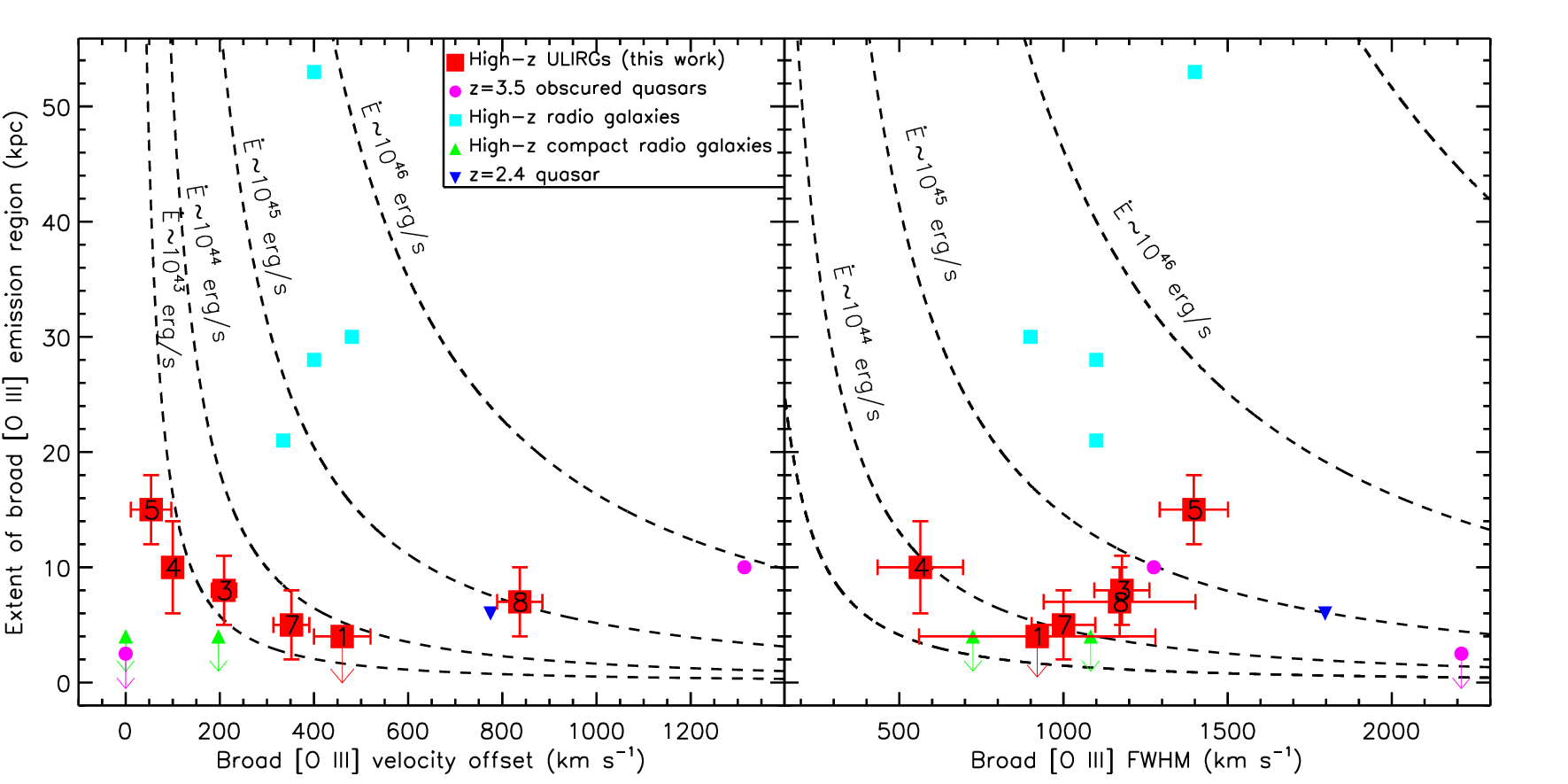,width=7in,angle=90}}
\caption{The projected spatial-extent of the broad [O~{\sc iii}]
  emission-line regions against
velocity offset from the systemic redshift ({\it left}) and FWHM ({\it
  right}) for our spatially resolved sources (see
  Table~\ref{Tab:OIIIprops} and Table~\ref{Tab:OIIIregions}). The
  numbers inside the symbols correspond to the source IDs. Also shown are other high-redshift sources with
  IFU data (symbols are the same as in Fig.~\ref{fig:Selection}). If two sides of an
  outflow are observed, half of the velocity difference between them
  is used, and the average FWHM of each side. For spatially
  unresolved data we use the seeing disk as an upper limit of the
  extent. For illustrative purposes, the dashed
  lines show the expected energy injection rates ($\dot{E}$) using a
  simple outflow model (see \S\ref{Sec:impact} for details).
    Strong evidence for extended outflows is not confirmed for source IDs
  1 and 4 (see \S\ref{Sec:NarrowLines}). Despite 3--4 magnitudes difference in radio luminosities, HzRGs and high-redshift
  ULIRGs containing AGN activity have comparable velocity offsets and
  emission-line widths and are both potentially dumping energy into their host galaxies at
 considerable rates ($\approx$10$^{43}$--10$^{46}$\,erg\,s$^{-1}$). 
  }
\label{fig:EvR}
\end{figure*}

\begin{figure}
\centerline{\psfig{figure=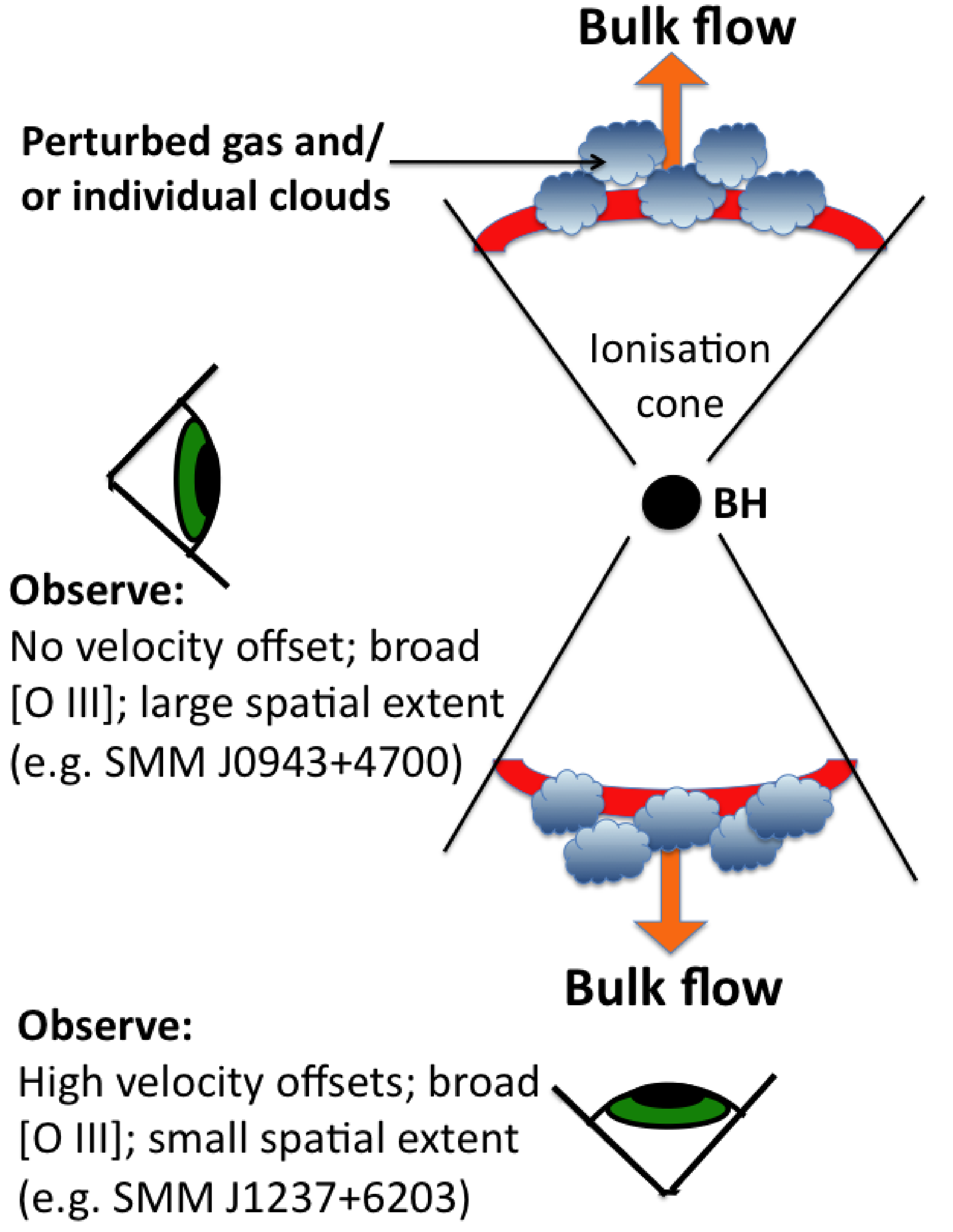,width=2.5in}}
\caption{A schematic diagram to illustrate a possible interpretation of the
  observations of the broad [O~{\sc iii}] emission lines. If a fast wind is
initially launched around the central BH it could sweep up clouds of
gas in a bi-polar outflow (as assumed by some models e.g., \citealt{DeBuhr2012};
\citealt{Zubovas2012}; see \S~\ref{Sec:impact} for more discussion). The
orientation with respect to the line of sight will determine the velocity offsets
and spatial extent observed; however, broad emission lines will be always be
observed. This could explain the diversity in the properties of our
observations. In addition, obscuring material in the host galaxy would result in parts
of the outflow not being visible which could explain why in some
sources we only observe either blueshifted or redshifted broad emission lines.}
\label{fig:schematic}
\end{figure}

We can use the measured properties of the broad [O~{\sc iii}] emission
to place constraints on the outflow properties. In Fig.~\ref{fig:EvR}
we plot the observed spatial extent of the broad
[O~{\sc iii}] emission lines against the velocity offset from the systemic redshifts of the host
galaxies (values are from the fits to the spectra shown in Fig.~\ref{fig:regions} and are provided in
Table~\ref{Tab:OIIIregions}).\footnote{\label{Footnote:SMM12} The source SMM\,J1237+6203 does
  not appear in this table. For this source we use a spatial extent of
  5\,kpc (see Fig.~\ref{fig:velcurves}) and the FWHM and velocity
  offset from the galaxy-integrated spectrum (Table~\ref{Tab:OIIIprops}). See the end of
  \S\ref{Sec:velprofiles} for details.} The dashed lines in Fig.~\ref{fig:EvR} represent estimated
energy injection rates which we discuss later. The broad
[O~{\sc iii}] emission-line properties ($\Delta
v\approx$50--850\,km\,s$^{-1}$; spatial extent $\approx$4--15\,kpc) can be explained by
the presence of an AGN-driven bi-polar outflow plus the
addition of obscuring material in the host galaxy. In Fig.~\ref{fig:schematic} we show a simple schematic diagram to attempt to
explain the properties and diversity of the broad [O~{\sc iii}]
emission lines. Hydrodynamical simulations
and analytical models predict that a powerful AGN-wind, launched from the central BH, could sweep up clouds of gas resulting in an
kpc-scale outflow (e.g., \citealt{DeBuhr2012};
\citealt{Zubovas2012} and references therein). Galaxy-wide outflows are thought
to be expanding bubbles on either side of the host galaxy, being forced into this shape by the denser
material in the centre of the host-galaxy (e.g.,
\citealt{Zubovas2012}; \citealt{FaucherGiguere2012}). As we are
observing the outflow through luminous [O~{\sc iii}] emission, the
{\it observed} properties of the outflow will also depend on the
size and orientation of the AGN ionisation cones (e.g., \citealt{Crenshaw2010}). Multiple clouds being caught
up in the bulk flow of the outflow (e.g., \citealt{Zubovas2012}) or
perturbed ionised gas would result in broad [O~{\sc iii}] emission
lines (e.g., \nocite{VillarMartin2011b} Villar-Mart\'in
et~al. 2011b). The `near-side' of
the outflow would therefore be observed as a broad and blueshifted
emission line and the `far-side' would be observed as broad and redshifted
emission (see Fig.~\ref{fig:schematic}). However, part of the outflow could be obscured by
material in the host galaxy, as is likely to be the case in clumpy
and dusty systems such as high-redshift ULIRGs (e.g.,
\citealt{Ivison2007}; \citealt{Tacconi2008};
\citealt{Engel2010}). High-redshift ULIRGs are kinematically
disturbed, potentially resulting in a complex distribution of
obscuring material, which, along with the relative orientation of the AGN, could explain why in
some cases we only see either a blueshifted (i.e., when only the
near-side of the outflow is observed) or redshifted broad
emission line (i.e., when only the far-side of the outflow is
observed; also see Fig. 6 and 7 in \citealt{Crenshaw2010}). 

The range in spatial extents and velocity
offsets of the implied outflows that we observe (Fig.~\ref{fig:EvR}) can also be explained using this simple model
(Fig.~\ref{fig:schematic}). The {\it observed} velocity offsets
and {\it observed} spatial extents will be highly dependent on the
orientation of the outflow with respect to the line of sight. If the
axis of the outflow is orientated along the line of sight, high
velocity offsets and a small spatial extent would be observed (e.g., this
could explain the properties of SMM\,J1636+4057). Conversely, if
the axis of the outflow is in the plane of the sky, a small velocity offset and
a large spatial extent would be observed (e.g., this could explain the
properties of SMM\,J0943+4700); see Appendix A for further discussion
on individual sources. In contrast, the observed FWHM will be less
dependent on the orientation. We emphasise that this is a very simplified model and other interpretations could also be
valid. For example, the observed emission-line properties will also be dependent on the
intrinsic properties of the outflow, such as their intrinsic velocity and
distance traveled from the BH.

We have argued that the measured velocity offsets observed may not
represent the true velocity of the outflow due to an orientation effect. We attempt to estimate corrected velocity offsets by taking the two sources which contain a known broad-line AGN (SMM\,J1237+6203
and SMM\,J1636+4057; see Appendix A for details). In these
sources the identification of the broad-line AGN component suggests
that the outflow could be primarily orientated
towards us and therefore the observed velocity offset is likely to be
close to the true outflow velocity. For these two sources we find
that the average of FWHM/$\Delta v$ is $\approx$\,2; reassuringly we also find a median
ratio of FWHM/$\Delta v\approx$\,2 for low-redshift broad-line
AGN that display [O~{\sc iii}] properties indicative of
outflows (data from Mullaney et~al. in prep). We therefore suggest that FWHM/2 may
be an adequate approximation of the true outflow rate, although subject to
large uncertainties. Using this as a diagnostic, we estimate bulk outflow
velocities in the range of $\approx$300--700\,km\,s$^{-1}$, which are
similar to the predicted kpc-scale outflow velocities predicted by hydrodynamical
simulations and analytical models of outflows
($\approx$300-1500\,km\,s$^{-1}$), driven by AGN in objects with similar
properties to the sources in our sample (i.e., ULIRGs, with $L_{\rm{AGN}}\approx$10$^{46}$\,erg\,s$^{-1}$; \citealt{DeBuhr2012};
\citealt{Zubovas2012}; \citealt{FaucherGiguere2012}).

\subsubsection{Estimating outflow energy rates and determining what is
  powering the outflows}
\label{Sec:energy}
To attempt to quantify the impact the outflows may be having on the
host galaxies we consider approaches to estimate the energy injection rates into the
ISM. We do not have good constraints on mass of the gas entrained in
the outflows or on the geometry of the outflow, making such estimates uncertain. For illustrative purposes, we consider the energy injection rates assuming an energy conserving bubble into a
uniform medium (e.g., \citealt{Heckman1990}; \citealt{Veilleux2005}; \citealt{Nesvadba2008};  and references therein) which gives the following relation:
\begin{equation}
\dot{E} \approx1.5\times10^{46}r^{2}_{10}v^{3}_{1000}n_{0.5}$\,erg\,s$^{-1},
\label{Eq:energy}
\end{equation}
where $r_{10}$ is taken to be half the extent of the observed broad
[O~{\sc iii}] emission (in units of 10\,kpc) and $v_{1000}$ are the velocity offsets in units
of 1000\,km\,s$^{-1}$ (see Table~\ref{Tab:OIIIregions}). The ambient density (ahead of the expanding bubble), $n_{0.5}$, is in units of
0.5\,cm$^{-3}$ for which we have assumed $n_{0.5}$=1.\footnote{This
  density assumption is based on the the values of $n_{0.5}$=1--4 used
  in e.g. \cite{Heckman1990} and \cite{Nesvadba2008}. This value
  can be indirectly estimated from the (post-shock) electron density (see
\citealt{Heckman1990} and references therein for further details),
  derived from the [S~{\sc ii}]$\lambda\lambda6171,6731$ emission-line doublet which is sensitive to
electron densities in the range 100-10$^{5}$\,cm$^{-3}$
(\citealt{Osterbrock1989}). We used a stack of the rest-frame optical
spectra of the high-redshift ULIRGs from \cite{Swinbank2004} to derive a flux ratio of $R_{{\rm{[S II]}}}=F_{\lambda 6716}/F_{\lambda
  6731}=1.1\pm0.3$. This is consistent with the HzRGs of
\citealt{Nesvadba2008} who find $R_{{\rm{[S II]}}}$=1.0--1.1, such
that the implied densities of HzRGs and high-redshift ULIRGs are
similar. We therefore use the same value of $n_{0.5}=1$ as in \cite{Nesvadba2008}.
Although we have no direct way of measuring $n_{0.5}$ this unknown will only vary
our order-of-magnitude estimates of the energy injection rates by a
factor of a few.}

In the left panel of Fig.~\ref{fig:EvR} we plot on tracks indicating fixed
energy injection rates using Equation~\ref{Eq:energy}. We have already discussed that
the observed velocity offsets are unlikely to be representative of the
intrinsic velocities and that FWHM/2 is plausibly a better estimate of the true outflow
velocity. Therefore, in the right panel of Fig.~\ref{fig:EvR} we plot the extent of the broad
[O~{\sc iii}] emission against FWHM and tracks of constant energy injection rates using
Equation~\ref{Eq:energy}, but replacing the velocity offset with FWHM/2. For the two
sources with a known broad-line AGN in this sample (SMM\,J1237+6203
and SMM\,J1636+4057) and the sources with broad-line AGN in the comparison samples (i.e. \citealt{Nesvadba2006};
\citealt{CanoDiaz2012}), the two diagnostics for the bulk outflow velocity give
energy injection rates that are consistent within a factor of a few. We establish that the spatially extended outflows in these high-redshift ULIRGs are potentially injecting energy into their host galaxies at
considerable rates of $\approx$(0.08--\,3)$\times$10$^{45}$\,erg\,s$^{-1}$, where
the range indicates the values for different sources. Over
a typical AGN lifetime of 30\,Myr (e.g., \citealt{Hopkins2005}) the total energy injection would be of
order $\approx$(0.8--28)$\times$10$^{59}$\,ergs; however, we note
that outflows could continue $\sim$10 times longer than the active time of the
BH (\citealt{King2011}) which would increase this energy injection by
another order of magnitude. These outflow energy injection rates are comparable
to the typical binding energy of the host galaxies in these systems of
$\approx$10$^{59}$\,ergs (assuming a spheroid mass of
$\approx$10$^{11}$\,M$_{\sun}$, a velocity dispersion $\sigma
\approx$200\,km\,s$^{-1}$ and $R_{\rm{e}}$=4\,kpc; e.g., \citealt{Binney1987}) and demonstrate that these outflows could unbind at least a fraction of the ISM from the
host galaxies.

\begin{figure*}
\centerline{\psfig{figure=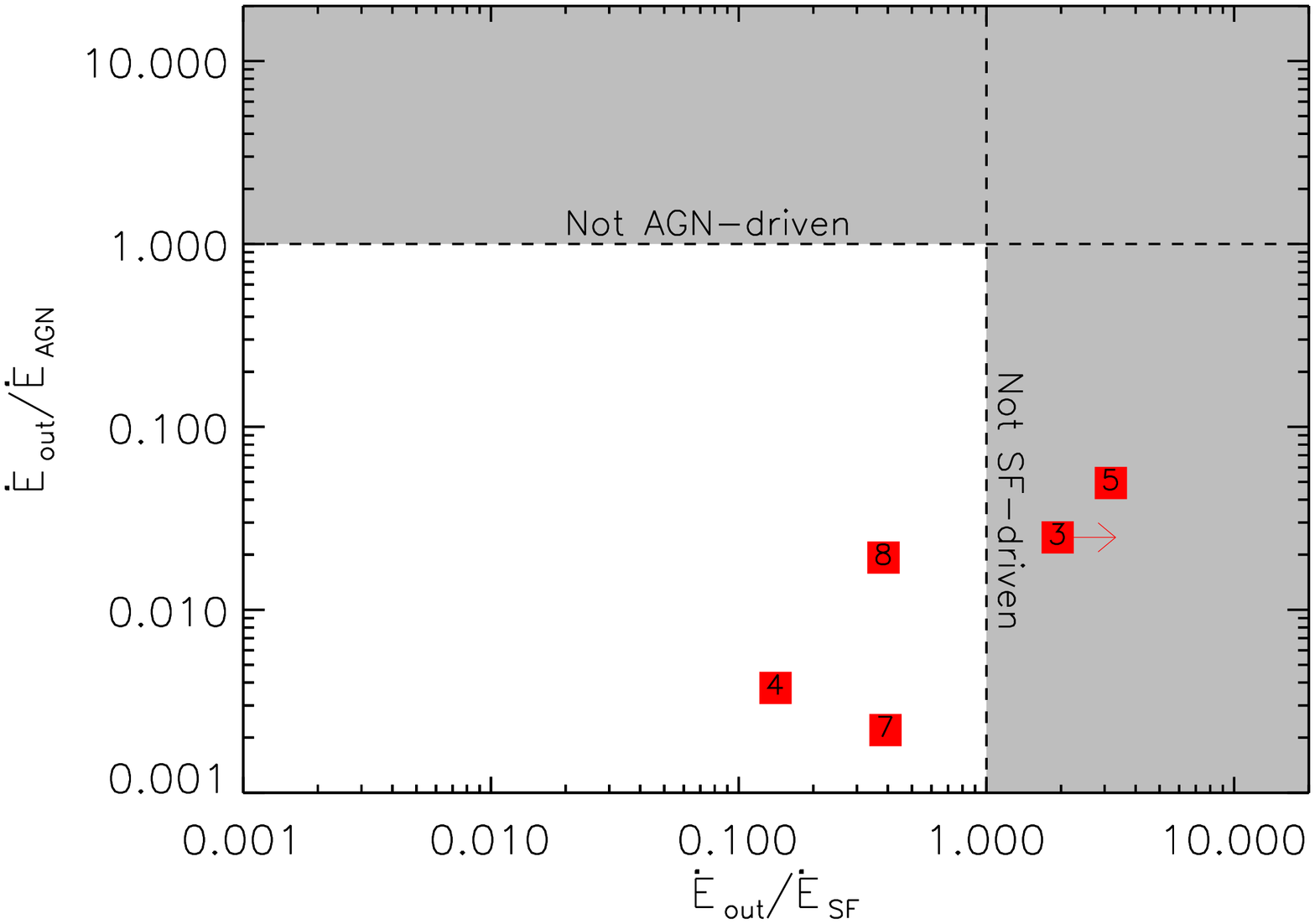,width=3.8in,angle=90}\psfig{figure=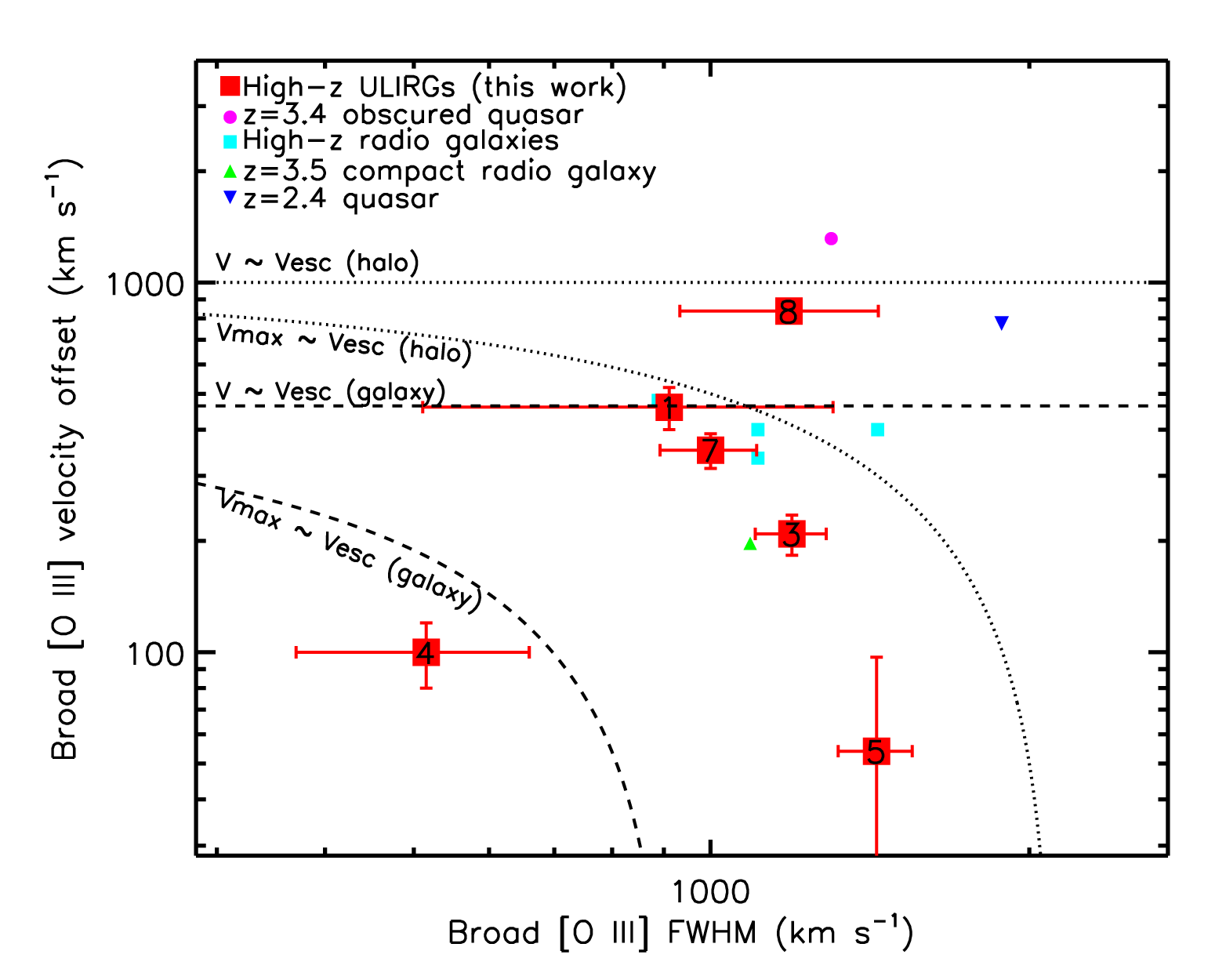,width=3.4in}}
\caption{{\it Left:} The ratio of the estimated outflow energy rate ($\dot{E}_{\rm{out}}$) to
  the derived input energy rate from AGN ($\dot{E}_{\rm{AGN}}$) against the ratio of the outflow energy rate to the
  predicted energy input rate from star formation ($\dot{E}_{\rm{SF}}$)
  for five of the six sources with spatially resolved [O~{\sc
    iii}] emission. For the sixth source we only have upper limits
  for both $\dot{E}_{\rm{AGN}}$ and $\dot{E}_{\rm{out}}$ and
  therefore have no constraints on the ratio of these quantities. The
  numbers inside the symbols correspond to the source IDs.
    We note that we lack strong evidence for extended outflows for source IDs
  1 and 4 (see \S\ref{Sec:NarrowLines}.) The shaded regions indicate where a $>$100\%
  coupling efficiency is required between the input energy and the gas to power the
  outflows, effectively ruling out such an input energy mechanism. The bolometric luminosity of the AGN appears to be
  energetically favourable to power all the outflows
  requiring only $\approx$0.2--5\% of the
  bolometric output to couple to the gas; however star formation cannot be ruled out
  in all cases. It should be noted that the values used rely heavily on
  assumptions and should be taken as illustrative only, with uncertainties
  at the order of magnitude level. Details of the derived values
  that are plotted are provided in \S\ref{Sec:impact} (also see
  Table~\ref{Tab:SEDpoints}). {\it Right:} Broad [O~{\sc iii}] emission-line velocity offset versus FWHM for
  the spatially resolved sources in our sample (values taken from
  regions of broad [O~{\sc iii}]; see
  Table~\ref{Tab:OIIIregions}). Also shown are other
  high-redshift sources with IFU data as described in Fig.~\ref{fig:Selection}. The
  maximum velocities ($v_{\rm{max}}=|v|+\frac{1}{2}$FWHM) of the gas are potentially high
enough to exceed the typical escape velocity of a typical massive galaxy
(10$^{11}$\,M$_{\sun}$ at 4\,kpc; dashed
curve). However, using simplistic arguments (see \S\ref{Sec:impact})
only one of our sources has a maximum outflow velocity that could
potentially unbind some of the gas from a typical galaxy halo
($\approx$10$^{13}$\,M$_{\sun}$ at $\approx$4\,kpc; dotted curve).}
\label{fig:Lbols}
\end{figure*}

We next consider what could be powering the outflows
based on energetic arguments. In Fig.~\ref{fig:Lbols} we compare the ratio of the outflow energy rates
with the potential energy input rates from: (1) the radiative output
of the AGN (i.e. the AGN bolometric luminosity; $L_{\rm{AGN}}$) and
(2) energy output from star formation
($\dot{E}_{\rm{SF}}$). We calculate outflow energy
injection rates using Equation 1, replacing $v$ with FWHM/2 and $r$
as half the spatial extent of the broad [O~{\sc iii}] region (see
Table~\ref{Tab:OIIIregions}; also see footnote~\ref{Footnote:SMM12}).  We have estimated
the bolometric luminosity of the AGN using SED modeling (described in
\S\ref{Sec:SEDfitting}; values given in
Table~\ref{Tab:SEDpoints}). We estimate the potential energy input
from star-formation using the star formation rates from
  our SED modelling (Table~\ref{Tab:SEDpoints}) and assuming that the mechanical energy injection
from supernovae and stellar winds would be at most
$\approx$7$\times$10$^{41}$($SFR$/$M_{\sun}$\,yr$^{-1}$)\,erg\,s$^{-1}$
(following \citealt{Leitherer1999} and
\citealt{Veilleux2005}).\footnote{ If we were to follow instead
  \cite{DallaVecchia2008} to estimate this energy input, the values would
  be a factor of $\approx$2 lower.  Conversely, radiation pressure from star-formation could
potentially contribute a comparable amount of pressure to stellar
and supernovae winds in ULIRGs (\citealt{Veilleux2005}). As we are
considering order of magnitude estimates only, these uncertainties
do not affect our conclusions.} We note that the assumptions made are highly
idealistic, and we have not, for example, considered the case of a momentum
conserving outflow. However, \cite{Nesvadba2006,Nesvadba2008} follow several methods to estimate
the energy injection in similar outflows in HzRGs and find that the different methods generally
agree on the order of magnitude level (also see \citealt{Veilleux2005} for further discussion on
the energetics of outflows and the potential energy input from star formation and
AGN activity).

Fig.~\ref{fig:Lbols} demonstrates that the radiative output of
  the AGN is energetically viable as the dominant power source of the outflows observed in
all of these sources. Assuming the bolometric luminosity provides an estimate of the initial energy
input from the AGN, so long as $\approx$0.2--5\% of this energy can
couple to the ISM, the radiative power of the AGN in these
systems is large enough to drive the observed outflows. This range of
coupling efficiencies is similar to the values required by many models to reproduce the
properties of local massive galaxies (e.g., \citealt{Springel2005};
\citealt{DiMatteo2005}; \citealt{DeBuhr2012}). In contrast, the required coupling
efficiencies for star formation are an order of magnitude
higher ($\approx$15\% to $\gg$100\%). For the three sources
with implied coupling efficiencies of $\approx$15--50\%  star formation
may play a significant role; however, based on the observed properties of the broad [O~{\sc iii}]
emission presented earlier (see \S\ref{Sec:NarrowLines}), and the
energetic arguments presented here, we tentatively suggest that the
AGN may play a dominant role in powering the outflows. Given the current quality of the available data, a more
detailed analysis is not yet warranted and the values of input energetics should be used with care.

\subsubsection{The potential fate of the outflowing gas}
\label{Sec:fate}
 Based on energetic arguments we have already demonstrated that the outflows we
observe are likely to unbind the gas from the host galaxies. However,
many models of galaxy formation require powerful AGN-driven outflows
to remove the gas completely from the host galaxy potentials to
efficiently truncate star formation and remove the fuel for future
accretion onto the BHs (e.g., \citealt{DiMatteo2005};
\citealt{Hopkins2008a}). Will the outflowing gas we observe be able to
escape the host galaxy and consequently remove the gas supply for
further star formation and future accretion? In Fig.~\ref{fig:Lbols}
we plot the broad [O~{\sc iii}] velocity offset against FWHM for our
spatially resolved sources and also plot other high-redshift AGN with
available IFU data from the literature. By taking an estimated galaxy
mass of 10$^{11}$\,M$_{\sun}$ (typical of these sources; e.g.,
\citealt{Engel2010}; \citealt{Wardlow2011}; \citealt{Hainline2011})
the escape velocity at a distance of 4\,kpc would be
$v_{\rm{esc}}\approx$460\,km\,s$^{-1}$ (see Fig.~\ref{fig:Lbols}). Considering the observed velocity offsets
($\Delta v$=50--850\,km\,s$^{-1}$), only the outflows in two sources
(SMM\,J1636+4057 and SMM\,J0217$-$0503) appear likely to exceed this
escape velocity. However, if we instead consider the the {\it maximum}
velocities of the emission-line gas ($v_{\rm{max}}\approx$400--1400\,km\,s$^{-1}$;
defined as $v_{\rm{max}}$ = ${|v|}$+FWHM/2; i.e.,
\citealt{Rupke2005a,Rupke2005b}) then the outflow velocities are found
to exceed the escape velocity (see dashed curve in
Fig.~\ref{fig:Lbols}) in all sources except RG\,J0332$-$2732,
suggesting that some fraction of the gas is able to escape the host
galaxy potential.

Unless the gas is completely removed from the host galaxy potential,
or is unable to cool on short timescales, significant star formation
and BH accretion may be able to recommence after the outflow
episode. We therefore consider a simplified approach to investigate if
the outflowing gas will be able to escape the dark-matter halo. Assuming a static dark matter halo with a mass of
$\approx$10$^{13}$\,M$_{\sun}$ (expected for these sources;
\citealt{Hickox2012}), a Navarro Frenk and White (NFW;
\citealt{Navarro1996}) density profile, and a point mass of
$\approx$10$^{11}$\,M$_{\sun}$ (i.e., the host galaxy), a particle
initially at $\approx$4\,kpc from the centre of the host galaxy would
need to be travelling at $v\approx$1000\,km\,s$^{-1}$ to escape beyond
the virial radius of the halo ($\approx$440\,kpc; M. Lovell
priv. comm.; see Fig.~\ref{fig:Lbols}). On the basis
of this simple model, where we assume the effects of gravity only,
particles travelling at $\approx$400-800\,km\,s$^{-1}$ at
$\approx$4\,kpc will return to the host galaxy within a few hundred
Myrs. From this analysis only two sources have maximum outflow
velocities that could potentially unbind some fraction of the gas from
the dark matter halo (Fig.~\ref{fig:Lbols}), which suggests that
the gas in the majority of the systems may eventually cool and
re-ignite another episode of AGN activity and star formation. However,
it has been proposed that {\it quasar mode} outflows, such as those
observed here, may be an effective way of ``pre-heating'' the gas at
early times prior to a mechanical-heating dominated phase (i.e.,\ the
so-called {\it radio mode}), which will then efficiently prevent the gas from
cooling (e.g., \citealt{vanDeVoort2011}; \citealt{McCarthy2011}). The
outflows we observe could be a means of energetically
pre-heating and ejecting gas at high redshift and therefore could be a
crucial stage in the evolution of massive galaxies. In this scenario these
galaxies will therefore be the progenitors of the radiatively weak,
low-redshift AGN in massive galaxies which are observed to have
mechanical heating that is preventing gas from cooling
(\citealt{Best2005,Best2006}; \citealt{Smolcic2009};
\citealt{Danielson2012}).


\section{Conclusions}
We have taken Gemini North-NIFS and VLT SINFONI IFU observations around the emission-line doublet [O~{\sc  iii}]$\lambda 4959,5007$ of
eight high-redshift ULIRGs that host AGN activity. Our aim was to
search for broad, high-velocity and spatially extended [O~{\sc iii}] emission as a
tracer of powerful AGN-driven outflows over galaxy-wide scales that are predicted by the
{\it quasar-mode} in galaxy evolution models. We have strong evidence that
such outflows exist in at least a fraction of the high-redshift
ULIRGs. Our main conclusions are the following:

\begin{enumerate}
\item We find narrow [O~{\sc iii}] emission lines (FWHM$\approx$ few hundred km\,s$^{-1}$) extended over galaxy-wide scales
  (up to $\approx$20\,kpc). By
  comparing these observations with
  high-redshift ULIRGs that
  are also undergoing intense starbursts and mergers but do not host
  significant AGN activity we show that the narrow [O~{\sc iii}]
  emission lines we observe (FWHM up to a few 100\,km\,s$^{-1}$) are
  consistent with tracing host
  galaxy dynamics and merger remnants (\S\ref{Sec:NarrowLines}).
\item In the four most luminous sources we find extremely broad [O~{\sc iii}]$\lambda\lambda$4959,5007 emission
  (FWHM$\approx$700--1400\,km\,s$^{-1}$) with high velocity-offsets
  (up to $\approx$850\,km\,s$^{-1}$), extended over
  $\approx$4--15\,kpc; i.e., the key signatures of galaxy-scale
  energetic outflows (\S\ref{Sec:NarrowLines}). The other four sources exhibit
  broad [O~{\sc iii}] emission but we have limited constraints on the
  spatial extent of the broad [O~{\sc iii}] emission due to lower quality data.
\item We apply a simple outflow model to show that the
  spatially extended outflows are potentially moving through the ISM with an energy rate of
  $\approx$(0.08--\,3)$\times$10$^{45}$\,erg\,s$^{-1}$,
  sufficient to unbind some of the gas from the host galaxy and
  potentially disrupting future accretion and star-formation in the
  host galaxy (\S\ref{Sec:energy}). These order of magnitude
    estimates are model dependent and rely on estimates of the
  density and therefore should be considered as illustrative only.
\item By estimating the energy injection rates from the AGN and star
  formation driven winds we find that the outflows require $\approx$0.2--5\% of the AGN luminosity to couple to the
  gas, while star formation require coupling
  efficiencies of 15\%--$\gg$100\% (\S\ref{Sec:energy}). We use a combination of energetic arguments and comparison to non-AGN
   systems to show that, although star formation may also play a significant role
   in some sources, the AGN activity is feasibly the dominant power source for driving the
   outflows in all sources.
\item The high maximum outflow velocities observed in these galaxies
  suggest that at least some fraction of the gas
  will be unbound from the host galaxies, but significant fractions of
  gas are very unlikely to be completely unbound from the galaxy
  haloes (\S\ref{Sec:fate}).
\end{enumerate}
The galaxies observed here are two to three orders
 of magnitude more common than the extreme high-redshift radio
 galaxies that are often associated with powerful AGN large-scale
 outflows at high redshift. We have shown that energetic large-scale outflows are present in
 at least a fraction of high-redshift ULIRGs that host AGN activity, which are believed to be the
 progenitors of today's massive elliptical galaxies. We have suggested that
 these outflows may be a crucial phase in the evolution of massive galaxies.


\subsection*{Acknowledgments}

We gratefully acknowledge support from the Science and Technology
Facilities Council (CMH; DMA; AMS; SA-Z; ADM). IRS acknowledges a
Leverhulme Senior Fellowship. FEB acknowledges support from Basel-CATA
(PFB-06/2007) and CONICYT-Chile (FONDECYT 1101024). We thank Mark Lovell for useful
discussions, Ric Davies for providing his sky-subtraction code. We thank the referee for detailed
comments. This work is based on observations carried out
with the Very Large Telescope of ESO (Programme ID:
087.A-0660(A)) and on observations obtained at the Gemini
Observatory (Programme IDs: GN-2008A-Q-58 and GN-2009B-Q-1), which is operated by the Association of Universities for
Research in Astronomy, Inc., under a cooperative agreement with the
NSF on behalf of the Gemini partnership: the National Science
Foundation (United States), the Particle Physics and Astronomy
Research Council (United Kingdom), the National Research Council
(Canada), CONICYT (Chile), the Australian Research Council
(Australia), CNPq (Brazil) and CONICET (Argentina). This publication
makes use of data products from the {\it Wide-field Infrared Survey
Explorer}, which is a joint project of the University of California,
Los Angeles, and the Jet Propulsion Laboratory/California Institute of
Technology, funded by the National Aeronautics and Space
Administration. This research has made use of data from HerMES project
(http://hermes.sussex.ac.uk/). HerMES is a {\it Herschel} Key Programme
utilising Guaranteed Time from the SPIRE instrument team, ESAC
scientists and a mission scientist. HerMES will be described in Oliver
et al. 2012, MNRAS. {\it Herschel} is an ESA space observatory with science instruments provided by European-led Principal Investigator consortia and with important participation from NASA.

\begin{table*}
\begin{center}
{\centerline{\sc Target SED Data and Derived Properties}}
\begin{tabular}{llcccccccccc}
\hline
\smallskip
ID & Source                                 &$S_{8}$           &$S_{24}$         &$S_{70}$ &$S_{250}$&$S_{350}$&$S_{500}$&$S_{850}$&$L_{\rm{IR,SF}}$                       & SFR    & $L_{\rm{AGN,SED}}$            \\
    &                                             &($\mu$Jy)      &(mJy)             &(mJy)     &(mJy)     & (mJy)      &(mJy)     &(mJy)     &(10$^{12}$L$_{\sun}$)&(M$_{\sun}$\,yr$^{-1}$)&(10$^{46}$\,erg\,s$^{-1}$)\\ 
(1) & (2)                                       &(3)                 &(4)                  &(5)         &(6)         &(7)          &(8)         &(9)          &(10)                  & (11)&(12)\\
\hline
1 & SMM\,J0217--0503              &44.7$\pm$2.3&0.49$\pm$0.05        &-            &36.3$\pm$5.8 &33.9$\pm$6.3 &21.7$\pm$6.8  &4.4$\pm$1.7 &4.5$\pm$0.8&780$\pm$160&$<$0.2 \\
2 & SMM\,J0302+0006               &63.4$\pm$6.8&0.45$\pm$0.05         &$<$13.6&-                       &-                      &-                       &4.4$\pm$1.3&2.0$\pm$0.3&340$\pm$50 &0.14$\pm$0.09 \\ 
3 & RG\,J0302+0010                  &-                     &$<$0.24               &-           &-                       &-                      &-                       &$<$4.5          &$<$2.1$\ddagger$      &$<$360 &2.0$\pm$0.2$^{\star\star}$\\
4 & RG\,J0332--2732                 &7.8$\pm$0.2  &0.94$\pm$0.09         &-            &27.5$\pm$5.8  &27.5$\pm$6.3 &20.3$\pm$6.8  &$<$3.6          &4.9$\pm$0.9&850$\pm$120&2.2$\pm$0.4 \\
5 & SMM\,J0943+4700$^{\star}$  &-                     &0.57$\pm$0.06         &-            &22.6$\pm$5.8  &26.5$\pm$6.3 &17.3$\pm$6.8  &8.7$\pm$1.5 &7.5$\pm$1.5&1300$\pm$200&5.8$\pm$0.2  \\
6 & SMM\,J1235+6215               &64.1$\pm$22  &0.63$\pm$0.06        &$<$4.1   &30.4$\pm$5.8 &37.4$\pm$6.3  &27.0$\pm$6.8  &8.3$\pm$2.5 &6.9$\pm$1.0&1200$\pm$170&0.7$\pm$0.4\\
7 & SMM\,J1237+6203               &(1.02$\pm$0.09)$\dagger$&(1.61$\pm$0.7)$\dagger$&$<$6.0    &$<$14        &$<$24              &$<$24          &5.3$\pm$1.7 &2.4$\pm$0.2&430$\pm$50&5.33$\pm$0.05 \\
8 & SMM\,J1636+4057               &66.8$\pm$6.8 &0.95$\pm$0.1         &$<$9.1    &44.4$\pm$5.8 &36.6$\pm$6.3  &29.0$\pm$6.8  &8.2$\pm$1.7&7.9$\pm$1.5    &1400$\pm$200&1.9$\pm$0.6\\
\hline
\hline
\end{tabular}
\caption{\label{Tab:SEDpoints}
{\protect\sc Notes:}\protect\\
Column (1): Source ID. 
Column (2): Source name. 
Column (3): 8$\mu$m flux densities and their uncertainties taken from: \protect\cite{Hainline2009}; \protect\cite{Cardamone2010}; or the Spitzer archive.
Column (4): 24$\mu$m flux densities from: \protect\cite{Ivison2007}; \protect\cite{Valiante2007}; \protect\cite{MenendezDelmestre2009}; Dickinson et~al. in prep; or the Spitzer archive. Uncertainties are assumed to be 10\% of the measured values.
Column (5): 70$\mu$m flux densities taken from \protect\cite{Hainline2009}.
Columns (6)--(8): The 350$\mu$m, 450$\mu$m and 500$\mu$m flux densities. This HerMES data was accessed through the HeDaM database (http://hedam.oamp.fr) operated by CeSAM and hosted by the Laboratoire d'Astrophysique de Marseille. The uncertainties quoted are the confusion noise for each band.
Column (9): 850\,$\mu$m flux densities and their uncertainties taken from: \protect\cite{Cowie2002}; \protect\cite{Chapman2004}; \protect\cite{Chapman2005}; \protect\cite{Clements2008}. 
Column (10): Derived IR luminosities (8--1000$\mu$m) from the star formation component of the SED fitting procedures described in \protect\S\ref{Sec:SEDfitting}.
Column (11): Star formation rates derived using the quoted $L_{\rm{IR,SF}}$ values and the relationship from \protect\cite{Kennicutt1998}.
Column (12): The estimated AGN bolometric luminosity calculated using the methods outlined in \protect\S\ref{Sec:SEDfitting}. $\ddagger$The value quoted is calculated using the radio-infrared correlation (assuming a $q=2.1$). This is an upper limit as there will be an unknown AGN contribution to the radio flux.
$^{\star\star}$This AGN bolometric luminosity is derived from the [O~{\sc iii}] luminosity (see \protect\S\ref{Sec:SEDfitting}).
$^{\star}$ For SMM\,J0943+4700 the flux densities have been corrected for an amplification
factor of 1.2. We note that the quoted flux densities will cover both radio counterparts to this source (see Appendix A5).
$\dagger$These are the 12$\mu$m and 22$\mu$m flux densities taken from the {\it Wide-field Infrared Survey Explorer} archive.
}
\end{center}
\end{table*}

\begin{table*}
\pagestyle{empty}
\begin{center}
\pagestyle{empty}
{\footnotesize
{\centerline{\sc Galaxy-integrated [O {\sc iii}]$\lambda 5007$
    Emission-line properties}}
\begin{tabular}{ll|ccc|ccc|cc}
\noalign{\smallskip}
\noalign{\smallskip}
           && \multicolumn{3}{c}{
             Component A.}
           & \multicolumn{3}{c}{
             Component B}
           & \\ 
\noalign{\smallskip}
\hline
\noalign{\smallskip}
ID & Source & $z_{\rm{[O III]}\rm{,A}}$ & 	FWHM$_{\rm{A}}$        & Flux$_{\rm{A}}$ 
& $z_{\rm{[O III]}\rm{,B}}$ & FWHM$_{\rm{B}}$       & Flux$_{\rm{B}}$
& $\Delta v$&Total [O~{\sc iii}] Luminosity
\\
  &                              &
  &(km\,s$^{-1}$)&(10$^{-16}$\,erg\,s$^{-1}$\,cm$^{-2}$) & &(km\,s$^{-1}$)&(10$^{-16}$\,erg\,s$^{-1}$\,cm$^{-2}$)&(km\,s$^{-1}$)&(10$^{42}$erg\,s$^{-1}$)\\
 (1)     &(2)     &(3)     &  (4)   & (5) & (6)  & (7)
           &(8) &(9) &(10)\\
\noalign{\smallskip}
\hline
1&SMM\,J0217--0503  & 2.0213[6]    & 540$\pm$170   &3.1$\pm$0.7    &-&-&-&-&9.3\\
2&SMM\,J0302+0006 &  1.4049[10]     & 900$\pm$300  &1.3$\pm$0.4    & - & - & -&-&1.6\\
3&RG\,J0302+0010    &  2.23861[12]   & 200$\pm$40    &0.87$\pm$0.17& 2.2419[7] & 1100$\pm$120   &2.7$\pm$0.4&$+300\pm$60&13.9\\	
4&RG\,J0332--2732    &  2.3145[5]     & 480$\pm$140   &2.3$\pm$0.5   &-&-&-&-&9.9\\
5&SMM\,J0943+4700$^{\star}$ &  3.3506[4]     & 430$\pm$110  &1.1$\pm$0.5    & 3.3503[8]& 1300$\pm$200 &2.6$\pm$0.4 &$+20\pm$60 &38.4\\
6&SMM\,J1235+6215 &  2.199[2]   & 1500$\pm$500& 1.2$\pm$0.4   & - &- &-  &- &4.5\\
7&SMM\,J1237+6203 &  2.07502[3]   & 234$\pm$10    &4.1$\pm$0.2    &2.0714[4] & 1000$\pm$100   &2.9$\pm$0.3 &$-350\pm$40 &22.5\\
8&SMM\,J1636+4057 &  2.38500[12] & 240$\pm$30    &1.3$\pm$0.3    &2.3817[18] &900$\pm$200  &1.0$\pm$0.4&$-290\pm$160 &16.8\\
   &&  & & & 2.4003[10]& 1100$\pm$180  &1.4$\pm$0.2&+1350$\pm$90 &\\
\hline
\hline
\end{tabular}
}
\caption{\label{Tab:OIIIprops}
{\protect\sc Notes:}\protect\\
The properties of the galaxy-integrated [O {\sc iii}]
  emission-line profiles shown in Fig.~\ref{Fig:int_specs}. Numbers in square
  brackets give the uncertainty in the last decimal place. Column (1):
  Source ID. Column (2): Source name. Columns (3)--(5): Properties of
  the narrowest Gaussian components (Component A): redshift, FWHM and
  flux respectively. Columns
  (6)--(8): Same as the previous three columns but for the broader
  Gaussian components (Component B). Column (9): The velocity offset 
  between the two Gaussian components. Column (10): The total [O~{\sc
    iii}] luminosity. The bottom
  line of the table gives the properties of the third Gaussian component which is fit to the profile of SMM\,J1636+4057. Flux and luminosity values are not
  extinction corrected. The quoted uncertainties are from the
    parameters of the emission-line fits. The true uncertainties on the fluxes are a
    factor or $\approx$2--3 higher, due to uncertainties in the
    absolute flux calibration. $^{\star}$The flux values for SMM\,J0943+4700 have been
  corrected for an amplification factor of 1.2. 
}

\end{center}
\end{table*}

\begin{landscape}
\begin{table}
\pagestyle{empty}
\begin{center}
\pagestyle{empty}
{\footnotesize
{\centerline{\sc Spatially Integrated Emission-Line Properties}}
\begin{tabular}{lll|ccc|cccc|cccc|c}
\noalign{\smallskip}
           &
           &
           & \multicolumn{3}{c}{
             [O~{\sc iii}] Component A.}
           & \multicolumn{3}{c}{
             [O~{\sc iii}] Component B}
           &
           & \multicolumn{3}{c}{
             H$\beta$}
           & \\ 
\noalign{\smallskip}
\hline
\noalign{\smallskip}
ID & Source & Region &$z_{\rm{[O III]}\rm{,A}}$ & FWHM$_{\rm{A}}$  &
Flux$_{\rm{A}}$                & $z_{\rm{[O III]}\rm{,B}}$ &
FWHM$_{\rm{B}}$ & Flux$_{\rm{B}}$ & $\Delta v$& $z_{\rm{H}\beta}$ &
FWHM$_{\rm{H}\beta}$ & Flux$_{\rm{H}\beta}$  &$\log$([O~{\sc
  iii}]/H$\beta$) &Linear Extent\\
(1)&(2)&(3)&(4)&(5)&(6)&(7)&(8)&(9)&(10)&(11)&(12)&(13)&(14)&(15)\\
\hline
\hline

1 & SMM\,J0217--0503 &BR &2.0229[11]  & 900$\pm$400& 0.42$\pm$0.15   &-&-&-&$-460\pm$60$\dagger$&-&-&$<$0.31&$>$0.13&$\le$4\\
     &                         & NR& 2.0209[2]     & 290$\pm$80   & 1.26$\pm$0.19    &-&-&-&-&-&-&$<$0.34&$>$0.57&11$\pm$4\\
 \noalign{\smallskip} 
3 & RG\,J0302+0010  &BR&2.23873[7] &180$\pm$20	& 0.53$\pm$0.06&2.2410[3] &1180$\pm$80 &2.59$\pm$0.16 &$+210\pm$30&2.2441[11]&780$\pm$170&0.34$\pm$0.10&0.97$\pm$0.13&8$\pm$3\\
       &                      &NR&2.23864[12]&200$\pm$30&0.17$\pm$0.02& - & - & -&-&-&-&$<$0.03&$>$0.73&6$\pm$3\\
 \noalign{\smallskip} 
4 & RG\,J0332--2732    &BR&2.3143[4] & 560$\pm$130   &0.81$\pm$0.16   &-&-&-&$-100\pm$20$\dagger$&-&-&$<$0.20&$>$0.61&10$\pm$4\\
     &                          &NR&2.3147[4] & 510$\pm$130   &0.62$\pm$0.12&-&-&-&-&-&-&$<$0.19&$>$0.52&11$\pm$4\\
 \noalign{\smallskip} 
5 & SMM\,J0943+4700$^{\star}$& BR &  3.3511[4]  &	340$\pm$70 &0.44$\pm$0.13   & 3.3503[5] & 1400$\pm$100 & 2.41$\pm$0.17&$-50\pm$40&3.3500[7]&430$\pm$120&0.30$\pm$0.09&0.97$\pm$0.12&15$\pm$3\\
      &                        & NR&  3.3493[2]&	400$\pm$40&0.48$\pm$0.04 & - & - & -&-&-&-&$<$0.07&$>$0.71&8$\pm$3\\
 \noalign{\smallskip} 
8 & SMM\,J1636+4057 & BR&2.3834[8]&1100$\pm$170&0.77$\pm$0.12&2.4023[7]&1220$\pm$160&1.07$\pm$0.14&$+1670\pm$100$\dagger\dagger$ &-&-&$<$0.11&$>$1.2&7$\pm$3 \\
        &                       & NR&2.38481[7] &257$\pm$14 &1.25$\pm$0.06 &- &- &- &- &-&-&$<$0.11&$>$1.1&20$\pm$3\\

\hline
\hline
\end{tabular}
}
\caption{\label{Tab:OIIIregions}
{\protect\sc Notes:}\protect\\
The properties of the [O~{\sc iii}] and H$\beta$ emission-line
profiles extracted from the sub-regions shown in
  Fig.~\ref{fig:regions}. Column (1): Source ID. Column (2): Source
  name. Column (3): The spatial region from which the spectra are
  extracted, either the broad region (BR) or narrow region (NR); see \S\ref{Sec:regions}. Columns (4)--(6): Properties of
  the narrowest [O~{\sc iii}] Gaussian components (Component A):
  redshift, FWHM (km\,s$^{-1}$) and
  flux (10$^{-16}$\,erg\,s$^{-1}$\,cm$^{-2}$) respectively. Columns (7)--(9): Same as the previous three columns but for the broader
  [O~{\sc iii}] Gaussian components (Component B). Column (10): The velocity offset (km\,s$^{-1}$)
  between the two Gaussian components. Columns (11)--(13): Properties of
  the H$\beta$ emission lines: redshift, FWHM (km\,s$^{-1}$) and
  flux (10$^{-16}$\,erg\,s$^{-1}$\,cm$^{-2}$)
  respectively. Column
  (14): The logarithm of the ratio of the total [O~{\sc iii}] and
  H$\beta$ fluxes. Column (15): The approximate
  linear-extent, along the most extended axis, of the defined regions (kpc). We take the size of seeing disks as the uncertainty on these
  measurements. All flux limits are 3$\sigma$ upper limits,
  which are calculated using the noise of the line-free continuum and the
  FWHM of the [O~{\sc iii}] emission line from the same region. Flux
  values are not corrected for dust extinction. The quoted uncertainties
  are from the parameters of the emission-line fits. The true
  uncertainties on the fluxes will be a
    factor or $\approx$2--3 higher, due to uncertainties in the
    absolute flux calibration.  $^{\star}$The flux values for SMM\,J0943+4700 have been
  corrected for an amplification factor of 1.2. $\dagger$ These are the velocity offsets to the H$\alpha$ emission in
  this region. $\dagger\dagger$ For analysis purposes we define
the velocity offset as half of this value, see
\S\ref{Sec:velprofiles} and Appendix A8 for details.}

\end{center}
\end{table}
\end{landscape}


\appendix

\section{Notes on individual sources}
\label{Sec:sources}

Here we provide background notes on each source from our sample and provide
some more details of our analysis on individual sources that is not required
for the overall discussion given in the main text. The 1.4\,GHz flux
densities are provided in Table~\ref{Tab:otherProps} and the
infrared and sub-mm flux densities are provided in
Table~\ref{Tab:SEDpoints}. All of the parameters
for the profile fits to the galaxy-integrated spectra
(see Fig.~\ref{Fig:int_specs}) and spectra extracted from different
regions (see Fig.~\ref{fig:regions}) are provided in
Table~\ref{Tab:OIIIprops} and Table~\ref{Tab:OIIIregions}
respectively. 

\subsection{SMM~J0217-0503}
This source was first identified as a sub-mm and radio source by
\cite{Coppin2006} and \cite{Simpson2006} respectively. IFU emission
around the H$\alpha$ emission line reveal at least two merging
components (\citealt{AlaghbandZadeh2012} ) over
$\approx$20--25\,kpc. Our IFU observations of the [O~{\sc
  iii}] emission line also reveal at least two systems separated by $\approx$3$^{\prime\prime}$
($\approx$25\,kpc; see Fig.~\ref{fig:velmaps}). The northern system has broad H$\alpha$ emission
in the central regions (FWHM$\approx$940\,km\,s$^{-1}$;
Alaghband-Zadeh et~al. priv. comm.) and
a high [N~{\sc iii}]/H$\alpha\approx$1.1--1.3 ratio (Alaghband-Zadeh
priv. comm). Along with our measured ratio [O~{\sc iii}]/H$\beta >$1.3
this suggests that this source is most likely hosting AGN activity
(e.g. ~\citealt{Kewley2006}). In Fig.~\ref{fig:regions} we show that
the [O~{\sc iii}] kinematics in this northern component also displays the
signatures of an outflow; i.e. a broad emission line
(FWHM=900$\pm$400\,km\,s$^{-1}$) which has a high velocity offset from
the H$\alpha$ emission ($\Delta v=-460\pm60$\,km\,s$^{-1}$;
Fig.~\ref{fig:velcurves}). Due to the complexity of blending between
  the [N~{\sc ii}] doublet and H$\alpha$ line, it is not possible to confirm
  if this outflowing component is also present in the H$\alpha$
  emission-line profile. The broad [O~{\sc iii}] emission in this northern component is spatially unresolved
($\le4.0$\,kpc) and has a modest signal-to-noise ratio such that we do
  not have strong evidence for a galaxy-wide outflow in this source. 

The southern system has H$\alpha$ emission extended over $\approx$10\,kpc with
kinematics which may be described by a rotating disk; however, the
double peaked emission line and a large gradient in the [N~{\sc ii}]/H$\alpha$
ratio from east to west could also indicate that this southern system has undergone a
recent merger (\citealt{AlaghbandZadeh2012}). Our observations of the
[O~{\sc iii}] emission line displays a redshift, line-width and
position that is only consistent with the western $\approx$5\,kpc of the
H$\alpha$ emission, with no [O~{\sc iii}] detected in the eastern side. The [O~{\sc iii}] emission is located where the
[N~{\sc ii}]/H$\alpha$ ratio is lowest suggesting a different chemical composition across the
galaxy, as opposed to excitation effects (e.g., \citealt{Kewley2006}). These observations are consistent with the scenario that this southern
system may have undergone a recent merger.

\subsection{SMM~J0302+0006}
This sub-mm source was first detected by \cite{Webb2003}. Optical and near-IR
spectra show a high [N {\sc ii}]/H$\alpha=$1.38$\pm$0.07 line ratio,
and {\it Spitzer} mid-infrared spectroscopic observations show an
excess in the mid-infrared continuum suggesting AGN
activity (\citealt{Swinbank2004}; \citealt{Chapman2005}; \citealt{MenendezDelmestre2009}). {\em HST}
imaging and H$\alpha$ IFU data show this source contains two
components separated by $\approx$11\,kpc (\citealt{Swinbank2006};
Men\'endez-Delmestre et~al. 2012). 

The galaxy-integrated [O~{\sc iii}] emission
we observe is broad (FWHM=900$\pm$300\,km\,s$^{-1}$;
Fig.~\ref{Fig:int_specs}) but is
spatially unresolved ($\le 3.6$\,kpc). However, due to the
low signal-to-noise ratio of the integrated spectrum we cannot rule out the existence of
low surface brightness extended emission. No H$\beta$ emission was
detected in the galaxy-integrated spectrum, with a 3$\sigma$ upper
limit on the flux of $F_{\rm{H}\beta}<8\times$10$^{-17}$\,erg\,s$^{-1}$\,cm$^{-2}$ (assuming the
line-width is the same as that measured for [O~{\sc iii}]).

\subsection{RG~J0302+0010}
This source was first identified by \cite{Chapman2004}. This source has rest-frame UV spectroscopy that reveals strong [C~{\sc IV}]
emission, which along with the the high [N~{\sc ii}]/H$\alpha$=1.13$\pm$0.4 line
ratio implies AGN activity (\citealt{Chapman2004}; \citealt{Swinbank2004}). Observations with long-slit
spectroscopy show that the [O~{\sc iii}] emission is broad and has an
asymmetric line profile (\citealt{Takata2006}).

The galaxy-integrated spectrum around the [O~{\sc iii}]$\lambda\lambda$4959,5007
emission-line doublet reveals both a narrow and redshifted broad
component (Fig.~\ref{Fig:int_specs}). The broad component is located in the central
$\approx$8\,kpc, where the line ratio [O~{\sc iii}]/H$\beta$=9$\pm$3
implies bright AGN activity (Fig.~\ref{fig:regions}). We confirmed the
broad emission line was spatially extended by integrating spectra in
two $0.6^{\prime\prime}\times0.6^{\prime\prime}$ bins, one in the
north of the emission-line region and one in the south. We fit these
spectra following the methods outlined in \S\ref{Sec:SEDfitting} and found that the broad
emission line has a velocity shear of 125$\pm$30\,km\,s$^{-1}$ which
is broadly consistent with the velocity gradient observed in Fig.~\ref{fig:velcurves}.

The broad [O~{\sc iii}] emission line is most likely due to an outflow with the near-side of the outflow being
obscured by dust (see \S\ref{Sec:impact} for a full discussion). Blueshifted broad components are more commonly associated
with outflows (e.g., \citealt{Heckman1981}) as opposed to the
redshifted broad component we observe here. However, redshifted broad
components are observed in a fraction of local AGN and can be
explained by the relative orientations of the AGN and obscuring
material in host galaxy (e.g., \citealt{Barth2008}; \citealt{Crenshaw2010}). In particular IFU observations of the
local interacting Seyfert galaxy LEDA 135736, identify a broad, redshifted outflow
associated with AGN activity, which could be a low redshift analogue of
this source (\citealt{Gerssen2009}).

\subsection{RG~J0332-2732}
IFU observations of this source reveals that H$\alpha$ is extended over $\approx$10\,kpc with velocity offsets of
$\Delta v\lesssim$\,200\,km\,s$^{-1}$ across the galaxy. The brightest region (Fig.~\ref{fig:regions}; Table~\ref{Tab:OIIIregions}) has a high [O~{\sc iii}]/H$\beta>$4.0 line ratio, which
along with the high [N~{\sc ii}]/H$\alpha$=0.8$\pm$0.1 line ratio (\citealt{AlaghbandZadeh2012}) suggests AGN activity (e.g.,~\citealt{Kewley2006}).

The [O~{\sc iii}] emission-line kinematics (Fig.~\ref{fig:velmaps}; Fig.~\ref{fig:velcurves}) appear to be tracing
similar kinematics to the H$\alpha$ emission line (\citealt{AlaghbandZadeh2012}) with similar line widths (FWHM$\approx$200-700\,km\,s$^{-1}$) and
velocity gradients ($\Delta v\lesssim$200\,km\,s$^{-1}$). The [O~{\sc iii}] emission is more extended to the north than
the H$\alpha$ emission and, along with the observation that the [N~{\sc
  ii}]/H$\alpha$ ratio is highest in the north (\citealt{AlaghbandZadeh2012}), suggests the presence of an AGN ionisation cone. 

Unlike the majority of our sample with spatially resolved data, we are
unable to fit multiple components to the emission line profile and are
unable to de-couple the
signatures of an outflow from the galaxy dynamics. However, we note that the emission lines in the
brightest regions display very large widths (FWHM up to
$\approx$700\,km\,s$^{-1}$; Fig.~\ref{fig:velcurves}), which
are broader than the narrow lines we associate with galaxy
dynamics in the other sources ($\approx$100--500\,km\,s$^{-1}$;
Fig.~\ref{fig:narrow_comps}). Although not as well defined as in the other
sources, we speculate that an outflow could be responsible for these large line
widths.

\subsection{SMM~J0943+4700}
This sub-mm source was first identified by \cite{Cowie2002} and is modestly lensed (amplification of
1.2). \cite{Ledlow2002} identified two radio counterparts separated by
$\approx$30\,kpc in projection and labelled them as H6 and H7. The latter shows narrower and fainter [O~{\sc iii}] emission than its companion but
dominates the sub-mm and CO emission, indicating that the
bulk of star formation is occurring in this component
(\citealt{Takata2006}; \citealt{Valiante2007}; \citealt{Engel2010};
\citealt{Riechers2011}). An excess in
the mid-infrared continuum observed in {\it Spitzer} mid-infrared
spectroscopy indicate the presence of AGN activity in this source
(\citealt{Valiante2007}) although both H6 and H7 are covered by this spectra. H6, the source observed in our observations,
has rest-frame UV emission-line widths, and line ratios indicative of
a narrow-line Seyfert 1 galaxy (\citealt{Ledlow2002}). Additionally
long-slit near-IR spectroscopy of H6 reveals [O~{\sc iii}] emission that
is broad and extended over $\approx$3$^{\prime\prime}$--4$^{\prime\prime}$
(\citealt{Takata2006}), indicative of an outflow.

Using our IFU observations we find that the [O~{\sc iii}] emission from this source is very extended
($\approx$20\,kpc) with a complex morphology and kinematic structure
(Fig.~\ref{fig:velmaps}). The north-western region shows modest [O~{\sc iii}] line-widths
(FWHM=200--300\,km\,s$^{-1}$) consistent with being due to galaxy
dynamics and merger remnants (Fig.~\ref{fig:narrow_comps}; see \S~\ref{Sec:NarrowLines}
for a discussion). In addition there is extremely broad [O~{\sc iii}]
emission (FWHM=1000--1400\,km\,s$^{-1}$) extended over $\approx$15\,kpc
(Fig.~\ref{fig:regions} and Fig.~\ref{fig:velcurves}), which we
attribute to an energetic outflow. Despite observing modest velocity offsets between the
narrow and broad [O~{\sc iii}] emission lines ($\Delta
v\le$150\,km\,s$^{-1}$; Fig.~\ref{fig:velcurves}), there is clearly extremely
turbulent ionised gas over a considerable extent. The zero velocity
offset between the broad and narrow [O~{\sc iii}] components in the spatially-integrated spectra
(Fig.~\ref{Fig:int_specs} and Fig.~\ref{fig:regions}) could
be explained if we are observing an outflow orientated in the plane of the
sky.

\subsection{SMM~J1235+6215}
This source was identified as a sub-mm bright galaxy by
\cite{Chapman2005}. Deep {\em Chandra} images shows that it hosts a heavily obscured X-ray luminous
AGN ($L_{0.5-8\rm{keV}}$=10$^{44.0}$\,erg\,s$^{-1}$;
$N_{\rm{H}}$=10$^{24}$\,cm$^{-2}$; \citealt{Alexander2005}). An excess in the mid-infrared continuum observed in {\it Spitzer} mid-infrared
spectroscopy further suggests the presence of AGN activity in this source
(\citealt{MenendezDelmestre2009}). IFU observations
reveal a bright compact ($\approx$3\,kpc) source of broad  H$\alpha$ emission
(FWHM $\gsim$1000\,km\,s$^{-1}$) attributed to AGN activity and
spatially offset ($\approx$0.5$^{\prime\prime}$) from the narrower H$\alpha$
emission (FWHM~$\approx$~500--800\,km\,s$^{-1}$), which is likely to be dominated by star-formation activity
(Men\'endez-Delmestre et~al. 2012). Long-slit near-IR spectroscopy
suggest that the [O~{\sc iii}] emission is extended over $\approx$~1$^{\prime\prime}$
($\approx$8\,kpc; \citealt{Takata2006}).

The galaxy-integrated spectrum around the [O~{\sc
  iii}]$\lambda\lambda$4959,5007 emission-line doublet reveals an
extremely broad emission line
(FWHM=1500$\pm$500\,km\,s$^{-1}$) which may suggest an outflow
(Fig.~\ref{Fig:int_specs}). \cite{Takata2006} find that this sources has
extended [O~{\sc iii}] emission over
$\approx$1$^{\prime\prime}$. Although we find tentative evidence that
the [O~{\sc iii}] emission is marginally extended ($\approx$0.8$^{\prime\prime}$;
7\,kpc) in our data we are unable to constrain the properties of the extended emission due to insufficient
signal-to-noise ratio of the data to fit the emission-line profiles.

We also detect H$\beta$ emission, at a redshift of  $z_{\rm{H}\beta}$=2.2031$\pm$0.0006, with a width of
FWHM$_{\rm{H}\beta}$=500$\pm$200\,km\,s$^{-1}$ and flux $F_{\rm{H\beta}}$=
0.9$\pm$0.2$\times$10$^{-16}$\,erg\,s$^{-1}$\,cm$^{-2}$, extended over
$\approx$1$^{\prime\prime}$ ($\approx$8\,kpc). The
extremely broad [O~{\sc iii}] emission we observe is offset from the
H$\beta$ emission $\Delta v \approx-400\pm200$\,km\,s$^{-1}$. Although there is weak evidence for an extended outflow in this
  source, deeper observations are required to confirm this.

\subsection{SMM~J1237+6203}
This source was identified as a sub-mm bright galaxy by
\cite{Chapman2005}. The bright optical counterpart ($R = 20.2$;
\citealt{Chapman2005}), X-ray luminosity ($L_{0.5-8\rm{keV}}=10^{44.3}$\,erg\,s$^{-1}$; \citealt{Alexander2005}) and
optical-near-IR spectroscopy that reveal broad emission lines (Ly$\alpha$;
N\,{\sc V}; C\,{\sc IV}; H$\gamma$--H$\alpha$; FWHM
$\approx$~2100-2700\,km\,s$^{-1}$; \citealt{Chapman2005};
\citealt{Takata2006}; \citealt{Coppin2008}), result in this
source being classified as a broad-line quasar, with a virial BH mass of
log($M_{\rm{BH}}$)=8.2\,M$_{\sun}$ (\citealt{Alexander2008}). 

IFU observations around redshifted [O~{\sc iii}] of SMM~J1237+6203 were first presented in
\cite{Alexander2010} where a complete discussion of the source is provided. We note the orientation of this source in
\cite{Alexander2010} is flipped in the east-west direction and the
correct orientation is given in this work (Fig.~\ref{fig:velmaps}). In addition we provide
updated [O~{\sc iii}] flux and luminosity measurements.

The galaxy-integrated spectrum around the [O~{\sc iii}] emission line
(Fig.~\ref{Fig:int_specs}) shows a bright narrow component with a prominent blue-wing. This type of profile is most commonly
interpreted as being the result of an AGN-driven outflow
(e.g., \citealt{Heckman1981}; \citealt{Nesvadba2008}). The velocity and FWHM profiles (Fig.~\ref{fig:velcurves}) show that
the narrow emission has a small velocity gradient of $\Delta
v\approx$200\,km\,s$^{-1}$ across the galaxy and is likely to be tracing
the host galaxy dynamics (also see \citealt{Alexander2010}). In contrast, the
broad [O~{\sc iii}] component (FWHM$\approx$1000\,km\,s$^{-1}$) is offset from the narrow component by $\approx -350$\,km\,s$^{-1}$, over $\approx$5\,kpc,
providing strong evidence for an outflow in this region. Although the broad component is only formally fit over $\approx$5\,kpc, low surface
brightness broad emission exists up to $\approx$8\,kpc in total extent (\citealt{Alexander2010}). 

We also identify H$\beta$ emission in the galaxy-integrated spectrum,
with a redshift of $z$=2.0754[2] and flux
$F_{\rm{H}\beta}=$1.1$\pm
0.4\times$10$^{-15}$\,erg\,s$^{-1}$\,cm$^{-2}$. The width of the
H$\beta$ emission line (FWHM=2030$\pm$70\,km\,s$^{-1}$) indicates the
presence of a broad-line region, which is consistent with that found
from H$\alpha$ observations of this source (\citealt{Coppin2008}).

\subsection{SMM~J1636+4057}
\label{Sec:N2}

\begin{figure*}
\centerline{\psfig{figure=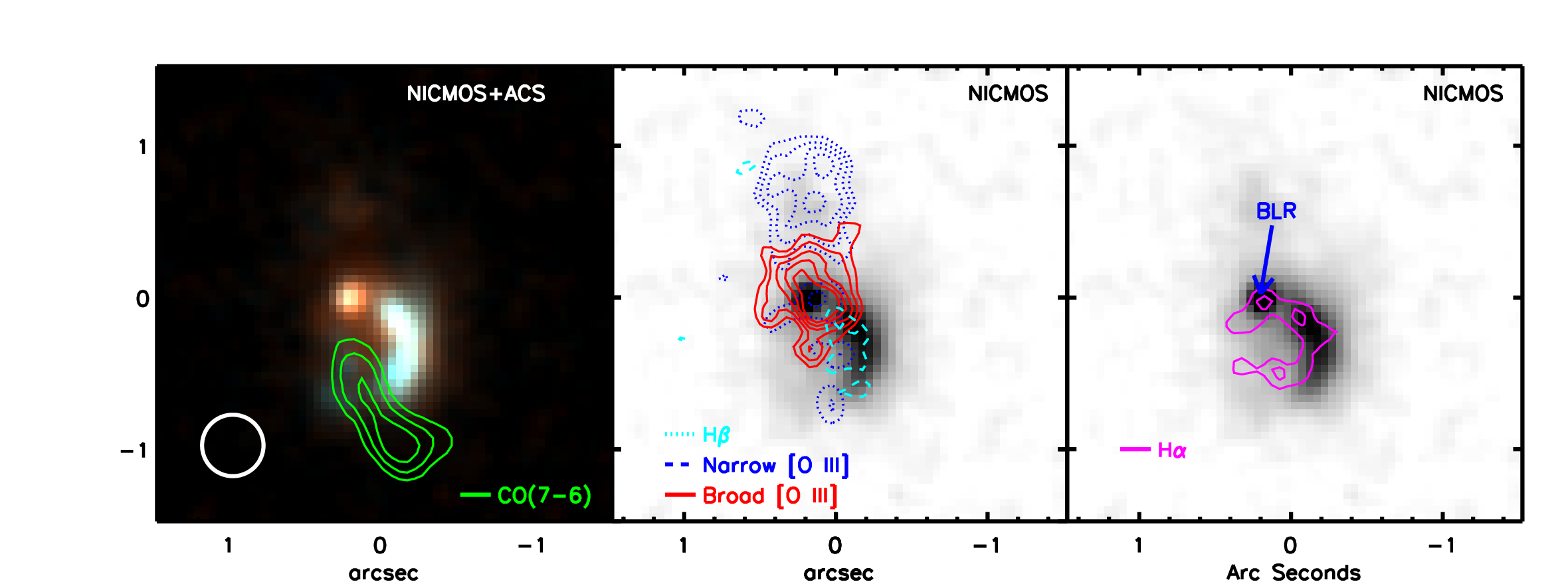,width=7.5in}}
\caption{{\it Left:} True colour $I_{814}H_{160}$ image of
  SMM~J1636+4057 from the $HST$ ACS and NICMOS imaging (Swinbank
  et~al. 2005). The contours
  denote the intensity of the velocity integrated CO(7--6) millimetre
  emission (\citealt{Tacconi2008}). The circle denotes the
  approximate seeing disk of our observations. {\it Middle:} $HST$ NICMOS $H_{160}$
  grey-scale image with contours
  overlaid of narrow band (wavelength collapsed) images from our IFU observations. The solid
  contours denote the extent/intensity of the broad [O~{\sc iii}]
  emission, which is associated with the red continuum seen in the colour image. The dotted contours denote the
  extent/intensity of the narrow [O~{\sc iii}] for which the
  majority is associated with faint emission observed to the
  north of the red continuum peak. The dashed contours denote the
  H$\beta$ emission which is spatially coincident with the blue
  arc in the $HST$ image. {\it Right:} The same image as in the central
  panel but with the contours from the IFU+adaptive optics H$\alpha$ observations of Men\'{e}ndez-Delmestre
  et~al. (2012) overlaid. There is extremely broad (FWHM$\approx$2700\,km\,s$^{-1}$) unresolved H$\alpha$
  emission, attributed to an AGN broad line region (BLR) and
  narrower spatially extended emission along the blue arc, which is attributed to
  star formation. The field-of-view of these H$\alpha$
  observations did not cover the region of narrow [O~{\sc iii}]
  emission seen in the north of the middle panel. North is up and east is left. 
}
\label{fig:SMMJ16_overlay}
\end{figure*}

This sub-mm source was first identified by \cite{Scott2002} and
\cite{Ivison2002}. {\it Hubble Space Telescope} ({\it HST}) ACS and NICMOS observations reveal a complex galactic morphology with at least
three merging or interacting components, covering $\approx$8--15\,kpc in
projection (\citealt{Swinbank2005}; see
Fig.~\ref{fig:SMMJ16_overlay}). An excess in
the mid-infrared continuum observed in the {\it Spitzer} mid-infrared
spectroscopy indicate the presence of AGN activity in this source
(\citealt{MenendezDelmestre2009}). Long-slit and IFU near-IR spectroscopy
reveal a broad-line AGN (broad spatially unresolved H$\alpha$ and
H$\beta$ emission with FWHM$\approx$2000--3000\,km\,s$^{-1}$) and narrow
H$\alpha$ emission traces star-formation along the UV bright arc
shown in the {\it HST} image (Fig.~\ref{fig:SMMJ16_overlay}; \citealt{Smail2003}; \citealt{Swinbank2005}; Men\'{e}ndez-Delmestre et~al. 2012). The broad-line AGN is found at $\Delta v\approx700$\,km\,s$^{-1}$
from the narrow H$\alpha$ emission, providing further evidence that
this system is undergoing a merger. The presence of a molecular gas cloud in the southern regions of the system is shown
through high and low excitation CO emission-line gas
(\citealt{Tacconi2008}; \citealt{Engel2010}; \citealt{Ivison2011}).
Previous long-slit observations show that the [O~{\sc iii}] emission is kinematically
complex and spatially extended ($\approx$1--2$''$) with broad
components (FWHM $\lesssim$~2000~km~s$^{-1}$; \citealt{Smail2003};
\citealt{Takata2006}), providing initial evidence for an energetic outflow.

Our galaxy-integrated spectrum around the [O~{\sc
  iii}]$\lambda\lambda$4959,5007 emission-line doublet (Fig.~\ref{Fig:int_specs}) reveals a prominent
narrow component and two broad components separated by $\Delta v =
1700\pm100$\,km\,s$^{-1}$. We are confident that the broad redshifted
emission line is also from [O~{\sc iii}] as \cite{Smail2003} also
identify [Ne~{\sc V}] and N~V emission lines at approximately this
redshift.

In Fig.~\ref{fig:velmaps} we show that the [O~{\sc iii}] emission-line morphology, velocity field and FWHM map from our IFU
observations. Narrow [O~{\sc iii}] emission
(100~$\lesssim$~FWHM~$\lesssim$~500\,km\,s$^{-1}$) is extended over
$\approx$1$^{\prime\prime}$ ($\approx$8\,kpc) to the north of the peak in surface brightness, associated with faint
infrared emission in the {\it HST} image, and fainter narrow emission is observed
$\approx$1$^{\prime\prime}$ to the south; this is consistent with the
long-slit observations of \cite{Smail2003} (also see Fig.~\ref{fig:SMMJ16_overlay}). The redshift of this narrow
[O~{\sc iii}] emission is consistent with the molecular gas and H$\alpha$ arc and
therefore is likely to be tracing merger remnants and galaxy dynamics
(e.g., Fig.~\ref{fig:velcurves}). We also detect faint H$\beta$ emission that
is spatially coincident with the star-forming arc (Fig.~\ref{fig:SMMJ16_overlay}).

The two broad [O~{\sc iii}] emission-line components (FWHM$\approx$1200\,km\,s$^{-1}$) are spatially coincident with the
broad H$\alpha$ emission line (Fig.~\ref{fig:regions} and
Fig.~\ref{fig:SMMJ16_overlay}) and are moderately spatially extended ($\approx$0.8$^{\prime\prime}$;
$\approx$7\,kpc). We verified that the broad emission lines were intrinsically
spatially extended by extracting spectra from two
$0.6^{\prime\prime}\times0.6^{\prime\prime}$ bins; one in the north
and one in the south of the observed broad [O~{\sc iii}] region. We found that that
the ratio of the two broad components changed from north to south. The
redshifted broad component dominates in the north and the blueshifted
broad component dominates in the south, which is also demonstrated in the
velocity map of this source (Fig.~\ref{fig:velmaps}). The broad [O~{\sc iii}] components are offset
from the broad H$\alpha$ emission line by $\Delta
v\approx\pm850$km\,s$^{-1}$ (Fig.~\ref{fig:velcurves};
\citealt{Swinbank2005}), indicating that we are
observing both sides of an outflow (e.g., \citealt{Nesvadba2008}). To
verify that the broad H$\alpha$ is due to a true broad-line region, as opposed to dynamics in the host galaxy, we attempted to re-fit the spectrum of
\cite{Swinbank2005} with a narrow-line Gaussian component plus two
additional Gaussian components with the same redshifts and line-widths of the broad [O~{\sc
  iii}] emission lines (e.g., similar to the methods used by \citealt{CanoDiaz2012} and Mullaney et~al. in
prep). We found that this did not adequately describe the H$\alpha$
emission-line profile and an additional broad component is still
required (with FWHM$\approx$2700\,km\,s$^{-1}$ and $z=2.393\pm0.005$; consistent
with \citealt{Swinbank2005}) providing
strong evidence of a true AGN broad-line region. 

The outflow we observe is spatially coincident with the AGN in this
source, and no outflow is observed in the region of unobscured
star-formation (the UV bright arc). We conclude that the broad [O~{\sc
  iii}] emission lines we see are due a bi-polar AGN-driven
outflow at $v\approx\pm850$\,km\,s$^{-1}$.

\subsection{SMM~J2217+0010}
This source was first identified by \cite{Smail2004} and
\cite{Chapman2005}. An excess in the mid-infrared continuum observed in the {\it Spitzer} mid-infrared
spectroscopy suggests the presence of AGN activity in this source
(\citealt{MenendezDelmestre2009}). The results of \cite{Takata2006}
suggest bright [O~{\sc iii}] emission ($F_{\rm{[O
    III]}}=$\,6.0~$\times$~10$^{-16}$\,erg\,s$^{-1}$\,cm$^{-2}$) and a
high [O~{\sc iii}]/H$\beta\approx$10 ratio further indicating AGN activity.
However in our observations [O~{\sc iii}]$\lambda\lambda 4959,5007$  emission was undetected. By spatially integrating over the extent of a seeing disk
and assuming an emission line with a FWHM of 500\,km\,s$^{-1}$, we
infer a 3$\sigma$ upper limit
on the [O~{\sc iii}] flux (luminosity) of $F_{\rm{[O
    III]}}<$\,4.6~$\times$~10$^{-17}$\,erg\,s$^{-1}$\,cm$^{-2}$
($L_{\rm{[O III]}}<2.6\times10^{42}$\,erg\,s$^{-1}$). This
upper limit is considerably lower than the value quoted in
\cite{Takata2006} which may arise due to the presence a strong sky-line at the observed wavelength of the [O~{\sc
  iii}]$\lambda$5007 emission line in this source, potentially resulting in an
excess measurement of the flux quoted in \cite{Takata2006}. 

\subsection{SMM~J2217+0017}
This source was identified as a sub-mm bright galaxy by
\cite{Tamura2009}. We did not detected [O~{\sc
  iii}]$\lambda\lambda 4959,5007$  emission. This source was detected in H$\alpha$ emission in the
same data cube (\citealt{AlaghbandZadeh2012} ), therefore the
non-detection is due to an [O~{\sc iii}] flux that is below our
detection threshold. Assuming an emission line of a FWHM of 350\,km\,s$^{-1}$
(to match the H$\alpha$ observations; \citealt{AlaghbandZadeh2012}) and
integrating over the spatial extent of the H$\alpha$ emission, we infer
a 3$\sigma$ upper limit on the [O~{\sc iii}] flux (luminosity) of $F_{\rm{[O
    III]}}<$\,2.6~$\times$~10$^{-17}$\,erg\,s$^{-1}$\,cm$^{-2}$ ($L_{\rm{[O~III]}}<$1.0$\times$10$^{42}$\,erg\,s$^{-1}$). 


\begin{thebibliography}{}

\bibitem[\protect\citeauthoryear{{Alaghband-Zadeh et~al.}}{{Alaghband-Zadeh
  et~al.}}{2012}]{AlaghbandZadeh2012}
{Alaghband-Zadeh et~al.} S.,  2012, ArXiv e-prints, 1205.5452

\bibitem[\protect\citeauthoryear{{Alatalo et~al.}}{{Alatalo
  et~al.}}{2011}]{Alatalo2011}
{Alatalo et~al.} K.,  2011, \apj, 735, 88

\bibitem[\protect\citeauthoryear{{Alexander}, {Bauer}, {Chapman}, {Smail},
  {Blain}, {Brandt} \& {Ivison}}{{Alexander} et~al.}{2005}]{Alexander2005}
{Alexander} D.~M.,  {Bauer} F.~E.,  {Chapman} S.~C.,  {Smail} I.,  {Blain}
  A.~W.,  {Brandt} W.~N.,    {Ivison} R.~J.,  2005, \apj, 632, 736

\bibitem[\protect\citeauthoryear{{Alexander} \& {Hickox}}{{Alexander} \&
  {Hickox}}{2011}]{Alexander2012}
{Alexander} D.~M.,  {Hickox} R.~C.,  2011, ArXiv e-prints, 1112.1949

\bibitem[\protect\citeauthoryear{{Alexander}, {Swinbank}, {Smail}, {McDermid}
  \& {Nesvadba}}{{Alexander} et~al.}{2010}]{Alexander2010}
{Alexander} D.~M.,  {Swinbank} A.~M.,  {Smail} I.,  {McDermid} R.,
  {Nesvadba} N.~P.~H.,  2010, \mnras, 402, 2211

\bibitem[\protect\citeauthoryear{{Alexander et~al.}}{{Alexander
  et~al.}}{2003}]{Alexander2003}
{Alexander et~al.} D.~M.,  2003, \aj, 125, 383

\bibitem[\protect\citeauthoryear{{Alexander et~al.}}{{Alexander
  et~al.}}{2008}]{Alexander2008}
{Alexander et~al.} D.~M.,  2008, \aj, 135, 1968

\bibitem[\protect\citeauthoryear{{Barth}, {Greene} \& {Ho}}{{Barth}
  et~al.}{2008}]{Barth2008}
{Barth} A.~J.,  {Greene} J.~E.,    {Ho} L.~C.,  2008, \aj, 136, 1179

\bibitem[\protect\citeauthoryear{{Best}, {Kaiser}, {Heckman} \&
  {Kauffmann}}{{Best} et~al.}{2006}]{Best2006}
{Best} P.~N.,  {Kaiser} C.~R.,  {Heckman} T.~M.,    {Kauffmann} G.,  2006,
  \mnras, 368, L67

\bibitem[\protect\citeauthoryear{{Best}, {Kauffmann}, {Heckman}, {Brinchmann},
  {Charlot}, {Ivezi{\'c}} \& {White}}{{Best} et~al.}{2005}]{Best2005}
{Best} P.~N.,  {Kauffmann} G.,  {Heckman} T.~M.,  {Brinchmann} J.,  {Charlot}
  S.,  {Ivezi{\'c}} {\v Z}.,    {White} S.~D.~M.,  2005, \mnras, 362, 25

\bibitem[\protect\citeauthoryear{Binney \& Tremaine}{Binney \&
  Tremaine}{1987}]{Binney1987}
Binney J.,  Tremaine S.,  1987, Galactic dynamics.
Princeton series in astrophysics, Princeton University Press

\bibitem[\protect\citeauthoryear{{Blustin et~al.}}{{Blustin
  et~al.}}{2003}]{Blustin2003}
{Blustin et~al.} A.~J.,  2003, \aap, 403, 481

\bibitem[\protect\citeauthoryear{{Bower}, {Benson}, {Malbon}, {Helly}, {Frenk},
  {Baugh}, {Cole} \& {Lacey}}{{Bower} et~al.}{2006}]{Bower2006}
{Bower} R.~G.,  {Benson} A.~J.,  {Malbon} R.,  {Helly} J.~C.,  {Frenk} C.~S.,
  {Baugh} C.~M.,  {Cole} S.,    {Lacey} C.~G.,  2006, \mnras, 370, 645

\bibitem[\protect\citeauthoryear{{Cano-D{\'{\i}}az}, {Maiolino}, {Marconi},
  {Netzer}, {Shemmer} \& {Cresci}}{{Cano-D{\'{\i}}az}
  et~al.}{2012}]{CanoDiaz2012}
{Cano-D{\'{\i}}az} M.,  {Maiolino} R.,  {Marconi} A.,  {Netzer} H.,  {Shemmer}
  O.,    {Cresci} G.,  2012, \aap, 537, L8

\bibitem[\protect\citeauthoryear{{Cardamone et~al.}}{{Cardamone
  et~al.}}{2010}]{Cardamone2010}
{Cardamone et~al.} C.~N.,  2010, \apjs, 189, 270

\bibitem[\protect\citeauthoryear{{Cattaneo} \& {Bernardi}}{{Cattaneo} \&
  {Bernardi}}{2003}]{Cattaneo2003}
{Cattaneo} A.,  {Bernardi} M.,  2003, \mnras, 344, 45

\bibitem[\protect\citeauthoryear{{Chapman}, {Blain}, {Smail} \&
  {Ivison}}{{Chapman} et~al.}{2005}]{Chapman2005}
{Chapman} S.~C.,  {Blain} A.~W.,  {Smail} I.,    {Ivison} R.~J.,  2005, \apj,
  622, 772

\bibitem[\protect\citeauthoryear{{Chapman}, {Smail}, {Blain} \&
  {Ivison}}{{Chapman} et~al.}{2004}]{Chapman2004}
{Chapman} S.~C.,  {Smail} I.,  {Blain} A.~W.,    {Ivison} R.~J.,  2004, \apj,
  614, 671

\bibitem[\protect\citeauthoryear{{Clements et~al.}}{{Clements
  et~al.}}{2008}]{Clements2008}
{Clements et~al.} D.~L.,  2008, \mnras, 387, 247

\bibitem[\protect\citeauthoryear{{Colina}, {Arribas} \& {Borne}}{{Colina}
  et~al.}{1999}]{Colina1999}
{Colina} L.,  {Arribas} S.,    {Borne} K.~D.,  1999, \apjl, 527, L13

\bibitem[\protect\citeauthoryear{{Colina}, {Arribas} \&
  {Monreal-Ibero}}{{Colina} et~al.}{2005}]{Colina2005}
{Colina} L.,  {Arribas} S.,    {Monreal-Ibero} A.,  2005, \apj, 621, 725

\bibitem[\protect\citeauthoryear{{Coppin}, {Chapin}, {Mortier}, {Scott},
  {Borys}, {Dunlop}, {Halpern} \& {Hughes}}{{Coppin} et~al.}{2006}]{Coppin2006}
{Coppin} K.,  {Chapin} E.~L.,  {Mortier} A.~M.~J.,  {Scott} S.~E.,  {Borys} C.,
   {Dunlop} J.~S.,  {Halpern} M.,    {Hughes} 2006, \mnras, 372, 1621

\bibitem[\protect\citeauthoryear{{Coppin et~al.}}{{Coppin
  et~al.}}{2008}]{Coppin2008}
{Coppin et~al.} K.,  2008, \mnras, 389, 45

\bibitem[\protect\citeauthoryear{{Cowie}, {Barger} \& {Kneib}}{{Cowie}
  et~al.}{2002}]{Cowie2002}
{Cowie} L.~L.,  {Barger} A.~J.,    {Kneib} J.-P.,  2002, \aj, 123, 2197

\bibitem[\protect\citeauthoryear{{Cox}, {Dutta}, {Di Matteo}, {Hernquist},
  {Hopkins}, {Robertson} \& {Springel}}{{Cox} et~al.}{2006}]{Cox2006}
{Cox} T.~J.,  {Dutta} S.~N.,  {Di Matteo} T.,  {Hernquist} L.,  {Hopkins}
  P.~F.,  {Robertson} B.,    {Springel} V.,  2006, \apj, 650, 791

\bibitem[\protect\citeauthoryear{{Crenshaw}, {Kraemer} \& {George}}{{Crenshaw}
  et~al.}{2003}]{Crenshaw2003}
{Crenshaw} D.~M.,  {Kraemer} S.~B.,    {George} I.~M.,  2003, \araa, 41, 117

\bibitem[\protect\citeauthoryear{{Crenshaw}, {Schmitt}, {Kraemer}, {Mushotzky}
  \& {Dunn}}{{Crenshaw} et~al.}{2010}]{Crenshaw2010}
{Crenshaw} D.~M.,  {Schmitt} H.~R.,  {Kraemer} S.~B.,  {Mushotzky} R.~F.,
  {Dunn} J.~P.,  2010, \apj, 708, 419

\bibitem[\protect\citeauthoryear{{Croton et~al.}}{{Croton
  et~al.}}{2006}]{Croton2006}
{Croton et~al.} D.~J.,  2006, \mnras, 365, 11

\bibitem[\protect\citeauthoryear{{Dalla Vecchia} \& {Schaye}}{{Dalla Vecchia}
  \& {Schaye}}{2008}]{DallaVecchia2008}
{Dalla Vecchia} C.,  {Schaye} J.,  2008, \mnras, 387, 1431

\bibitem[\protect\citeauthoryear{{Danielson}, {Lehmer}, {Alexander}, {Brandt},
  {Luo}, {Miller}, {Xue} \& {Stott}}{{Danielson} et~al.}{2012}]{Danielson2012}
{Danielson} A.~L.~R.,  {Lehmer} B.~D.,  {Alexander} D.~M.,  {Brandt} W.~N.,
  {Luo} B.,  {Miller} N.,  {Xue} Y.~Q.,    {Stott} J.~P.,  2012, \mnras,
  p.~2616

\bibitem[\protect\citeauthoryear{{Davies}}{{Davies}}{2007}]{Davies2007}
{Davies} R.~I.,  2007, \mnras, 375, 1099

\bibitem[\protect\citeauthoryear{{Debuhr}, {Quataert} \& {Ma}}{{Debuhr}
  et~al.}{2012}]{DeBuhr2012}
{Debuhr} J.,  {Quataert} E.,    {Ma} C.-P.,  2012, \mnras, 420, 2221

\bibitem[\protect\citeauthoryear{{Di Matteo}, {Springel} \& {Hernquist}}{{Di
  Matteo} et~al.}{2005}]{DiMatteo2005}
{Di Matteo} T.,  {Springel} V.,    {Hernquist} L.,  2005, \nat, 433, 604

\bibitem[\protect\citeauthoryear{{Dimitrijevi{\'c}}, {Popovi{\'c}}, {Kova{\v
  c}evi{\'c}}, {Da{\v c}i{\'c}} \& {Ili{\'c}}}{{Dimitrijevi{\'c}}
  et~al.}{2007}]{Dimitrijevi2007}
{Dimitrijevi{\'c}} M.~S.,  {Popovi{\'c}} L.~{\v C}.,  {Kova{\v c}evi{\'c}} J.,
  {Da{\v c}i{\'c}} M.,    {Ili{\'c}} D.,  2007, \mnras, 374, 1181

\bibitem[\protect\citeauthoryear{{Elvis et~al.}}{{Elvis
  et~al.}}{1994}]{Elvis1994}
{Elvis et~al.} M.,  1994, \apjs, 95, 1

\bibitem[\protect\citeauthoryear{{Engel et~al.}}{{Engel
  et~al.}}{2010}]{Engel2010}
{Engel et~al.} H.,  2010, \apj, 724, 233

\bibitem[\protect\citeauthoryear{{Faucher-Giguere} \&
  {Quataert}}{{Faucher-Giguere} \& {Quataert}}{2012}]{FaucherGiguere2012}
{Faucher-Giguere} C.-A.,  {Quataert} E.,  2012, ArXiv e-prints, 1204.2547

\bibitem[\protect\citeauthoryear{{Feruglio}, {Maiolino}, {Piconcelli}, {Menci},
  {Aussel}, {Lamastra} \& {Fiore}}{{Feruglio} et~al.}{2010}]{Feruglio2010}
{Feruglio} C.,  {Maiolino} R.,  {Piconcelli} E.,  {Menci} N.,  {Aussel} H.,
  {Lamastra} A.,    {Fiore} F.,  2010, \aap, 518, L155+

\bibitem[\protect\citeauthoryear{{Fu} \& {Stockton}}{{Fu} \&
  {Stockton}}{2009}]{Fu2009}
{Fu} H.,  {Stockton} A.,  2009, \apj, 690, 953

\bibitem[\protect\citeauthoryear{{Ganguly} \& {Brotherton}}{{Ganguly} \&
  {Brotherton}}{2008}]{Ganguly2008}
{Ganguly} R.,  {Brotherton} M.~S.,  2008, \apj, 672, 102

\bibitem[\protect\citeauthoryear{{Gerssen}, {Wilman}, {Christensen}, {Bower} \&
  {Wild}}{{Gerssen} et~al.}{2009}]{Gerssen2009}
{Gerssen} J.,  {Wilman} D.~J.,  {Christensen} L.,  {Bower} R.~G.,    {Wild} V.,
   2009, \mnras, 393, L45

\bibitem[\protect\citeauthoryear{{Gibson et~al.}}{{Gibson
  et~al.}}{2009}]{Gibson2009}
{Gibson et~al.} R.~R.,  2009, \apj, 692, 758

\bibitem[\protect\citeauthoryear{{Gofford et~al.}}{{Gofford
  et~al.}}{2011}]{Gofford2011}
{Gofford et~al.} J.,  2011, \mnras, 414, 3307

\bibitem[\protect\citeauthoryear{{Greene} \& {Ho}}{{Greene} \&
  {Ho}}{2005}]{Greene2005}
{Greene} J.~E.,  {Ho} L.~C.,  2005, \apj, 627, 721

\bibitem[\protect\citeauthoryear{{Greene}, {Zakamska}, {Ho} \&
  {Barth}}{{Greene} et~al.}{2011}]{Greene2011}
{Greene} J.~E.,  {Zakamska} N.~L.,  {Ho} L.~C.,    {Barth} A.~J.,  2011, \apj,
  732, 9

\bibitem[\protect\citeauthoryear{{Hainline}, {Blain}, {Smail}, {Alexander},
  {Armus}, {Chapman} \& {Ivison}}{{Hainline} et~al.}{2011}]{Hainline2011}
{Hainline} L.~J.,  {Blain} A.~W.,  {Smail} I.,  {Alexander} D.~M.,  {Armus} L.,
   {Chapman} S.~C.,    {Ivison} R.~J.,  2011, \apj, 740, 96

\bibitem[\protect\citeauthoryear{{Hainline}, {Blain}, {Smail}, {Frayer},
  {Chapman}, {Ivison} \& {Alexander}}{{Hainline} et~al.}{2009}]{Hainline2009}
{Hainline} L.~J.,  {Blain} A.~W.,  {Smail} I.,  {Frayer} D.~T.,  {Chapman}
  S.~C.,  {Ivison} R.~J.,    {Alexander} D.~M.,  2009, \apj, 699, 1610

\bibitem[\protect\citeauthoryear{{Heckman}, {Armus} \& {Miley}}{{Heckman}
  et~al.}{1990}]{Heckman1990}
{Heckman} T.~M.,  {Armus} L.,    {Miley} G.~K.,  1990, \apjs, 74, 833

\bibitem[\protect\citeauthoryear{{Heckman}, {Kauffmann}, {Brinchmann},
  {Charlot}, {Tremonti} \& {White}}{{Heckman} et~al.}{2004}]{Heckman2004}
{Heckman} T.~M.,  {Kauffmann} G.,  {Brinchmann} J.,  {Charlot} S.,  {Tremonti}
  C.,    {White} S.~D.~M.,  2004, \apj, 613, 109

\bibitem[\protect\citeauthoryear{{Heckman}, {Miley}, {van Breugel} \&
  {Butcher}}{{Heckman} et~al.}{1981}]{Heckman1981}
{Heckman} T.~M.,  {Miley} G.~K.,  {van Breugel} W.~J.~M.,    {Butcher} H.~R.,
  1981, \apj, 247, 403

\bibitem[\protect\citeauthoryear{{Helou}, {Soifer} \& {Rowan-Robinson}}{{Helou}
  et~al.}{1985}]{Helou1985}
{Helou} G.,  {Soifer} B.~T.,    {Rowan-Robinson} M.,  1985, \apjl, 298, L7

\bibitem[\protect\citeauthoryear{{Hickox et~al.}}{{Hickox
  et~al.}}{2012}]{Hickox2012}
{Hickox et~al.} R.~C.,  2012, \mnras, 421, 284

\bibitem[\protect\citeauthoryear{{Holt}, {Tadhunter} \& {Morganti}}{{Holt}
  et~al.}{2008}]{Holt2008}
{Holt} J.,  {Tadhunter} C.~N.,    {Morganti} R.,  2008, \mnras, 387, 639

\bibitem[\protect\citeauthoryear{{Hopkins}, {Hernquist}, {Cox}, {Di Matteo},
  {Robertson} \& {Springel}}{{Hopkins} et~al.}{2006}]{Hopkins2006}
{Hopkins} P.~F.,  {Hernquist} L.,  {Cox} T.~J.,  {Di Matteo} T.,  {Robertson}
  B.,    {Springel} V.,  2006, \apjs, 163, 1

\bibitem[\protect\citeauthoryear{{Hopkins}, {Hernquist}, {Cox} \& {Kere{\v
  s}}}{{Hopkins} et~al.}{2008}]{Hopkins2008a}
{Hopkins} P.~F.,  {Hernquist} L.,  {Cox} T.~J.,    {Kere{\v s}} D.,  2008,
  \apjs, 175, 356

\bibitem[\protect\citeauthoryear{{Hopkins}, {Hernquist}, {Martini}, {Cox},
  {Robertson}, {Di Matteo} \& {Springel}}{{Hopkins} et~al.}{2005}]{Hopkins2005}
{Hopkins} P.~F.,  {Hernquist} L.,  {Martini} P.,  {Cox} T.~J.,  {Robertson} B.,
   {Di Matteo} T.,    {Springel} V.,  2005, \apjl, 625, L71

\bibitem[\protect\citeauthoryear{{Hopkins}, {Richards} \&
  {Hernquist}}{{Hopkins} et~al.}{2007}]{Hopkins2007a}
{Hopkins} P.~F.,  {Richards} G.~T.,    {Hernquist} L.,  2007, \apj, 654, 731

\bibitem[\protect\citeauthoryear{{Ibar}, {Ivison}, {Best}, {Coppin}, {Pope},
  {Smail} \& {Dunlop}}{{Ibar} et~al.}{2010}]{Ibar2010}
{Ibar} E.,  {Ivison} R.~J.,  {Best} P.~N.,  {Coppin} K.,  {Pope} A.,  {Smail}
  I.,    {Dunlop} J.~S.,  2010, \mnras, 401, L53

\bibitem[\protect\citeauthoryear{{Ivison}, {Papadopoulos}, {Smail}, {Greve},
  {Thomson}, {Xilouris} \& {Chapman}}{{Ivison} et~al.}{2011}]{Ivison2011}
{Ivison} R.~J.,  {Papadopoulos} P.~P.,  {Smail} I.,  {Greve} T.~R.,  {Thomson}
  A.~P.,  {Xilouris} E.~M.,    {Chapman} S.~C.,  2011, \mnras, 412, 1913

\bibitem[\protect\citeauthoryear{{Ivison}, {Smail}, {Papadopoulos}, {Wold},
  {Richard}, {Swinbank}, {Kneib} \& {Owen}}{{Ivison} et~al.}{2010}]{Ivison2010}
{Ivison} R.~J.,  {Smail} I.,  {Papadopoulos} P.~P.,  {Wold} I.,  {Richard} J.,
  {Swinbank} A.~M.,  {Kneib} J.-P.,    {Owen} F.~N.,  2010, \mnras, 404, 198

\bibitem[\protect\citeauthoryear{{Ivison et~al.}}{{Ivison
  et~al.}}{2002}]{Ivison2002}
{Ivison et~al.} R.~J.,  2002, \mnras, 337, 1

\bibitem[\protect\citeauthoryear{{Ivison et~al.}}{{Ivison
  et~al.}}{2007}]{Ivison2007}
{Ivison et~al.} R.~J.,  2007, \mnras, 380, 199

\bibitem[\protect\citeauthoryear{{Kennicutt}
  Jr.}{{Kennicutt}}{1998}]{Kennicutt1998}
{Kennicutt} Jr. R.~C.,  1998, \araa, 36, 189

\bibitem[\protect\citeauthoryear{{Kewley}, {Groves}, {Kauffmann} \&
  {Heckman}}{{Kewley} et~al.}{2006}]{Kewley2006}
{Kewley} L.~J.,  {Groves} B.,  {Kauffmann} G.,    {Heckman} T.,  2006, \mnras,
  372, 961

\bibitem[\protect\citeauthoryear{{King}}{{King}}{2005}]{King2005}
{King} A.,  2005, \apjl, 635, L121

\bibitem[\protect\citeauthoryear{{King}, {Zubovas} \& {Power}}{{King}
  et~al.}{2011}]{King2011}
{King} A.~R.,  {Zubovas} K.,    {Power} C.,  2011, \mnras, 415, L6

\bibitem[\protect\citeauthoryear{{Kov{\'a}cs}, {Chapman}, {Dowell}, {Blain},
  {Ivison}, {Smail} \& {Phillips}}{{Kov{\'a}cs} et~al.}{2006}]{Kovacs2006}
{Kov{\'a}cs} A.,  {Chapman} S.~C.,  {Dowell} C.~D.,  {Blain} A.~W.,  {Ivison}
  R.~J.,  {Smail} I.,    {Phillips} T.~G.,  2006, \apj, 650, 592

\bibitem[\protect\citeauthoryear{{Le Tiran}, {Lehnert}, {van Driel}, {Nesvadba}
  \& {Di Matteo}}{{Le Tiran} et~al.}{2011}]{LeTiran2011}
{Le Tiran} L.,  {Lehnert} M.~D.,  {van Driel} W.,  {Nesvadba} N.~P.~H.,    {Di
  Matteo} P.,  2011, \aap, 534, L4

\bibitem[\protect\citeauthoryear{{Ledlow}, {Smail}, {Owen}, {Keel}, {Ivison} \&
  {Morrison}}{{Ledlow} et~al.}{2002}]{Ledlow2002}
{Ledlow} M.~J.,  {Smail} I.,  {Owen} F.~N.,  {Keel} W.~C.,  {Ivison} R.~J.,
  {Morrison} G.~E.,  2002, \apjl, 577, L79

\bibitem[\protect\citeauthoryear{{Lehnert}, {Nesvadba}, {Le Tiran}, {Di
  Matteo}, {van Driel}, {Douglas}, {Chemin} \& {Bournaud}}{{Lehnert}
  et~al.}{2009}]{Lehnert2009}
{Lehnert} M.~D.,  {Nesvadba} N.~P.~H.,  {Le Tiran} L.,  {Di Matteo} P.,  {van
  Driel} W.,  {Douglas} L.~S.,  {Chemin} L.,    {Bournaud} F.,  2009, \apj,
  699, 1660

\bibitem[\protect\citeauthoryear{{Leitherer}, {Schaerer}, {Goldader},
  {Gonz{\'a}lez Delgado}, {Robert}, {Kune}, {de Mello}, {Devost} \&
  {Heckman}}{{Leitherer} et~al.}{1999}]{Leitherer1999}
{Leitherer} C.,  {Schaerer} D.,  {Goldader} J.~D.,  {Gonz{\'a}lez Delgado}
  R.~M.,  {Robert} C.,  {Kune} D.~F.,  {de Mello} D.~F.,  {Devost} D.,
  {Heckman} T.~M.,  1999, \apjs, 123, 3

\bibitem[\protect\citeauthoryear{{Lutz}, {Maiolino}, {Spoon} \&
  {Moorwood}}{{Lutz} et~al.}{2004}]{Lutz2004}
{Lutz} D.,  {Maiolino} R.,  {Spoon} H.~W.~W.,    {Moorwood} A.~F.~M.,  2004,
  \aap, 418, 465

\bibitem[\protect\citeauthoryear{{Maiolino et~al.}}{{Maiolino
  et~al.}}{2012}]{Maiolino2012}
{Maiolino et~al.} R.,  2012, ArXiv e-prints,1204.2904

\bibitem[\protect\citeauthoryear{{Matsuoka}}{{Matsuoka}}{2012}]{Matsuoka2012}
{Matsuoka} Y.,  2012, \apj, 750, 54

\bibitem[\protect\citeauthoryear{{McCarthy}, {Schaye}, {Bower}, {Ponman},
  {Booth}, {Dalla Vecchia} \& {Springel}}{{McCarthy}
  et~al.}{2011}]{McCarthy2011}
{McCarthy} I.~G.,  {Schaye} J.,  {Bower} R.~G.,  {Ponman} T.~J.,  {Booth}
  C.~M.,  {Dalla Vecchia} C.,    {Springel} V.,  2011, \mnras, 412, 1965

\bibitem[\protect\citeauthoryear{{Men{\'e}ndez-Delmestre
  et~al.}}{{Men{\'e}ndez-Delmestre et~al.}}{2009}]{MenendezDelmestre2009}
{Men{\'e}ndez-Delmestre et~al.} K.,  2009, \apj, 699, 667

\bibitem[\protect\citeauthoryear{{Men{\'e}ndez-Delmestre
  et~al.}}{{Men{\'e}ndez-Delmestre et~al.}}{2012}]{MenendezDelmestre2012}
{Men{\'e}ndez-Delmestre et~al.} K.,  2012, ApJ submitted

\bibitem[\protect\citeauthoryear{{Miller}, {Fomalont}, {Kellermann},
  {Mainieri}, {Norman}, {Padovani}, {Rosati} \& {Tozzi}}{{Miller}
  et~al.}{2008}]{Miller2008}
{Miller} N.~A.,  {Fomalont} E.~B.,  {Kellermann} K.~I.,  {Mainieri} V.,
  {Norman} C.,  {Padovani} P.,  {Rosati} P.,    {Tozzi} P.,  2008, \apjs, 179,
  114

\bibitem[\protect\citeauthoryear{{Morrison}, {Owen}, {Dickinson}, {Ivison} \&
  {Ibar}}{{Morrison} et~al.}{2010}]{Morrison2010}
{Morrison} G.~E.,  {Owen} F.~N.,  {Dickinson} M.,  {Ivison} R.~J.,    {Ibar}
  E.,  2010, \apjs, 188, 178

\bibitem[\protect\citeauthoryear{{Mullaney}, {Alexander}, {Goulding} \&
  {Hickox}}{{Mullaney} et~al.}{2011}]{Mullaney2011}
{Mullaney} J.~R.,  {Alexander} D.~M.,  {Goulding} A.~D.,    {Hickox} R.~C.,
  2011, \mnras, 414, 1082

\bibitem[\protect\citeauthoryear{{Navarro}, {Frenk} \& {White}}{{Navarro}
  et~al.}{1996}]{Navarro1996}
{Navarro} J.~F.,  {Frenk} C.~S.,    {White} S.~D.~M.,  1996, \apj, 462, 563

\bibitem[\protect\citeauthoryear{{Nesvadba}, {Lehnert}, {De Breuck}, {Gilbert}
  \& {van Breugel}}{{Nesvadba} et~al.}{2007}]{Nesvadba2007}
{Nesvadba} N.~P.~H.,  {Lehnert} M.~D.,  {De Breuck} C.,  {Gilbert} A.,    {van
  Breugel} W.,  2007, \aap, 475, 145

\bibitem[\protect\citeauthoryear{{Nesvadba}, {Lehnert}, {De Breuck}, {Gilbert}
  \& {van Breugel}}{{Nesvadba} et~al.}{2008}]{Nesvadba2008}
{Nesvadba} N.~P.~H.,  {Lehnert} M.~D.,  {De Breuck} C.,  {Gilbert} A.~M.,
  {van Breugel} W.,  2008, \aap, 491, 407

\bibitem[\protect\citeauthoryear{{Nesvadba}, {Lehnert}, {Eisenhauer},
  {Gilbert}, {Tecza} \& {Abuter}}{{Nesvadba} et~al.}{2006}]{Nesvadba2006}
{Nesvadba} N.~P.~H.,  {Lehnert} M.~D.,  {Eisenhauer} F.,  {Gilbert} A.,
  {Tecza} M.,    {Abuter} R.,  2006, \apj, 650, 693

\bibitem[\protect\citeauthoryear{{Nesvadba}, {Polletta}, {Lehnert}, {Bergeron},
  {De Breuck}, {Lagache} \& {Omont}}{{Nesvadba} et~al.}{2011}]{Nesvadba2011}
{Nesvadba} N.~P.~H.,  {Polletta} M.,  {Lehnert} M.~D.,  {Bergeron} J.,  {De
  Breuck} C.,  {Lagache} G.,    {Omont} A.,  2011, \mnras, 415, 2359

\bibitem[\protect\citeauthoryear{{Nesvadba et~al.}}{{Nesvadba
  et~al.}}{2007}]{Nesvadba2007b}
{Nesvadba et~al.} N.~P.~H.,  2007, \apj, 657, 725

\bibitem[\protect\citeauthoryear{{Osterbrock}}{{Osterbrock}}{1989}]{Osterbrock1989}
{Osterbrock} D.~E.,  1989, {Astrophysics of gaseous nebulae and active galactic
  nuclei}.
University Science Books

\bibitem[\protect\citeauthoryear{{Page}, {Carrera}, {Stevens}, {Ebrero} \&
  {Blustin}}{{Page} et~al.}{2011}]{Page2011}
{Page} M.~J.,  {Carrera} F.~J.,  {Stevens} J.~A.,  {Ebrero} J.,    {Blustin}
  A.~J.,  2011, \mnras, 416, 2792

\bibitem[\protect\citeauthoryear{{Rafferty}, {Brandt}, {Alexander}, {Xue},
  {Bauer}, {Lehmer}, {Luo} \& {Papovich}}{{Rafferty}
  et~al.}{2011}]{Rafferty2011}
{Rafferty} D.~A.,  {Brandt} W.~N.,  {Alexander} D.~M.,  {Xue} Y.~Q.,  {Bauer}
  F.~E.,  {Lehmer} B.~D.,  {Luo} B.,    {Papovich} C.,  2011, \apj, 742, 3

\bibitem[\protect\citeauthoryear{{Reeves}, {O'Brien} \& {Ward}}{{Reeves}
  et~al.}{2003}]{Reeves2003}
{Reeves} J.~N.,  {O'Brien} P.~T.,    {Ward} M.~J.,  2003, \apjl, 593, L65

\bibitem[\protect\citeauthoryear{{Richards et~al.}}{{Richards
  et~al.}}{2006}]{Richards2006}
{Richards et~al.} G.~T.,  2006, \apjs, 166, 470

\bibitem[\protect\citeauthoryear{{Riechers}, {Hodge}, {Walter}, {Carilli} \&
  {Bertoldi}}{{Riechers} et~al.}{2011}]{Riechers2011}
{Riechers} D.~A.,  {Hodge} J.,  {Walter} F.,  {Carilli} C.~L.,    {Bertoldi}
  F.,  2011, \apjl, 739, L31

\bibitem[\protect\citeauthoryear{{Robson}}{{Robson}}{1996}]{Robson1996}
{Robson} I.,  1996, {Active galactic nuclei}.
Wiley

\bibitem[\protect\citeauthoryear{{Rupke}, {Veilleux} \& {Sanders}}{{Rupke}
  et~al.}{2005a}]{Rupke2005a}
{Rupke} D.~S.,  {Veilleux} S.,    {Sanders} D.~B.,  2005a, \apjs, 160, 87

\bibitem[\protect\citeauthoryear{{Rupke}, {Veilleux} \& {Sanders}}{{Rupke}
  et~al.}{2005b}]{Rupke2005b}
{Rupke} D.~S.,  {Veilleux} S.,    {Sanders} D.~B.,  2005b, \apjs, 160, 115

\bibitem[\protect\citeauthoryear{{Rupke} \& {Veilleux}}{{Rupke} \&
  {Veilleux}}{2011}]{Rupke2011}
{Rupke} D.~S.~N.,  {Veilleux} S.,  2011, \apjl, 729, L27+

\bibitem[\protect\citeauthoryear{{Sanders}, {Soifer}, {Elias}, {Madore},
  {Matthews}, {Neugebauer} \& {Scoville}}{{Sanders} et~al.}{1988}]{Sanders1988}
{Sanders} D.~B.,  {Soifer} B.~T.,  {Elias} J.~H.,  {Madore} B.~F.,  {Matthews}
  K.,  {Neugebauer} G.,    {Scoville} N.~Z.,  1988, \apj, 325, 74

\bibitem[\protect\citeauthoryear{{Scott et~al.}}{{Scott
  et~al.}}{2002}]{Scott2002}
{Scott et~al.} S.~E.,  2002, \mnras, 331, 817

\bibitem[\protect\citeauthoryear{{Silk} \& {Rees}}{{Silk} \&
  {Rees}}{1998}]{Silk1998}
{Silk} J.,  {Rees} M.~J.,  1998, \aap, 331, L1

\bibitem[\protect\citeauthoryear{{Silverman et~al.}}{{Silverman
  et~al.}}{2008}]{Silverman2008}
{Silverman et~al.} J.~D.,  2008, \apj, 679, 118

\bibitem[\protect\citeauthoryear{{Simpson et~al.}}{{Simpson
  et~al.}}{2006}]{Simpson2006}
{Simpson et~al.} C.,  2006, \mnras, 372, 741

\bibitem[\protect\citeauthoryear{{Simpson et~al.}}{{Simpson
  et~al.}}{2012}]{Simpson2012}
{Simpson et~al.} C.,  2012, \mnras, 421, 3060

\bibitem[\protect\citeauthoryear{{Smail}, {Chapman}, {Blain} \&
  {Ivison}}{{Smail} et~al.}{2004}]{Smail2004}
{Smail} I.,  {Chapman} S.~C.,  {Blain} A.~W.,    {Ivison} R.~J.,  2004, \apj,
  616, 71

\bibitem[\protect\citeauthoryear{{Smail}, {Chapman}, {Ivison}, {Blain},
  {Takata}, {Heckman}, {Dunlop} \& {Sekiguchi}}{{Smail}
  et~al.}{2003}]{Smail2003}
{Smail} I.,  {Chapman} S.~C.,  {Ivison} R.~J.,  {Blain} A.~W.,  {Takata} T.,
  {Heckman} T.~M.,  {Dunlop} J.~S.,    {Sekiguchi} K.,  2003, \mnras, 342, 1185

\bibitem[\protect\citeauthoryear{{Smol{\v c}i{\'c} et~al.}}{{Smol{\v c}i{\'c}
  et~al.}}{2009}]{Smolcic2009}
{Smol{\v c}i{\'c} et~al.} V.,  2009, \apj, 696, 24

\bibitem[\protect\citeauthoryear{{Springel}, {Di Matteo} \&
  {Hernquist}}{{Springel} et~al.}{2005}]{Springel2005}
{Springel} V.,  {Di Matteo} T.,    {Hernquist} L.,  2005, \mnras, 361, 776

\bibitem[\protect\citeauthoryear{{Swinbank}, {Chapman}, {Smail}, {Lindner},
  {Borys}, {Blain}, {Ivison} \& {Lewis}}{{Swinbank}
  et~al.}{2006}]{Swinbank2006}
{Swinbank} A.~M.,  {Chapman} S.~C.,  {Smail} I.,  {Lindner} C.,  {Borys} C.,
  {Blain} A.~W.,  {Ivison} R.~J.,    {Lewis} G.~F.,  2006, \mnras, 371, 465

\bibitem[\protect\citeauthoryear{{Swinbank}, {Smail}, {Chapman}, {Blain},
  {Ivison} \& {Keel}}{{Swinbank} et~al.}{2004}]{Swinbank2004}
{Swinbank} A.~M.,  {Smail} I.,  {Chapman} S.~C.,  {Blain} A.~W.,  {Ivison}
  R.~J.,    {Keel} W.~C.,  2004, \apj, 617, 64

\bibitem[\protect\citeauthoryear{{Swinbank et~al.}}{{Swinbank
  et~al.}}{2005}]{Swinbank2005}
{Swinbank et~al.} A.~M.,  2005, \mnras, 359, 401

\bibitem[\protect\citeauthoryear{{Tacconi et~al.}}{{Tacconi
  et~al.}}{2008}]{Tacconi2008}
{Tacconi et~al.} L.~J.,  2008, \apj, 680, 246

\bibitem[\protect\citeauthoryear{{Takata}, {Sekiguchi}, {Smail}, {Chapman},
  {Geach}, {Swinbank}, {Blain} \& {Ivison}}{{Takata} et~al.}{2006}]{Takata2006}
{Takata} T.,  {Sekiguchi} K.,  {Smail} I.,  {Chapman} S.~C.,  {Geach} J.~E.,
  {Swinbank} A.~M.,  {Blain} A.,    {Ivison} R.~J.,  2006, \apj, 651, 713

\bibitem[\protect\citeauthoryear{{Tamura}, {Kohno}, {Nakanishi}, {Hatsukade},
  {Iono}, {Wilson}, {Yun} \& {Takata}}{{Tamura} et~al.}{2009}]{Tamura2009}
{Tamura} Y.,  {Kohno} K.,  {Nakanishi} K.,  {Hatsukade} B.,  {Iono} D.,
  {Wilson} G.~W.,  {Yun} M.~S.,    {Takata} T.,  2009, \nat, 459, 61

\bibitem[\protect\citeauthoryear{{Tombesi}, {Cappi}, {Reeves} \&
  {Braito}}{{Tombesi} et~al.}{2012}]{Tombesi2012}
{Tombesi} F.,  {Cappi} M.,  {Reeves} J.~N.,    {Braito} V.,  2012, \mnras, 422,
  L1

\bibitem[\protect\citeauthoryear{{Trump et~al.}}{{Trump
  et~al.}}{2006}]{Trump2006}
{Trump et~al.} J.~R.,  2006, \apjs, 165, 1

\bibitem[\protect\citeauthoryear{{Valiante}, {Lutz}, {Sturm}, {Genzel},
  {Tacconi}, {Lehnert} \& {Baker}}{{Valiante} et~al.}{2007}]{Valiante2007}
{Valiante} E.,  {Lutz} D.,  {Sturm} E.,  {Genzel} R.,  {Tacconi} L.~J.,
  {Lehnert} M.~D.,    {Baker} A.~J.,  2007, \apj, 660, 1060

\bibitem[\protect\citeauthoryear{{van de Voort}, {Schaye}, {Booth} \& {Dalla
  Vecchia}}{{van de Voort} et~al.}{2011}]{vanDeVoort2011}
{van de Voort} F.,  {Schaye} J.,  {Booth} C.~M.,    {Dalla Vecchia} C.,  2011,
  \mnras, 415, 2782

\bibitem[\protect\citeauthoryear{{Veilleux}, {Cecil} \&
  {Bland-Hawthorn}}{{Veilleux} et~al.}{2005}]{Veilleux2005}
{Veilleux} S.,  {Cecil} G.,    {Bland-Hawthorn} J.,  2005, \araa, 43, 769

\bibitem[\protect\citeauthoryear{{Veilleux}, {Kim} \& {Sanders}}{{Veilleux}
  et~al.}{1999}]{Veilleux1999}
{Veilleux} S.,  {Kim} D.-C.,    {Sanders} D.~B.,  1999, \apj, 522, 113

\bibitem[\protect\citeauthoryear{{Villar-Mart{\'{\i}}n}, {Humphrey}, {Delgado},
  {Colina} \& {Arribas}}{{Villar-Mart{\'{\i}}n}
  et~al.}{2011}]{VillarMartin2011b}
{Villar-Mart{\'{\i}}n} M.,  {Humphrey} A.,  {Delgado} R.~G.,  {Colina} L.,
  {Arribas} S.,  2011, \mnras, 418, 2032

\bibitem[\protect\citeauthoryear{{Villar-Mart{\'{\i}}n}, {Tadhunter},
  {Humphrey}, {Encina}, {Delgado}, {Torres} \&
  {Mart{\'{\i}}nez-Sansigre}}{{Villar-Mart{\'{\i}}n}
  et~al.}{2011}]{VillarMartin2011a}
{Villar-Mart{\'{\i}}n} M.,  {Tadhunter} C.,  {Humphrey} A.,  {Encina} R.~F.,
  {Delgado} R.~G.,  {Torres} M.~P.,    {Mart{\'{\i}}nez-Sansigre} A.,  2011,
  \mnras, 416, 262

\bibitem[\protect\citeauthoryear{{Wardlow et~al.}}{{Wardlow
  et~al.}}{2011}]{Wardlow2011}
{Wardlow et~al.} J.~L.,  2011, \mnras, 415, 1479

\bibitem[\protect\citeauthoryear{{Webb et~al.}}{{Webb et~al.}}{2003}]{Webb2003}
{Webb et~al.} T.~M.,  2003, \apj, 587, 41

\bibitem[\protect\citeauthoryear{{Westmoquette}, {Clements}, {Bendo} \&
  {Khan}}{{Westmoquette} et~al.}{2012}]{Westmoquette2012}
{Westmoquette} M.,  {Clements} D.,  {Bendo} G.,    {Khan} S.,  2012, ArXiv
  e-prints, 1205.0203

\bibitem[\protect\citeauthoryear{{Willott}, {Rawlings}, {Blundell} \&
  {Lacy}}{{Willott} et~al.}{1998}]{Willott1998}
{Willott} C.~J.,  {Rawlings} S.,  {Blundell} K.~M.,    {Lacy} M.,  1998,
  \mnras, 300, 625

\bibitem[\protect\citeauthoryear{{Zheng}, {Xia}, {Mao}, {Wu} \& {Deng}}{{Zheng}
  et~al.}{2002}]{Zheng2002}
{Zheng} X.~Z.,  {Xia} X.~Y.,  {Mao} S.,  {Wu} H.,    {Deng} Z.~G.,  2002, \aj,
  124, 18

\bibitem[\protect\citeauthoryear{{Zubovas} \& {King}}{{Zubovas} \&
  {King}}{2012}]{Zubovas2012}
{Zubovas} K.,  {King} A.,  2012, \apjl, 745, L34

\end{thebibliography}
\end{document}